\documentclass[final]{article}

\usepackage{textcomp}
\usepackage{graphicx}
\usepackage{soul}
\usepackage{a4wide}
\usepackage{graphicx}
\usepackage{authblk}
\usepackage{amssymb}
\usepackage{epstopdf}
\usepackage[sans]{dsfont}
\usepackage[applemac]{inputenc}
\usepackage[english]{babel}
\usepackage{latexsym}
\usepackage{subfigure}
\usepackage{color}
\usepackage{float}
\usepackage{amsmath}
\usepackage{amsfonts}
\numberwithin{equation}{section}
\usepackage{amsthm}
\usepackage[numbers]{natbib}
\usepackage[bookmarks=true,colorlinks=true,linkcolor={blue},urlcolor={blue}, citecolor={blue},pdfstartview={XYZ null null 1.22}]{hyperref}%

\makeindex             
\newcommand{\be}{\begin{equation}}
\newcommand{\ee}{\end{equation}}

\newcommand{\epsi}{\varepsilon}
\newcommand{\ba}{\begin{array}}
\newcommand{\ea}{\end{array}}
\newcommand{\pM}{\left(\begin{array}}
\newcommand{\Mp}{\end{array}\right)}
\newcommand{\RR}{ {\bf R}}

\newcommand{\R}{\mathbb{R}}

\def\reall{\mathfrak{I}}
\def\bqa{\begin{eqnarray}}
\def\eqa{\end{eqnarray}}

\def\N{\mathbb{N}}
\def\R{\mathbb{R}}

\def\pw{w{'}}
\def\pv{w{'}_*}
\def\v{w_*}
\def\knu{\kappa}

\def\Ball{ {\mathcal {B}}}

\def\RR{\mathbb R}

\def\cm{c_{\textrm{max}}}
\def\ps{\p_{\infty}}
\def\Ur{\textcolor{black}{V_r}}
\def\Ua{\textcolor{black}{V_a}}
\def\p{\rho}
\def\gd{U}
\def\N{\mathcal N}
\def\L{\mathcal L}

\def\tw{v}
\def\tv{v_*}
\def\ptw{v'}
\def\ptv{v'_*}
\def\ppw{w''}
\def\pptw{v''}

\definecolor{lorange}{rgb}{1.0, 0.31, 0.0}
\definecolor{mauve}{rgb}{0.88, 0.69, 1.0}

\newtheorem{thm}{Theorem}

\newtheorem{proposition}[thm]{Proposition} 
\newtheorem{remark}[thm]{Remark} 
\newtheorem{alg}{Algorithm}

\begin{document}
\title{Recent advances in opinion modeling:\\
 control and social influence}
\author{Giacomo Albi\thanks{TUM, Boltzmannstra\ss e 3, Garching (M\"unchen), Germany ({\tt giacomo.albi@ma.tum.de})}, Lorenzo Pareschi \thanks{University of Ferrara, via Machiavelli 35, Ferrara, Italy ({\tt lorenzo.pareschi@unife.it})}, Giuseppe Toscani \thanks{University of Pavia, via Ferrata 1, Pavia, Italy ({\tt giuseppe.toscani@unipv.it})}, Mattia Zanella \thanks{University of Ferrara,  via Machiavelli 35, Ferrara, Italy ({\tt mattia.zanella@unife.it})}}
%
\maketitle

\abstract{We survey some recent developments on the mathematical modeling of opinion dynamics. After an introduction on opinion modeling through interacting multi-agent systems described by partial differential equations of kinetic type, we focus our attention on two major advancements: optimal control of opinion formation and influence of additional social aspects, like conviction and number of connections in social networks, which modify the agents' role in the opinion exchange process.
}

\vskip 1cm

\section{Preliminaries}
We introduce some of the essential literature on the opinion formation, by emphasizing the role of the kinetic description. New problems recently treated in the scientific community are outlined. Then, the  mathematical description of the core ideas of kinetic models for opinion formation are presented in details.

\subsection{Introduction}
The statistical physics approach to social phenomena is currently attracting much interest, as indicated by the huge  and rapidly increasing number of papers and monographies based on it \cite{BAMT, CFL, NPT, PTa}.  In this rapidly increasing field of research, because of its pervasiveness in everyday life, the process  of opinion formation is nowadays one of the most studied application of mathematics to social dynamics \cite{BN, BA, BS2, DW, HK, T}. 

Along this survey, we focus on some recent advances in opinion formation modeling, which aims in coupling the process of opinion exchange with other aspects, which are closely related to the process itself, and takes into account the dependence on new variables which are usually neglected, mainly in reason of the mathematical difficulties that the introduction of further dimensions add to the models. 

These new aspects are deeply connected and range from opinion leadership and opinion control, to the role of conviction and the interplay between complex networks and the spreading of opinions. In fact, leaders are recognized to be important since they can exercise control over public opinion. It is a concept that goes back to Lazarsfeld et al. \cite{Laz}. In the course of their study of the presidential elections in the USA in 1940, it was found interpersonal communication to be much more influential than direct media effects. In \cite{Laz}  a theory of a two-step flow of communication was formulated, where so-called opinion leaders who are active media users select, interpret, modify, facilitate and finally transmit information from the media to less active parts of the population.  It is clear that various principal questions arise, mainly linked to this two-step flow of communication. The first one is related to the ability to effectively exercise a control on opinion and to the impact of modern communication systems, like social networks, to the dynamics of opinions. The second is related to the fact that the less active part of the population is in general adapting to leaders opinion only partially. Indeed, conviction plays a major role in this process, by acting as a measurable resistance to the change of opinion.

These enhancements will be modeled using the toolbox of classical kinetic theory \cite{PTa}. Within this choice, one will be able to present an almost uniform picture of opinion dynamics, starting from few simple rules.
 The kinetic model of reference was introduced by one of the authors in 2006 \cite{T}, and was subsequently generalized in many ways (see \cite{NPT,PTa} for recent surveys). 
The building block of kinetic models are pairwise interactions. In classical opinion formation, interactions among agents are usually described in terms of few relevant concepts, represented respectively by compromise and self-thinking. Once fixed in binary interactions, the microscopic rules are responsible of the formation of coherent structures. 

The remarkably simple compromise
process describes mathematically the way in which pairs of agents reach a fair compromise after exchanging opinions. The rule of compromise has been intensively studied  \cite{BN,BNKR,BNKVR,DAWF,HK,WDA}. The second one is the self-thinking process, which allows individual agents to change their opinions in an unpredictable way. It is usually mathematically described in terms of some random variable \cite{BN,T}. The resulting kinetic models are sufficiently general to take into
account a large variety of human behaviors, and to reproduce in many cases explicit steady profiles, from which one can easily elaborate information on the opinion behavior \cite{ANP,BS1,BS2,BS3,BMS}. 
For the sake of completeness, let us mention that many other models with analogous properties have been introduced and studied so far \cite{BS,Bis,BA,CdCT,GGS,GSGLB,GZ,LCCC,PVZ,Sen,SWS,VKR,W}.

Kinetic models have been also the basis for suitable generalizations, in which the presence  of leaders and their effect of the opinion dynamics has been taken into account \cite{AP1,BD,CP1,CP2,CKF,DMPW,DW}. Also, the possibility to establish an effective control on opinion, both through an external media or through the leaders' action, has attracted the interest of the research community \cite{AHP,APZa,BFY}. 
The methods here are strongly connected to analogous studies in crowd dynamics and flocking phenomena \cite{APb,APZb,BW,CFPT,FS}. 

Further, the effect of conviction in the formation of opinion started to be studied. While in general conviction is assumed to appear as a {static} parameter in the opinion dynamics \cite{C,CA,MT1,WDA}, in \cite{BrTo} conviction has been assumed to follow a proper evolution in the society on the basis of interactions with an external background. Recently, a similar approach has been used to model the effect of competence and the so-called equality bias phenomena \cite{PVZ}.  
This point of view was previously applied to the study of the formation of knowledge in \cite{PTb}, as a starting point to investigate its role in wealth distribution. Indeed, many of the aforementioned models share a common point of view with the statistical approach to distribution of wealth \cite{ChaCha,CPT,NPT,PTb}.  

More recently, in reason of their increasing relevance in modern societies, the statistical mechanics of opinion formation started to be applied to extract information from complex social networks \cite{Ace,AB,ASBS,BA,BAJ,HR1,HR2,Patt,WS}. In these models the number of connections of the agents play a major role in characterizing the dynamic \cite{APZc,APZd,Das,DL,HLL}. In \cite{APZc} the model links the graph evolution modeled by a discrete connection distribution dynamic with the spreading of opinion along the network.

Before starting our survey, it is essential to outline the peculiar aspects od the microscopic details of the binary interactions which express the microscopic change of opinion. Indeed, these interactions differ in many aspects from the usual  binary
interactions considered in classical kinetic theory of rarefied gases.
The first difference is that opinion is usually
identified with a continuum variable which can take values in a bounded interval. 
Second, the post-interaction opinions are not a linear transformation of
the pre-interaction ones. Indeed, it is realistic to assume that people
with a neutral opinion is more willing to change it, while the opposite
phenomenon happens with people which have extremal opinions. 

Once the details of the pairwise interactions are fixed, the explicit form of the bilinear kinetic equation of Boltzmann type follows \cite{PTa}. One of the main consequences of the kinetic description is that it constitute a powerful starting point to obtain, in view of standard asymptotic techniques, continuous mean-field models  with a reduced complexity, which maintain most of the physical properties of the underlying Boltzmann equation. The main idea, is to consider important only interactions which are \emph{grazing}, namely interactions in which the opinion variable does not change in a sensible way, while at the same time the frequency of the interactions is assumed to increase. This asymptotic limit (hereafter called quasi-invariant opinion limit) leads to partial differential
equations of Fokker-Planck type for the distribution of opinion among individuals, that in many cases allow for an analytic study. 

The rest of the survey is organized as follows. In the remaining part of Section 1 we describe the basic kinetic model for opinion formation. It represents the building block for the binary opinion dynamic which is used in the subsequent Sections.  
Next in Section 2 we deal with control problems for opinion dynamics. First by considering an external action which forces the agents towards a desired state and subsequently by introducing a leaders' population which acts accordingly to a prescribed optimal strategy. Here we start from the optimal control problem for the corresponding microscopic dynamic and approximate it through a finite time horizon or model predictive control technique. This permits to embed the feedback control directly into the limiting kinetic equations. Section 3 is then devoted to the modeling through multivariate distribution functions where the agents' opinion is coupled with additional variables. Specifically we consider the case where conviction is also an evolving quantity playing a role in the dynamic and the case where agents interact over an evolving social network accordingly to their number of connections. Some final remarks are contained in the last Section and details on numerical methods are given in a separate Appendix.

%
%
%

\subsection{Kinetic modelling}\label{standard}

On the basis of statistical mechanics, to construct a model for opinion formation the fundamental assumption is that agents are indistinguishable \cite{PTa}. An
agent's \emph{state} at any instant of time $t\geq0$ is completely characterized by
his opinion $w \in [-1,1]$ , where
$-1$ and $1$ denote  two (extreme) opposite opinions. 

The unknown is the density (or distribution function) $f = f(w, t)$,
where $w \in I = [-1, 1]$ and the time $t \ge 0$, whose time evolution is
described, as shown later, by a kinetic equation of Boltzmann type.

The precise meaning of the density $f$ is the following. Given the population to
study, if the opinions are defined on a sub-domain $D \subset \reall$ , the integral
 \[
 \int_D f(w,t)\, dw
 \]
represents the \emph{number} of individuals with opinion included in $D$ at time $t>
0$. It is assumed that the density function is normalized to $1$, that is
 \[
 \int_I f(w,t) \, dw = 1.
 \]
As always happens when dealing with a kinetic problem in which the variable belongs to
a bounded domain, this choice introduces supplementary mathematical difficulties in
the correct definition of binary interactions. In fact, it is essential to consider
only interactions that do not produce opinions outside the allowed interval, which
corresponds to imposing that the extreme opinions cannot be crossed. This crucial
limitation emphasizes the difference between the present {\em social} interactions,
where not all outcomes are permitted, and the classical interactions between
molecules, or, more generally,  the wealth trades (cf. \cite{PTa}, Chapter 5), where
the only limitation for trades was to insure that the post-collision wealths had to be
non-negative.

In order to build a realistic model, this severe limitation has to be coupled
with a reasonable physical interpretation of the process of opinion forming. In other
words, the impossibility of crossing the boundaries has to be a by-product of  good
modeling of binary interactions.

From a microscopic viewpoint, the binary interactions in \cite{T} were described by
the rules
\begin{equation}\begin{aligned}\label{ch6:trade_rule}
& \pw =  w - \eta P(w,w_*)( w- \v) + \xi D(w), \\
&\null \\[-.25cm]
 & \pv   =  \v - \eta P(\v,w)( \v-w) +  \xi_* D(w_*) .
\end{aligned}\end{equation}

In \eqref{ch6:trade_rule}, the pair $(w,\v)$, with $w, \v \in I$, denotes the
opinions of two arbitrary individuals before the interaction, and $(\pw,\pv)$ their
opinions after exchanging information between each other and with the exterior. The
coefficient $\eta\in (0,1/2)$ is a given constant, while $\xi$ and $\xi_*$
are random variables with the same distribution, with zero mean and variance
$\varsigma^2$, taking values on a set $\Ball \subseteq \R$.  The constant $\eta$ and
the variance $\varsigma^2$ measure respectively the compromise propensity and the degree
of spreading of opinion due to diffusion, which describes possible changes of opinion
due to personal access to information (self-thinking). Finally, the functions
$P(\cdot,\cdot)$ and $D(\cdot)$ take into account the local relevance of compromise and
diffusion for given opinions. 


Let us describe in detail the interaction on the right-hand side of
\eqref{ch6:trade_rule}. The first part is related to the compromise  propensity of the
agents,  and the last contains the diffusion effects due to individual deviations from
the average behavior.  The presence of both the functions $P(\cdot,\cdot)$ and $D(\cdot)$ is
linked to the hypothesis that openness to change of opinion is linked to the opinion
itself, and decreases as one gets closer to extremal opinions. This corresponds to the
natural idea that extreme opinions are more difficult to change. Various realizations
of these functions can be found in \cite{PTa,T}. In all cases, however, we assume that
both $P(w,\v)$ and $D(w)$ are non-increasing with respect to $|w|$, and in addition $0 \le
P(w,\v) \le 1$, $0 \le D(w) \le 1$. Typical examples are given by $P(w,w_*) = 1-|w|$ and $D(w) =\sqrt{1-w^2}$.

In the absence of the diffusion contribution ($\xi, \xi_*\equiv 0$),
\eqref{ch6:trade_rule} implies
  \begin{equation}\begin{aligned}\label{ch6:tr+-}
 & \pw+\pv = w+\v  + \eta (w-\v) \left( P(w,\v) - P(\v,w) \right),\\
\\[-.25cm]
 & \pw-\pv = \left(1- \eta( P(w,\v) + P(\v,w))\right)(w-\v).
 \end{aligned}\end{equation}
Thus, unless the function $P(\cdot,\cdot)$ is assumed constant, $P=1$, the  {\em
mean opinion} is not conserved and it can increase or decrease depending on the opinions
before the interaction. If $P(\cdot,\cdot)$ is assumed constant, the conservation law is
reminiscent of analogous conservations which take place in kinetic theory. In such a
situation, thanks to the upper bound on the coefficient $\eta$, equations
(\ref{ch6:trade_rule}) correspond to a granular-gas-like interaction \cite{PTa} where
the stationary state is a Dirac delta centered on the average opinion. This behavior
is a consequence of the fact that, in a single interaction, the compromise propensity
implies that the difference of opinion is diminishing, with $|\pw-\pv| =
(1-2\eta)|w-\v|$. Thus, all agents  in the society will end up with exactly the same
opinion. 

We remark, moreover, that, in the absence of diffusion, the lateral bounds are not
violated, since
 \begin{equation}\begin{aligned}\label{ch6:gran-rule}
\pw & =  & (1 - \eta P(w,\v))w + \eta P(w,\v) \v, \\
\\[-.25cm]
\pv  & =&   (1 - \eta P(\v,w) ) \v + \eta P(\v,w)w,
 \end{aligned}\end{equation}
 imply
 \[
 \max\left\{|\pw|, |\pv|\right\} \le  \max\left\{|w|,
 |\v|\right\}.
 \]

Let  $f(w,t)$ denote the distribution of opinion $w \in I$ at time $t \ge 0$. The
time evolution of $f$ is recovered as a balance between bilinear gain and loss of
opinion terms, described in weak form by the integro-differential equation of
Boltzmann type
\begin{equation}\begin{aligned}\label{ch6:weak boltz}
&\frac d{dt}\int_{I} \varphi(w)f(w,t)\,dv  = (Q(f,f),\varphi) =  \\
 &\qquad\qquad\lambda\left\langle \int_{I^2}   ( \varphi(\pw)+ \varphi(\pv)-\varphi(w)-\varphi(\v))  f(w)
f(\v)d w d \v \right\rangle,
\end{aligned}\end{equation}
where $(\pw,\pv)$ are the post-interaction opinions generated by the pair $(w,\v)$ in
\eqref{ch6:trade_rule}, $\lambda$ represents a constant rate of interaction and the brackets $\langle \cdot \rangle$ denote the expectation with respect to the random variables $\xi$ and $\xi_*$.

Equation \eqref{ch6:weak boltz} is consistent with the fact that a
suitable choice of the function $D(\cdot)$ in \eqref{ch6:trade_rule} coupled with a
small support $\Ball$ of the random variables implies that both $|\pw| \le 1$ and
$|\pv| \le 1$. 
We do not insist here on further details on the evolution properties of the solution, by leaving them to the next Sections, where these properties are studied for the particular problems.

\section{Optimal control of consensus}\label{sec2}
Different to the classical approach where individuals are assumed to freely interact and exchange opinions with each other, here we are particularly interested in such problems in a constrained setting. We consider feedback type controls for the resulting process and present kinetic models 
including those controls. This can be used to study the influence on the system dynamics to enforce emergence of non spontaneous desired asymptotic states. 

Two relevant situations will be explored, first a distributed control, which models the action of an external force acting as a {\em policy maker}, like the effects of the media \cite{AHP}, next an indirect internal control, where we assume that the control corresponds to the strategies of {\em opinion leaders}, aiming to influence the consensus of the whole population \cite{APZa}.
In order to characterize the kinetic structure of the optimal control of consensus dynamics, we will start  to derive it as a feedback control from a general optimal control problem for the corresponding microscopic model, and thereafter we will connect it to the binary dynamics.

\subsection{Control by an external action}
\label{sec:AHP}
We consider the microscopic evolution of the opinions of $N$ agents, where  each agent's opinion $w_i \in I$, $I=[-1,1]$, $i=1,\ldots,N$ evolves according the following first order dynamical system
\begin{equation}\label{eq:pbm}
\begin{aligned}
&\dot{w}_i=\frac{1}{N}\sum_{j=1}^{N}P(w_i,w_j)(w_j-w_i) + u,\qquad\qquad\qquad w_i(0) = w_{0,i},
\end{aligned}
\end{equation}
where  $P(\cdot,\cdot)$ has again the role of the compromise function defined in \eqref{ch6:trade_rule}. The control $u=u(t)$ models the action of an external agent, e.g. a {\em policy maker} or {social media}. We assume that it is the solution of the following optimal control problem
\begin{equation}\label{eq:pbc}
\begin{aligned}
&u=\arg\min_{u\in\mathcal{U}} J (u):=\frac{1}{2}\int_0^T\frac{1}{N}\sum_{j=1}^N\left((w_j-w_d)^2+ \knu u^2\right)ds,\quad u(t)\in [u_L,u_R],
\end{aligned}
\end{equation}
with  $\mathcal{U}$ the space of the admissible controls. 
In formulation \eqref{eq:pbc} a  quadratic cost functionals  with a penalization parameter  $\knu>0$ is considered, and the value $w_d$ represents the desired opinion state. We refer to \cite{ABCK,AHP,CFPT, FS} for further discussion on the analytical and numerical studies on this class of problems. The additional constraints on the pointwise values of $u(t)$ given by $u_L$ and $u_R$, are necessary in order to preserve the bounds of $w_i\in I$ (see \cite{CBA,MiMa}).

\subsubsection{Model predictive control of the microscopic dynamics}
\label{sec:feedback}
In general, for large values of $N$, standard methods for the solution of problems of type \eqref{eq:pbm}--\eqref{eq:pbc} over the full time interval $[0,T]$ stumble upon prohibitive computational costs due to the nonlinear constraints.

In what follows we sketch an approximation method for the solution of \eqref{eq:pbm}--\eqref{eq:pbc}, based on {\em model predictive control} (MPC), which furnishes a suboptimal control by an iterative solution over a sequence of finite time steps, but, nonetheless, it allows an explicit representation of the control strategy \cite{AHP,CBA,MRRS, MaMi}.

Let us consider the time sequence $0=t_0<t_1<\ldots<t_M=T$, a discretization of the time interval $[0,T]$, where  $\Delta t = t_{n}-t_{n-1}$, for all $n=1,\ldots, M $ and $t_M = M\Delta t $. Then we assume the control to be constant on every interval $[t_n,t_{n+1}]$, and defined as a piecewise function, as follows
\begin{align}\label{eq:mpcctrl}
\bar{u}(t) = \sum_{n=0}^{M-1} \bar{u}^n\chi_{[t_n,t_{n+1}]}(t),
\end{align} 
where $\chi(\cdot)$ is the characteristic function of the interval $[t_n,t_{n+1}]$.
We consider a full discretization of the optimal control problem \eqref{eq:pbm}-\eqref{eq:pbc}, through a forward Euler scheme, and we solve on  every time frame $[t_n, t_n+\Delta t]$, the  reduced  optimal control problem
\begin{equation}\label{eq:disc_functional}
\min_{\bar{u}\in\bar{\mathcal{U}}}J_{\Delta t}(\bar{u}):= \frac{1}{2N}\sum_{j=1}^N (w_j^{n+1} -w_d)^2 +\frac{\knu}{2} \int_{t_n}^{t_{n+1}}\bar{u}^2 dt,
\end{equation}
subject to 
\begin{equation}
\begin{aligned}\label{eq:Fwmicr}
w_i^{n+1} &= w_{i}^n +  \frac{\Delta t}{N}\sum_{j=1}^N P(w^n_i,w^n_j)(w_j^n-w_i^n)+ \Delta t\bar{u}^n,\qquad
w_i^n &= w_i(t_n),
\end{aligned}
\end{equation}
for all $i=1,\ldots, N$, and $\bar{u}$ in the space of the admissible controls $\bar{\mathcal{U}}\subset \mathcal{U}$.
 Note that since the control $\bar{u}$ is a constant value over the time interval $[t_n,t_n+\Delta t]$, and $w^{n+1}$  depends linearly on $\bar{u}^n$ through \eqref{eq:Fwmicr}, the discrete optimal control problem \eqref{eq:disc_functional}  reduces to
\begin{equation}\label{eq:disc_functional1}
J_{\Delta t}(\bar{u}^n) =  \frac{1}{2N}\sum_{j=1}^N (w_j^{n+1}(\bar{u}^n)-w_d)^2 +\Delta t \frac{\knu}{2} (\bar{u}^n)^2.
\end{equation}
Thus, in order to find the minimizer of \eqref{eq:disc_functional}, it is sufficient to compute the derivative of \eqref{eq:disc_functional1} with returns us the optimal value expressed as follows
\begin{align}\label{eq:IC}
U^n= -\frac{1}{\knu+ \Delta t}\left(\frac{1}{N} \sum_{j=1}^N\left(w^n_j-w_d\right)+\frac{\Delta t}{N^2} \sum_{j,k}P(w^n_j,w^n_k)(w^n_k-w^n_j)\right).
\end{align}
Expression \eqref{eq:IC} furnishes a feedback control for the full discretized problem, which can be plugged as an {\em instantaneous control} into  \eqref{eq:Fwmicr}, obtaining the following constrained system
\begin{equation}
\begin{aligned}\label{eq:Fwmicr_constr}
w_i^{n+1} &= w_{i}^n +  \frac{\Delta t}{N}\sum_{j=1}^N P(w^n_i,w^n_j)(w_j^n-w_i^n)+ \Delta t U^n,\qquad
w_i^n &= w_i(t_n).
\end{aligned}
\end{equation}
A more general derivation can be obtained through a discrete Lagrangian approach for the optimal control problem \eqref{eq:disc_functional}--\eqref{eq:Fwmicr}, see \cite{AHP}. 
\begin{remark}
We remark that the scheme \eqref{eq:Fwmicr_constr} furnishes a suboptimal solution w.r.t. the original control problem. In particular if $P(\cdot,\cdot)$ is symmetric, only the average of the system is controlled. Let us set $w_d = 0$, and  $m^n = \sum_{i=1}^N w_i^n/N $. Summing on $i=1,\ldots,N$ equation  \eqref{eq:Fwmicr_constr} we have 
\begin{equation}
\begin{aligned}
m^{n+1}&= m^n-\frac{\Delta t}{\knu+\Delta t}m^n 
             = \left(1-\frac{\Delta t}{\knu+\Delta t}\right)^nm^0,
\end{aligned}
\end{equation}
which implies $m^\infty=0$. Thus, while the feedback control is able to control the mean of the system, it does not to assure the global consensus. We will see in the next Section how the introduction of a binary control depending on the pairs permits to recover the global consensus.
\end{remark}

\subsubsection{Binary Boltzmann control}
Following Section \ref{standard}, we consider now a kinetic model for the  evolution of  the density $f=f(w,t)$ of agents with opinion $w \in I=[-1,1]$  at time $t\geq 0$,  such that  the total mass is normalized to one. The evolution can be derived by considering the change in time of $f(w,t)$ depending on the interactions among the individuals of the binary type \eqref{ch6:trade_rule}.  In order to derive such Boltzmann description we follow the approach in \cite{APb, FH}.
We consider the model predictive control system \eqref{eq:Fwmicr_constr} in the simplified case of only two interacting agents, numbered $i$ and $j$. Their opinions are modified according to
\begin{equation}\label{eq:DBin}
\begin{aligned}
&w^{n+1}_i=w_i^n+\frac{\Delta t}{2}P(w_i^n,w_j^n)(w_j^n-w^n_i)+\frac{\Delta t}{2} \gd(w_i^n,w_j^n),\\
&w^{n+1}_j=w_j^n+\frac{\Delta t}{2}P(w_i^n,w_j^n)(w_i^n-w^n_j)+\frac{\Delta t}{2} \gd(w_j^n,w_i^n),\\
\end{aligned}
\end{equation}
where the feedback control term $\gd(w^n_i,w^n_j)$ is derived from \eqref{eq:IC} and yields
\begin{equation}
\begin{split}\label{eq:Econtrol}
\frac{\Delta t}{2}  \gd(w^n_i,w^n_j) =& -\frac{1}{2}\frac{\Delta t}{\knu+\Delta t} \left((w^{n}_j-w_d)+(w^{n}_i-w_d))\right)\\
&-\frac{1}{4}\frac{\Delta t^2}{\knu+\Delta t} \left(P^n_{ij}-P^n_{ji}\right)(w_j^n-w_i^n),
\end{split}
\end{equation}
having defined $P^n_{ij} = P(w^n_i,w^n_j)$. This formulation can be written as a binary Boltzmann dynamics
\begin{equation}
\begin{aligned}\label{eq:micr_kin1}
\pw = \,& w        + \eta P(w,w_*)(w-w_*)    +   \eta \gd(w,w_*)    +  \xi D(w),       
\\
\pw_* =\, & w_* + \eta P(w_*,w)(w_*-w)   +  \eta \gd(w_*,w)  +  \xi_*D(w_*).
\end{aligned}
\end{equation}
All quantities in \eqref{eq:micr_kin1} are defined as in \eqref{ch6:trade_rule}. 
The control $\gd(\cdot,\cdot)$, which is not present in \eqref{ch6:trade_rule}, acts as a forcing term  to steer consensus, or, in other words, it models the action of promoting the emergence of a desired status.

Thus we can associate the binary dynamics in \eqref{eq:DBin} to the original dynamics in \eqref{eq:micr_kin1}. Choosing $\eta = \Delta t/2$, the control term for the arbitrary pair $(w,w_*)$ reads
\begin{align}\label{eq:Econtrol2}
\eta \gd(w,w_*)      = \frac{2\eta}{\knu+2\eta} (K(w,w_*)+\eta H(w,w_*)),
\end{align}
where 
\begin{align}
&K(w,w_*) = \frac{1}{2}((w_d-w)+(w_d-w_*)), \\
&H(w,w_*) = \frac{1}{2} \left(P(w,w_*)-P(w_*,w)\right)(w-\v).
\end{align}
Note that $K(w,w_*)$  and $H(w,w_*)$ are both symmetric, which follows directly by \eqref{eq:pbm}--\eqref{eq:pbc}, since $u$ is the same for every agent. Embedding the control dynamics into \eqref{eq:micr_kin1} we obtain the following binary constrained interaction
\begin{equation}
\begin{aligned}\label{eq:BinD}
\pw& = w + \eta P(w,w_*)(w_*-w)+\beta (K(w,w_*)+\eta H(w,w_*)) +\xi D(w),\\
\pw_*&=w_* + \eta P(w_*,w)(w-w_*)+\beta (K(w,w_*)+\eta H(w,w_*)) +\xi_* D(w_*),\\
\end{aligned}
\end{equation}
with $\beta$ defined as follows
\begin{align}\label{eq:beta}
\beta := \frac{2\eta}{\knu+2\eta}.
\end{align}

In the absence of diffusion, from \eqref{eq:BinD} it follows that 
\begin{equation}
\begin{split}
\pw+\pv &= w+\v + \eta(P(w,\v)-P(\v,w))(\v-w)\\
 &\quad + 2\beta(K(w,w_*)+\eta H(w,w_*))\\
&=w+\v - 2\eta H(w,\v)+ 2\beta(K(w,w_*)+\eta H(w,w_*))\\
&= 2w_d - \left(1-\beta \right)(\knu+2\eta)\gd(w,\v) = 2w_d-\knu \gd(w,\v),\label{eq:meanop}
\end{split}
\end{equation}
thus in general the mean opinion is not conserved. 
Observe that the computation of the relative distance between opinions $|\pw-\pv|$ is equivalent to \eqref{ch6:tr+-}, since the subtraction cancels the control terms out, giving the inequality
\begin{align}
&|\pw-\pv|=\left(1-\eta (P(w,w_*)+P(w_*,w)\right)|w-w_*|\leq(1-2\eta)|w-w_*|,
\end{align}
which tells that the relative distance in opinion between two agents diminishes after each interaction~\cite{T}. 
In presence of noise terms, we should assure that the binary dynamics \eqref{eq:BinD} preserves the boundary, i.e. $\pw,\pv\in I$. An important role in this is played by functions $ P(\cdot,\cdot),\,D(\cdot)$, as stated by the following proposition.

\begin{proposition}\label{proposition:bounds}
Let assume that there exist $p>0$ and $m_C$ such that $p\leq P(w,\v)\leq 1$ and $m_C=\min\left\{(1-w)/D(w),D(w)\neq 0\right\}$. Then, provided 
\begin{align}\label{eq:assumption}
 &\beta\leq\eta p,\qquad\Theta\in\left(-m_C\left(\eta-\frac{\beta}{2}\right),m_C\left(\eta-\frac{\beta}{2}\right)\right),
\end{align}
are satisfied,  the binary interaction \eqref{eq:BinD} preserves the bounds, i.e. the post-interaction opinions $\pw,\pv$ are contained in $I=[-1,1]$.
\end{proposition}
\noindent
\textbf{Proof.} We refer to \cite{AHP,T} for a detailed proof.

\begin{remark}
Observe that, from the modeling viewpoint, noise is seen as an external term which can not be affected by the control dynamics. A different strategy is to account the action of the noise at the level of the microscopic dynamics \eqref{eq:pbm} and proceed with the optimization. This will lead to a different binary interaction with respect to \eqref{eq:BinD}, where the control influences also the action of the noise.
\end{remark}

\subsubsection{Main properties of the Boltzmann description}\label{sec:Boltzmann_proprieties}

In general the time evolution of the density $f(w,t)$ is found by resorting to a Boltzmann equation of type \eqref{ch6:weak boltz}, where the collisions are now given by \eqref{eq:BinD}.  In weak form we have
\begin{align}\label{eq:MBoltz}
(Q(f,f),\varphi)=\frac{\lambda}{2}\left\langle\int_{I^2}(\varphi(\pw)+\varphi(\v)-\varphi(w)-\varphi(v))f(w)f(\v) dw d\v\right\rangle,
\end{align}
where we omitted the time dependence for simplicity. Therefore  
the total opinion, obtained taking $\varphi(w)=1$, is preserved in time. This is the only conserved quantity of the process. 
Choosing $\varphi(w)=w$, we obtain the evolution of the average opinion, thus we have
\begin{align}\label{eq:B2}
\frac{d}{dt}\int_I wf(w,t)dw=\frac{\lambda}{2}\left\langle\int_{I^2} \left(\pw+\pv-w-\v\right)f(w,t)f(v,t)~dw~d\v\right\rangle.
\end{align}
Indicating the average opinion as  $m(t)=\int_I w f(w,t) \,dw,$ using \eqref{eq:meanop} we get
\begin{equation}
\begin{split}\label{eq:media}
\frac{dm(t)}{dt}=&
\lambda\beta(w_d-m(t))\\
&+\lambda\eta(1-\beta)\int_{I^2}(P(w,\v)-P(\v,w))\v f(\v)f(w)~dw\,d\v.
\end{split}
\end{equation}
Since $0\leq P(w,\v)\leq 1$, $|P(w,\v)-P(\v,w)|\leq1$, we can bound the derivative from below and above
\begin{align*}
\lambda\beta w_d-\lambda(\beta+\eta(1-\beta))m(t)\leq~ \frac{d}{dt}m(t)~\leq\lambda\beta w_d-\lambda(\beta-\eta(1-\beta))m(t).
\end{align*}
Note that in the limit $t\to\infty$, the average $m(t)$ converges to the desired state $w_d$, if 
$\beta-\eta(1-\beta)>0$. This implies the following restriction
$\knu<2.$
A similar analysis can be performed for the second moment $\varphi(w) = w^2$, showing the decay of the energy for particular choices of the interaction potential (cf. \cite{T,AHP,APZa} for further details on the proprieties of the moment of \eqref{ch6:weak boltz}).

\begin{remark}
In the symmetric case, $P(v,w)=P(w,v)$, equation \eqref{eq:media} is solved explicitly as
\begin{align}\label{eq:ave}
m(t)=\left(1-e^{-\lambda\beta t}\right)w_d+m(0)e^{-\lambda\beta t}
\end{align}
which, as expected, in the limit $t\to\infty$ converges to $w_d$, for any choice of the parameters.
\end{remark}

\subsubsection{The quasi--invariant opinion limit}

We will now introduce some asymptotic limit of the kinetic equation. The main idea is to scale interaction frequency and strength, $\lambda$ and $\eta$ respectively, diffusion $\varsigma^2$ at the same time, in order to maintain at any level of scaling the memory of the microscopic interactions \eqref{eq:BinD}. This approach is refereed to as \emph{quasi--invariant opinion} limit~\cite{T, FPTT, VILL}. 
Given $\epsi>0$, we consider the following scaling
\begin{align}\label{eq:scale}
\eta=\epsi,\quad
\lambda = \frac{1}{\epsi},\quad
\varsigma=\sqrt{\epsi}\sigma,\quad\beta = \frac{2\epsi}{\kappa +2\epsi}.
\end{align}



The ratio  $\varsigma^2/\eta=\sigma$ is of paramount importance in order to show in the limit the contribution of both the compromise propensity $\eta$ and the diffusion $\varsigma^2$. Other scalings lead to diffusion dominated $(\varsigma^2/\eta\to \infty)$ or compromise dominated $(\varsigma^2/\eta\to 0)$ equations.
In the sequel we show through formal computations how this approach leads to a Fokker--Planck equation type \cite{R}. We refer to \cite{T} for details and rigorous derivation.

After scaling, equation  \eqref{eq:MBoltz} reads
\begin{align}\label{eq:Boltz2}
\frac{d}{dt}\int_I \varphi(w)f(w,t)dw=\frac{1}{\epsi}\left\langle\int_{I^2} \left(\varphi(\pw)-\varphi(w)\right)f(w,t)f(v,t)~dw~dv\right\rangle,
\end{align}
while the scaled binary interaction dynamics \eqref{eq:BinD} is given by
\begin{equation}\label{eq:BinScale}
\pw -w=\epsi P(w,\v)(\v-w)  + \frac{2\epsi}{\kappa+2\epsi}K(w,\v)+\xi D(w)+O(\epsi^2).
\end{equation}
\noindent

In order to recover the limit for $\epsi\to 0$ we consider the second-order Taylor expansion of $\varphi$ around $w$,
\begin{align}\label{eq:Tay2}
\varphi(\pw)-\varphi(w)=(\pw-w)\partial_w\varphi(w)+\frac{1}{2}(\pw-w)^2\partial^2_w\varphi(\tilde{w})
\end{align}
where for some $0\leq\vartheta\leq1$ ,
$$\tilde{w}=\vartheta \pw+(1-\vartheta)w.$$ 
Therefore the approximation of the interaction integral in \eqref{eq:MBoltz} reads
\begin{equation}\label{eq:limBoltz}
\begin{aligned}
\lim_{\epsi\to0}&\frac{1}{\epsi}
\left\langle\int_{I^2}\left(\pw-w\right)\partial_w\varphi(w)f(w)f(\v)~dwd\v\right.\\
&\qquad\qquad\qquad\left.+\frac{1}{2}\left(\pw-w\right)^2\partial^2_w\varphi(w)f(w)f(\v)~dwd\v\right\rangle+R(\epsi).
\end{aligned}
\end{equation}
The term $R(\epsi)$  indicates the  remainder of the Taylor expansion and is such that
\begin{equation}
R(\epsi)=\frac{1}{2\epsi}\left\langle\int_{I^2}\left(\pw-w\right)^2(\partial^2_w\varphi(\tilde{w})-\partial^2_w\varphi(w))f(w)f(\v)~dwd\v\right\rangle.
\end{equation}
Under suitable assumptions on the function space of $\varphi$ and $\xi$ the remainder converges to zero as soon as $\epsi\to 0$ (see \cite{T}). 
Thanks to \eqref{eq:BinScale} the limit operator of \eqref{eq:limBoltz} is the following
\begin{align*}
&\int_{I^2}\left(P(w,\v)(\v-w)  + \frac{2}{\kappa}K(w,\v)\right)\partial_w\varphi(w)f(w)f(\v)~dwd\v\\
&\qquad\qquad+\frac{\sigma^2}{2}\int_{I} D(w)^2\partial^2_w\varphi(w)f(w)~dw.
\end{align*} 
Integrating back by parts the last expression, and supposing that the border terms vanish, we obtain the following Fokker--Planck equation 
\begin{equation}
\begin{aligned}\label{eq:FP1}
\frac{\partial}{\partial t} f &+\frac{\partial}{\partial w}\mathcal{P}[f](w)f(w)+ \frac{\partial}{\partial w}\mathcal{K}[f](w)f(w)~dv=\frac{\sigma^2}{2}\frac{\partial^2}{\partial w^2}(D(w)^2f(w)),
\end{aligned}
\end{equation}
where
\begin{align*}
&\mathcal{P}[f](w)=\int_I P(w,v)(v-w)f(v)~dv,\\
&\mathcal{K}[f](w)=\frac{2}{\kappa}\int_I K(w,v)f(v)~dv=\frac{1}{\kappa}\left((w_d-w)+(w_d-m)\right).
\end{align*}
As usual, $m(t)=\int_I wf(w,t)dw$ indicates the mean opinion.

\subsubsection{Stationary solutions}\label{sec:Steady}
One of the advantages of the Fokker--planck description is related to the possibility to identify analytical steady states. 
In this section we will look for steady solutions of the Fokker--Planck model \eqref{eq:FP1}, for particular choices of the microscopic interaction of the Boltzmann dynamics.

The stationary solutions, say $f_\infty(w)$, of  \eqref{eq:FP1} satisfy the  equation
\begin{equation}
\begin{aligned}\label{eq:FP2s}
\frac{\partial}{\partial w}\mathcal{P}[f](w)f(w)+ \frac{\partial}{\partial w}\mathcal{K}[f](w)f(w)~dv=\frac{\sigma^2}{2}\frac{\partial^2}{\partial w^2}(D(w)^2f(w)).
\end{aligned}
\end{equation}
As shown in \cite{APZa,APZc,T}, equation \eqref{eq:FP2s} can be analytically solved under some simplifications.
In general solutions to \eqref{eq:FP2s}  satisfy the ordinary differential equation
\begin{equation}\label{eq:odestat}
\dfrac{d f}{d w}=\left(\frac{2}{\sigma^2}\frac{\mathcal{P}[f](w)+\mathcal{K}[f](w)}{D(w)^2}-2\frac{D'(w)}{D(w)}\right)f.
\end{equation}
Thus
\begin{equation}\label{eq:finf}
f_{}(w)= \frac{C_0}{D(w)^2} \exp\left\{\dfrac{2}{\sigma^2}\int^w\left(\frac{\mathcal{P}[f](v)+\mathcal{K}[f](v)}{D(v)^2}\right)\,dv\right\},
\end{equation}
where $C_0$ is a normalizing constant. 

Let us consider the simpler case in which $P(w,v)=1$. Then the average opinion $m(t)$ evolves according to 
\begin{align}\label{eq:FPave}
m(t)=\left(1-e^{-2/\kappa t}\right)w_d+e^{-2/\kappa t}m(0),
\end{align}
which is obtained from the scaled equation \eqref{eq:Boltz2} through the quasi-invariant opinion limit (cf. also equation \eqref{eq:ave} for a comparison).

\noindent
In absence of control, i.e. for $\kappa\to\infty$, the mean opinion  is conserved, and the steady solutions of \eqref{eq:FP1} satisfy the differential equation \cite{T} 
\begin{align}\label{eq:edo}
\partial_w(D(w)^2f)=\frac{2}{\sigma^2}\left(m-w\right)f.
\end{align}
In presence of the control the mean opinion is in general  not conserved in time, even if, from \eqref{eq:FPave}, it is clear that $m(t)$ converges exponentially in time to $w_d$. Consequently
\begin{align}\label{eq:edo2}
&\partial_w(D(w)^2f)=\frac{2}{\sigma^2}\left(1+\frac{1}{\kappa}\right)(w_d-w)f.
\end{align}
Let us consider as diffusion function $D(w)=(1-w^2)$. Therefore the solution of \eqref{eq:edo} takes the form
\begin{equation}
\begin{split}\label{eq:sol}
f_\infty(w)=&~\frac{C_{m,\sigma}}{(1-w^2)^{2}}\left(\frac{1+w}{1-w}\right)^{m/(2\sigma^2)}\exp\left\{-\frac{1-m w}{\sigma^2\left(1-w^2\right)}\right\}\\=&~\frac{C_{m,\sigma}}{(1-w^2)^2}S_{m,\sigma^2}(w),
\end{split}
\end{equation}
where $C_{m,\sigma}$ is such that the mass of $f_\infty$ is equal to one. This solution is regular, and thanks to the presence of the exponential term  $f(\pm1)=0$. Moreover, due to the general non symmetry of $f$, the initial opinion distribution reflects on the steady state through the mean opinion.
The dependence on $\kappa$ can be rendered explicit. It gives
\begin{align}\label{eq:sol1}
f_\infty^\kappa(w)& = \frac{C_{w_d,\sigma,\kappa}}{(1-w^2)^2}\left(S_{w_d,\sigma}(w)\right)^{1+1/\kappa},
\end{align}
with $C_{w_d,\sigma,\kappa}$ the normalization constant.

\begin{figure}[htb]
\centering
\includegraphics[scale=0.3]{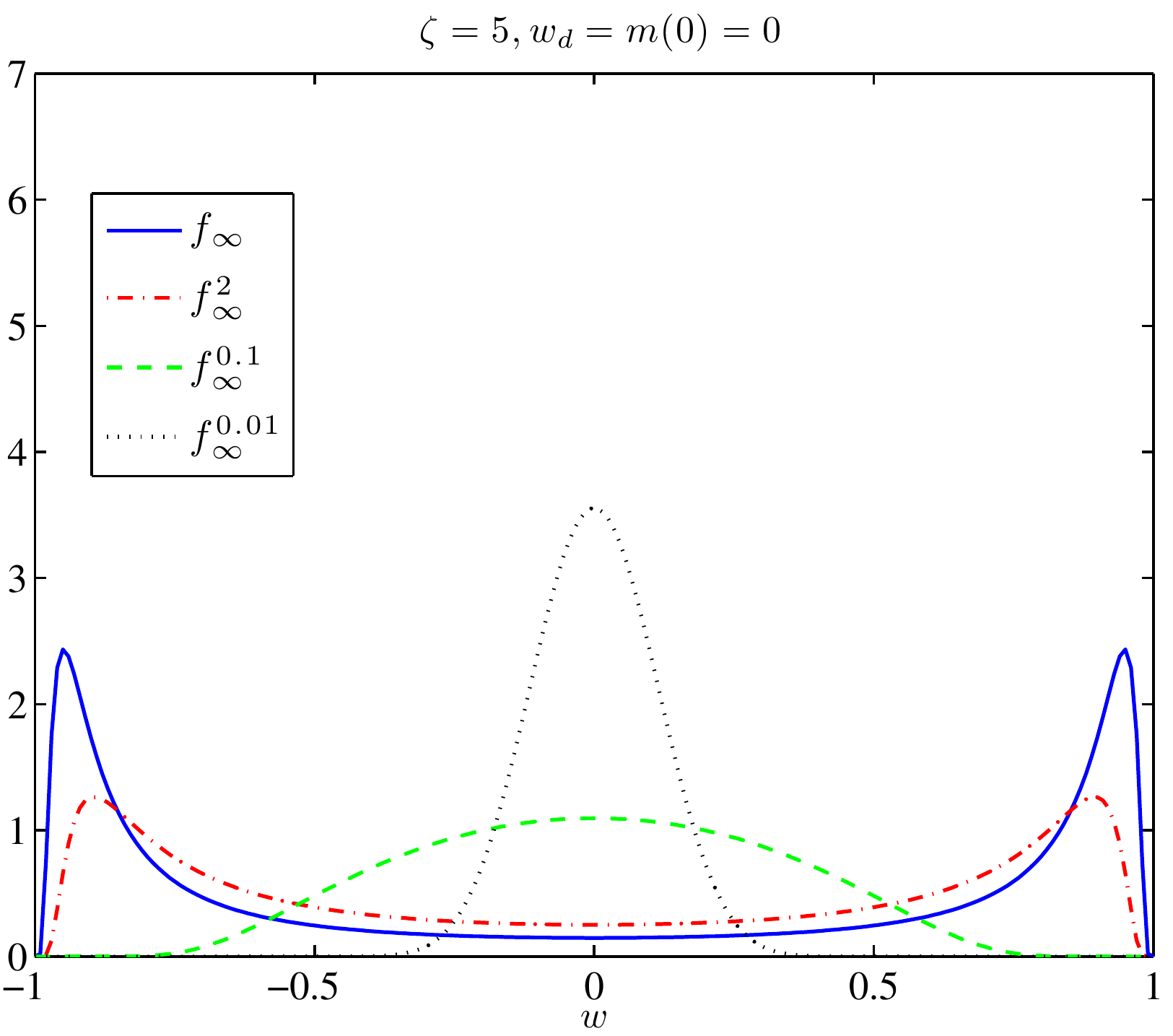}
\includegraphics[scale=0.3]{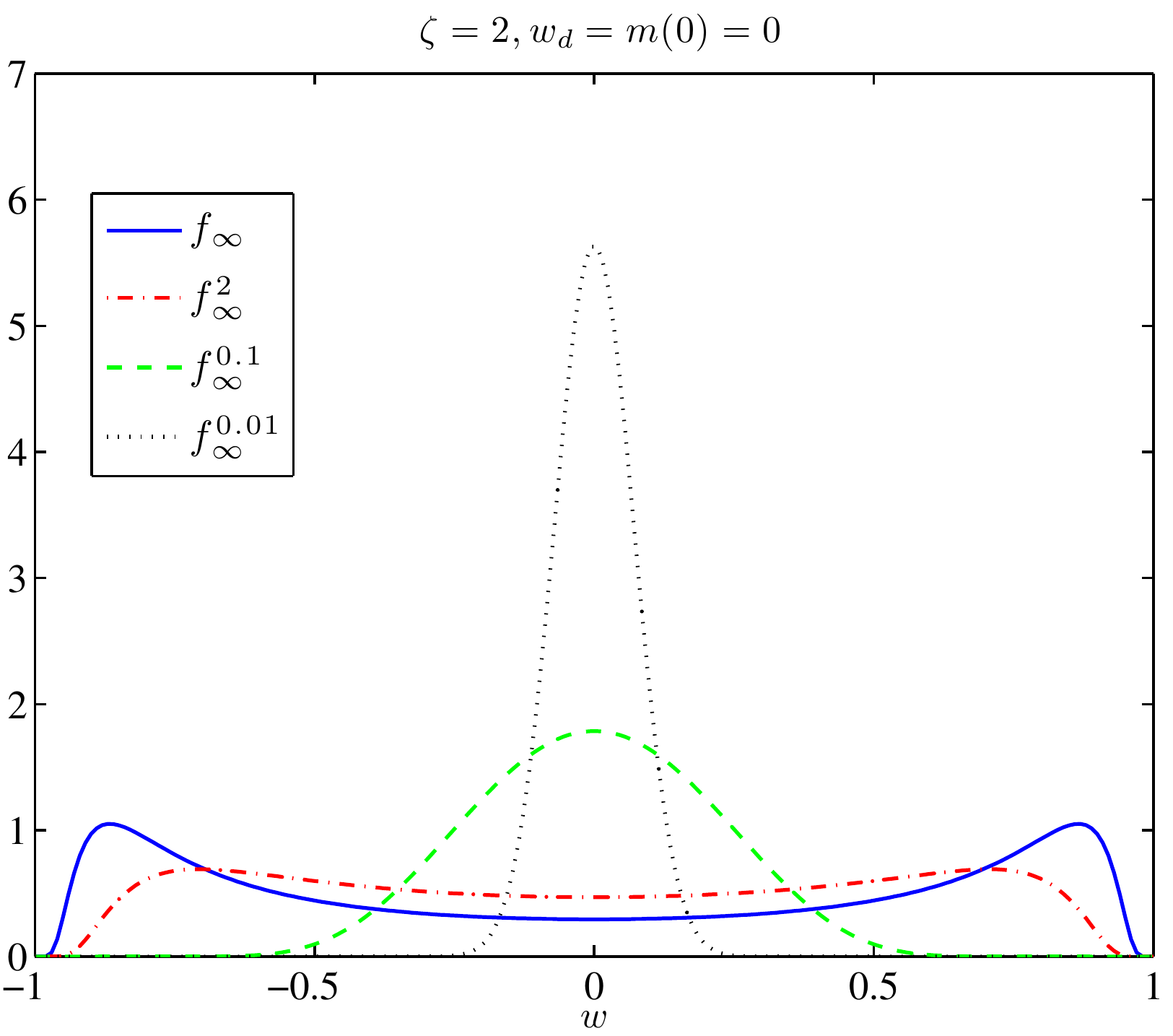}
\caption{Continuous line and dashed lines represent respectively the steady solutions $f_\infty$ and $f_\infty^\kappa$. On the left $w_d=m(0)=0$ with diffusion parameter $\zeta=\sigma^2=5$, on the right  $w_d=m(0)=0$ with diffusion parameter $\zeta=\sigma^2=2$.
In both cases the steady solution changes from a bimodal distribution to an unimodal distribution around $w_d$.}\label{fig:F1}
\end{figure}

We plot in Figure \ref{fig:F1} the steady profile $f_\infty$ and $f_\infty^\kappa$ for different choice of the parameters $\kappa$ and $\sigma$. The initial average opinion $m(0)$ is taken equal to the desired opinion $w_d$. In this way we can see that for $\kappa\to \infty$ the steady profile of \eqref{eq:edo2} approaches the one of \eqref{eq:edo}.  On the other hand small values of $\kappa$ give the desired distribution concentrated around $w_d$.
It is remarkable that in general  we can not switch from $f_\infty$ to $f_\infty^\kappa$ only acting  on the parameter $\kappa$, since the memory on the initial average opinion is lost for any $\kappa>0$.
We refer to \cite{AHP,T} for further discussion about stationary solutions of \eqref{eq:FP2s}.

\subsubsection{Numerical Tests}\label{sec:Num} 
Our goal is to investigate the action of the control dynamic at the mesoscopic level. We solve directly the kinetic equation \eqref{eq:MBoltz} obeying the binary interaction \eqref{eq:BinScale}, for small value of the scale parameter $\epsi>0$.
We  perform the numerical simulations using the Monte Carlo methods developed in \cite{APb,PTa}.

\subsubsection*{Sznajd's model}
\begin{figure}[htb]
\centering
\includegraphics[scale=0.3]{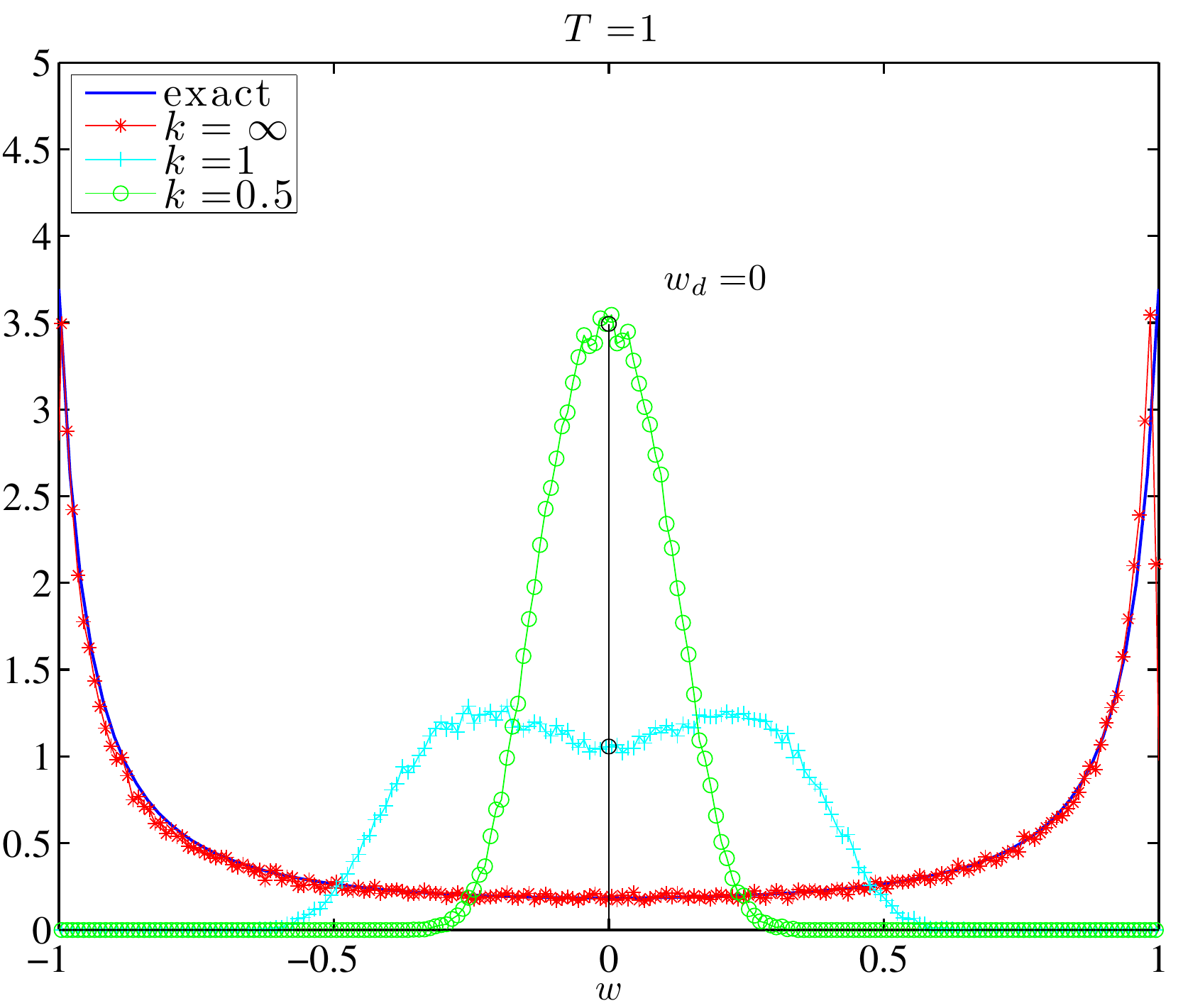}
\includegraphics[scale=0.3]{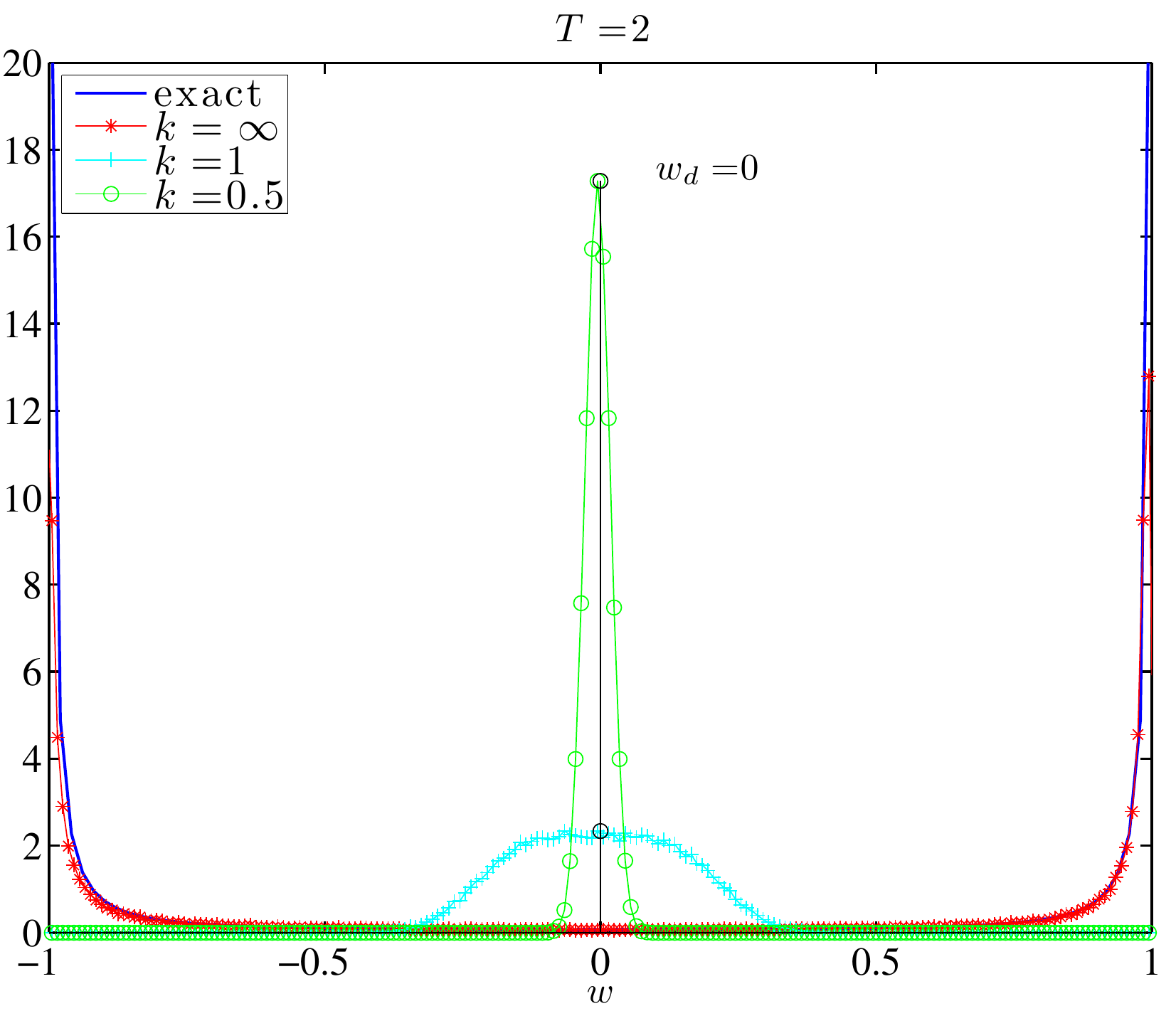}
\caption{Solution profiles at time $T=1$,  and $T=2$, for uncontrolled $\kappa=\infty$, mildly controlled case $\kappa=1$ , strong controlled case $\kappa=0.1$. Desired state is set to $w_d=0$.}\label{fig:F4b}
\end{figure}

Our first example refers to the mean-field Sznajd's model~\cite{SWS, ANP}
\begin{align}\label{eq:MFSz}
\partial_t f =\gamma\partial_w\left(w(1-w^2)f)\right),
\end{align}
corresponding to equation \eqref{eq:FP1} in the uncontrolled case without diffusion. It is obtained choosing $P(w,v)=1-w^2$ and assuming that the mean opinion $m(t)$ is always zero.
In~\cite{ANP} authors showed for $\gamma=1$ \emph{concentration} of the profile around zero, and  conversely for $\gamma=-1$ a \emph{separation} phenomena, namely concentration around $w=1$ and $w=-1$, by showing that explicit solutions are computable.
We  approximate the mean-field dynamics \label{eq:MFSz} in the \emph{separation} case, $\gamma=-1$, through the binary interaction \eqref{eq:BinScale}, sampling $N_s=1\times10^5$ agents, with scaling parameter $\epsi=0.005$.
In Figure \ref{fig:F4b} we simulate the evolution of $f(w,t)$ in the time interval $[0~2]$, starting from the uniform distribution on $I$, $f_0(w)=1/2$, in three different cases: uncontrolled ($\kappa=\infty$), mild control ($\kappa=1$) and strong control ($\kappa=0.1$). In the controlled cases
 the distribution is forced to converge to the desired state $w_d=0$. 

\subsubsection*{Bounded confidence model}
\begin{figure}[htb]
\centering
\includegraphics[scale=0.3]{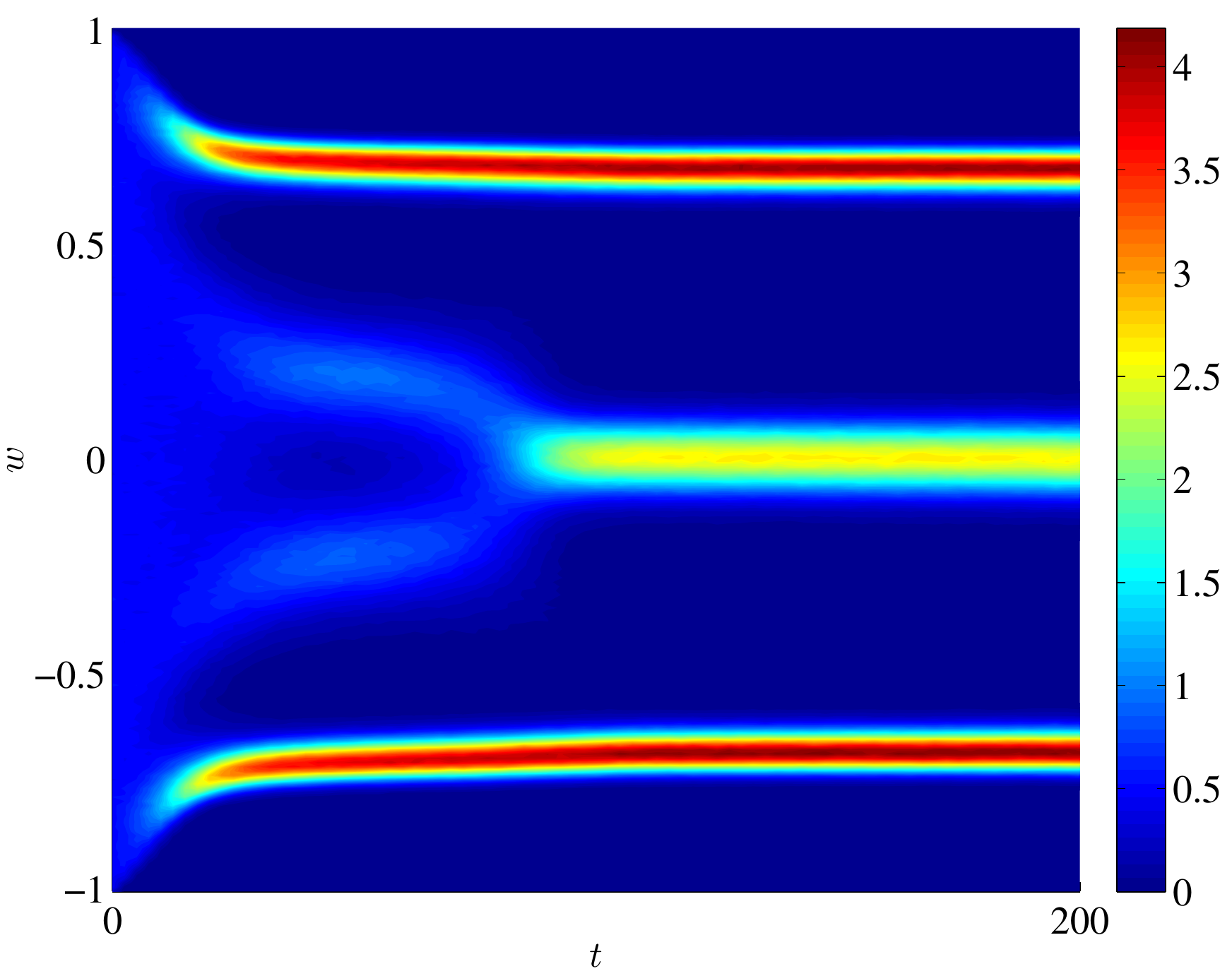}
\includegraphics[scale=0.3]{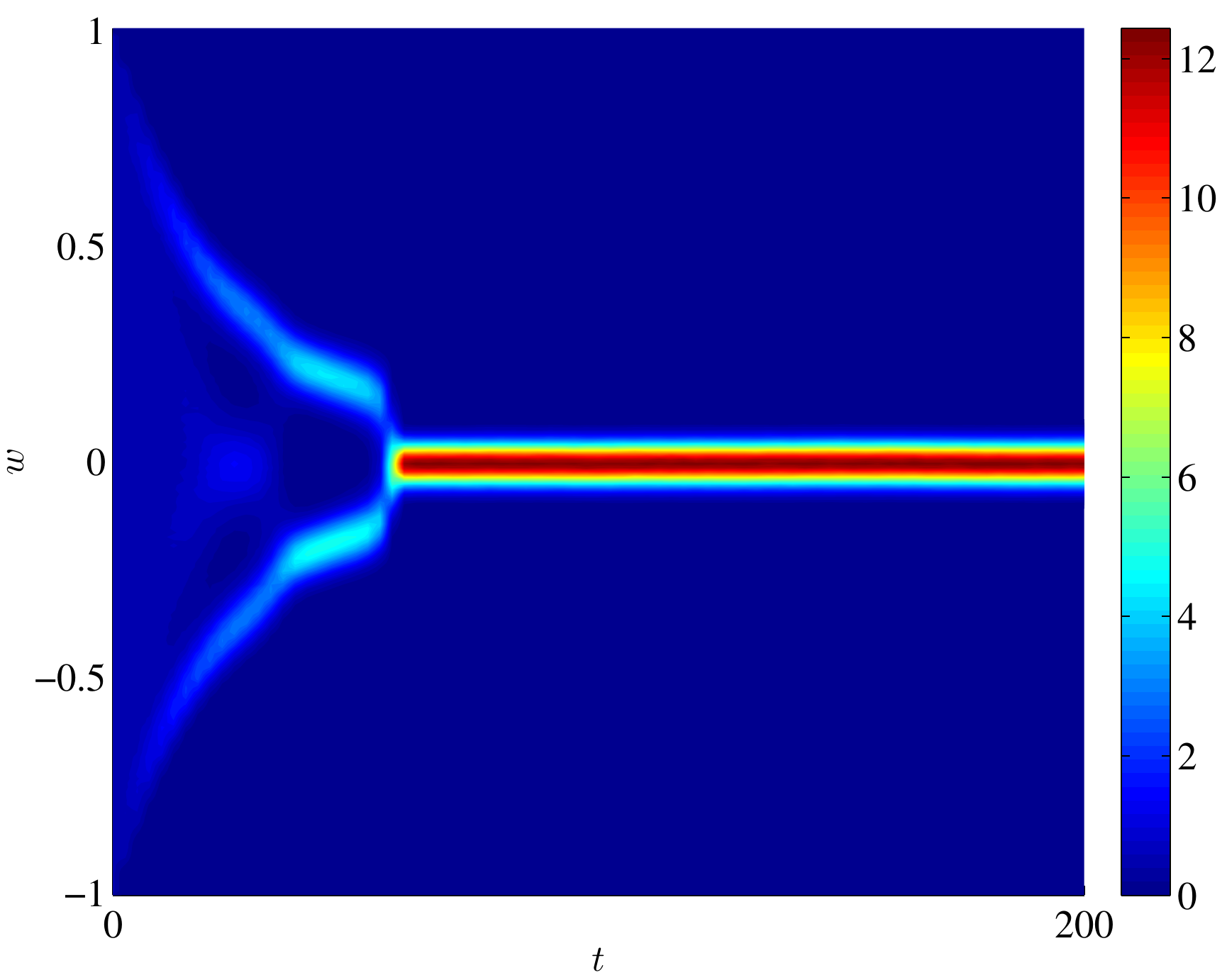}

\caption{On the right the penalization of the control parameter is $\knu=5\times 10^3$ on the left $\knu=5$. Evolution of the kinetic density,  using $N_s=2\times10^5$ sample particles on a $200\times400$ grid. Binary interactions \eqref{eq:BinScale} performed with $\epsi=0.01$ and $\varsigma=0.01$, $\Delta = 0.2$.   }\label{fig:F5}
\end{figure}

We consider now the \emph{bounded confidence model} introduced in \cite{HK}, where every agent interacts only within  a certain level of confidence. This can be modelled through the potential function 
$$P(w,v)=\chi(|w-v|\leq\Delta), \quad \Delta < 2.$$
In Figure~\ref{fig:F5}, we simulate the dynamics of the agents starting from a uniform distribution of the opinions on the interval $I=[-1,1]$. The binary interaction \eqref{eq:BinScale} refers to a diffusion parameter $\sigma=0.01$ and $\epsi=0.05$. Here $N_s = 2\times10^5$. The bounded confidence parameter is  $\Delta=0.2$, and we consider both cases (without control and with control), letting the system evolve in the time interval $\left[0~ T\right]$, with $T=200$. 
The figure to the left refers to the uncontrolled  case, where three mainstream opinions emerge. On the right the presence of the control with $\knu=5$  leads the opinions to concentrate around the desired opinion $w_d=0$.

\subsection{Control through leadership}
Several studies have been recently focused on the control of a large population through the action of a small portion of individuals, typically identified as leaders  \cite{ABCK,AP1, BFRS, FPR}. 
In this section we are interested in the opinion formation process of a followers' population steered by the action of a leaders' group. 
At a microscopic level we suppose to have a population of $N_F$ followers and $N_L$ leaders. Their dynamics is modelled as follows
\begin{align}
&\dot{w}_{i}=\dfrac{1}{N_F}\displaystyle\sum_{j=1}^{N_F}P\left(w_{i},w_{j}\right)\left(w_{j}-w_{i}\right)+\dfrac{1}{N_{L}}\sum_{h=1}^{N_{L}}S\left(w_{i},\tw_{h}\right)\left(\tw_{h}-w_{i}\right), \,\,  w_{i}\left(0\right)=w_{i,0}, 
\label{follower_dynamic}
\\
&\dot{\tw}_{k}=\dfrac{1}{N_L}\displaystyle\sum_{h=1}^{N_L}R\left(\tw_{k},\tw_{h}\right)\left(\tw_{h}-\tw_{k}\right)+u,\qquad\qquad\qquad\qquad\qquad\,\,\,\,\,\,\, \tw_{k}\left(0\right)=\tw_{k,0}, 
\label{leader_dynamic}
\end{align}
where $w_i,v_k \in I$, $I=[-1,1]$ for all $i=1,\ldots,N_F$ and $k=1,\ldots,N_L$ are the followers'  and leaders'  opinions. As in the previous section, $P(\cdot,\cdot),\, S(\cdot,\cdot)$ and  $R(\cdot,\cdot)$ are given \emph{compromise functions}, measuring the relative importance of the interacting agent in the consensus dynamics. Leaders' strategy is driven by a suitable control $u$, which minimizes the  functional
\begin{align}\label{JL_cost}
&J(u)=\dfrac{1}{2}\int_{0}^{T}\left(\dfrac{\psi}{N_L}\sum_{h=1}^{N_L}(\tw_h-w_d)^2+\dfrac{\mu}{N_L}\sum_{h=1}^{N_L}(\tw_h-m_F)^2\right)dt+\dfrac{\knu}{2}\int_0^{T}{{u^2}}dt,
\end{align}
where 
$T$ represents the final time horizon, $w_d$ is the desired opinion and $m_F$ is the average opinion of the followers group at time $t\geq{0}$, and $\psi,\mu >0$ are such that $\psi+\mu =1$. Therefore, leaders' behavior is driven by a suitable control strategy based on the interplay between the desire to force followers towards a given state, {\em radical behavior} ($\psi\approx 1$), and the necessity to keep a position close to the mean opinion of the followers in order to influence them {\em populistic behavior} ($\mu\approx 1$).


Note that, since the optimal control problem acts only over the leader dynamics we can approximate its solution by a model predictive control approximation as in Section \ref{sec:AHP}. Next we can build the corresponding constrained binary Boltzmann dynamic following \cite{APZa}.  

\subsubsection{Boltzmann constrained dynamics }
\label{sec:Boltzmann}

To derive the system of kinetic equation we introduce a density distribution of followers  $f_F(w,t)$ and leaders $f_L(\tw,t)$ depending on the opinion variables $w,\tw\in I$ and time $t\geq0$, see \cite{APZa,DMPW}.
It is assumed that the densities of the followers and the leaders satisfy
is
 \[
 \int_I f_F(w,t) \, dw = 1,
\qquad
 \int_I f_L(\tw,t) \, d\tw = \rho \leq 1.
 \] 
The kinetic model can be derived by considering the change in time of $f_F(w,t)$ and $f_L(\tw,t)$ depending on the interactions with the other individuals and on the leaders' strategy. This change depends on the balance between the gain and loss due to the binary interactions. 
Starting by the pair of opinions $(w,\v)$ and $(\tw,\tv)$, respectively the opinions of two followers and two leaders, the  post-interaction opinions are computed according to three dynamics: $a)$ the interaction between  two followers;  $b)$ the interaction between a follower and a leader;  $c)$  the interaction between  two leaders.
\begin{itemize}
\item[$a)$]
We assume that the opinions $(\pw, \pv)$ in the follower-follower interactions obey to the rule
\begin{equation}\begin{cases}\label{followerfollower_binary}
\pw=w+\eta P(w,\v)(\v-w)+\xi D_{F}(w), \\
\pv=\v+\eta P(\v,w)(w-\v)+\xi_* D_{F}(\v),
\end{cases}\end{equation}
where as usual $P(\cdot,\cdot)$ is the compromise function, and the diffusion variables $\xi,\xi_*$ are realizations of a random variable with zero mean, finite variance $\varsigma_F^2$. The noise influence is weighted by the function ${D}_F(\cdot)$, representing the local relevance of diffusion for a given opinion, and such that $0\leq {D}_{F}(\cdot)\leq 1$.
\item[$b)$]
The leader-follower interaction is described for every agent from the leaders group. Since the leader do not change opinion, we have
\begin{equation}\begin{cases}\label{followerleader_binary_1}
\ppw=w+\eta S(w,\tw)(\tw-w)+\zeta D_{FL}(w) \\
\pptw=\tw
\end{cases}\end{equation}

where $S(\cdot,\cdot)$ is the communication function and $\zeta$ a random variable with zero mean and finite variance ${\varsigma}_{FL}^2$, weighted again by the function $D_{FL}(\cdot)$.
\item[$c)$]
Finally, the post-interaction opinions $(\ptw,\ptv)$ of two leaders are given by
\begin{equation}\begin{split}\begin{cases}\label{binary_dynamics_leader0}
\ptw=\tw + \eta R(\tw,\tv)(\tv-\tw) + \eta \gd(\tw,\tv;m_F)+\theta D_L(\tw)\\
\ptv=\tv  + \eta R(\tv,\tw)(\tw-\tv)  + \eta \gd(\tw,\tv;m_F)+\theta_* D_L(\tv),\\
\end{cases}\end{split}\end{equation}  
where $R(\cdot,\cdot)$ is the compromise function and, similar to the previous dynamics, $\theta,\theta_*$ are random variables with zero mean and finite variance ${\varsigma}_{L}^2$, weighted by $D_L(\cdot)$. 
Moreover the leaders' dynamics include the feedback control, derived from \eqref{JL_cost} with the same approach of Section \ref{sec:feedback}. In this case the feedback control accounts for the average values of the followers' opinion 
\begin{align}m_F(t) = \int_I w f_F(w,t)\,dw,\end{align}
and it is defined as 
\begin{equation}
\begin{aligned}\label{u_partial1}
\eta\gd(\tw,\tv;m_F) =&\beta\left[ K(\tw,\tv;m_F)+ \eta H(\tw,\tv)\right].
\end{aligned}
\end{equation}
In \eqref{u_partial1}, $\beta$ has the same form of \eqref{eq:beta} and 
\begin{align}
&K(\tw,\tv;m_F)  = \frac{\psi}{2}\left((w_d-\tw)+(w_d-\tv)\right)+ \frac{\mu}{2}\left((m_F-\tw)+(m_F-\tv)\right),\label{KK}\\
&H(\tw,\tv) =  \frac{1}{2}(R(\tw,\tv)-R(\tv,\tw))(\tw-\tv).\label{HH}
\end{align}
Note that the control term, $K(\tw,\tv;m_F)$ depends on two contributions, a steering force towards the desired state $w_d$ and one towards the average opinion of the followers $m_F$, weighted respectively by the parameters $\psi$ and $\mu$, such that $\psi+\mu=1$. 
\end{itemize}

\subsubsection{Boltzmann--type modeling}
Following  \cite{PTa}, for a suitable choice of test functions $\varphi$ we can describe the evolution of $f_F(w,t)$ and $f_L(t)$ via a system of integro-differential equations of Boltzmann type 
\begin{equation}\begin{cases} \vspace{2 mm}\label{boltzmann_equation}
\dfrac{d}{dt}\displaystyle\int_I\varphi(w)f_F(w,t)dw=\left(Q_F(f_F,f_F),\varphi \right)+\left(Q_{FL}(f_F,f_L),\varphi\right), &\\
\dfrac{d}{dt}\displaystyle\int_I\varphi(\tw)f_L(\tw,t)d\tw=\left(Q_L(f_L,f_L),\varphi \right).&
\end{cases}\end{equation}
The operators $Q_F,Q_{FL}$ and $Q_L$ account for the binary exchange of opinions. Under the assumption
%
 that the interaction parameters are such that $|\pw|,|\ppw|,|\ptw|\le 1$ the action of the Boltzmann operators on a (smooth) function $\varphi$ can be written as
\begin{equation}\label{Q_F}
\left(Q_F(f_F,f_F),\varphi\right)=\lambda_F\left<\int_{I^2}(\varphi(\pw)-\varphi(w))f_F(w,t)f_F(v,t)dwdv\right>,
\end{equation}
\begin{equation}\label{Q_FL}
\left(Q_{FL}(f_F,f_L),\varphi\right)=\lambda_{FL}\left< \int_{I^2}(\varphi(\ppw)-\varphi(w))f_F(w,t)f_L(\tv,t)dwd\tv\right>,
\end{equation}
\begin{equation}\label{Q_L}
\left(Q_L(f_L,f_L),\varphi\right)=\lambda_L\left<\int_{I^2}(\varphi(\ptw)-\varphi(\tw))f_L(\tw,t)f_L(\tv,t)d\tw\,d\tv\right>,
\end{equation}
where $\lambda_F,\lambda_{FL},\lambda_L>0$ are constant relaxation rates and, as before, $\langle\,\cdot\,\rangle$ denotes the expectation taken with respect to the random variables characterizing the noise terms.

%
To study the evolution of the average opinions, we can take $\varphi(w)=w$ in \eqref{boltzmann_equation}. In general this leads to a complicated nonlinear system \cite{APZa}, however in the simplified situation of $P$ and $R$ symmetric and $S\equiv 1$ we obtain the following closed system of differential equations for the mean opinions 
\begin{equation}
\left\lbrace
\begin{split}\label{mean_system_hp}
\displaystyle\dfrac{d}{dt}m_L(t)&=\tilde{\eta}_L\psi\beta(w_d-m_L(t))+\tilde{\eta}_L \mu \beta(m_F(t)-m_L(t))\\[+.25cm]
\displaystyle\dfrac{d}{dt}m_F(t)&=\tilde{\eta}_{FL} \alpha (m_L(t)-m_F(t)),
\end{split}
\right.\end{equation}
where we introduced the notations $\tilde{\eta_L}=\rho\eta_L$, $\tilde{\eta}_{FL}=\rho~ \eta_{FL}$ and $m_L(t)=\displaystyle \dfrac{1}{\rho}\int_{I} vf_L(v,t)dv$.

Straightforward computations show that the exact solution of the above system has the following structure
\begin{equation}
\left\lbrace
\begin{aligned}\label{mean_soultion_hp}
m_L(t)&=  C_1 e^{-|\lambda_1| t}+C_2 e^{-|\lambda_2| t}+w_d\\[+.25cm]
m_F(t)&=  C_1\left(1+\frac{\lambda_1}{\beta\mu\tilde{\eta}_L}\right)e^{-|\lambda_1| t}
+C_2\left(1+\frac{\lambda_2}{\beta\mu\tilde{\eta}_L}\right)e^{-|\lambda_2| t}+w_d
\end{aligned}
\right.
\end{equation}
where $C_1,C_2$ depend on the initial data $m_F(0),m_L(0)$ in the following way
\begin{align*}
C_1=&-\frac{1}{\lambda_1-\lambda_2}\left( (\beta\tilde{\eta}_Lm_L(0)  + \lambda_2)m_L(0) - \mu\beta\tilde{\eta}_L m_F(0)-(\lambda_2+\beta\tilde{\eta}_L\psi)w_d\right)\\
C_2=& \quad\frac{1}{\lambda_1-\lambda_2}\left((\beta\tilde{\eta}_Lm_L(0)  + \lambda_1)m_L(0) - \mu\beta\tilde{\eta}_L m_F(0)-(\lambda_1+\beta\tilde{\eta}_L\psi)w_d\right)\end{align*}
with 
\begin{equation*}
\lambda_{1,2}=-\frac{1}{2}\left(\alpha\tilde{\eta}_{FL}+\beta\tilde{\eta}_{L}\right)\pm\frac{1}{2}\sqrt{(\alpha\tilde{\eta}_{FL}+\beta\tilde{\eta}_{L})^2-4\psi\alpha\beta\tilde{\eta}_L\tilde{\eta}_{FL}}.
\end{equation*}
Note that $\lambda_{1,2}$ are always negative, this assures that the contribution of the initial averages, $m_L(0),m_F(0)$, vanishes as soon as time increases and the mean opinions of leaders and followers converge towards the desired state $w_d$. Moreover, in absence of diffusion, it can be shown that the corresponding variance vanishes \cite{APZa}, i.e. under the above assumptions the steady state solutions have the form of a Dirac delta centered in the target opinion $w_d$. 

\subsubsection{Fokker-Planck Modeling}\label{sec:FP}
Once more, the study of the large-time behavior of the kinetic equation \eqref{boltzmann_equation} will take advantage by passing to a Fokker-Planck description. Therefore, similarly to Section \ref{sec:AHP}, we consider the {quasi--invariant opinion} limit~\cite{APZa,PTa,T}, introducing the parameter $\varepsilon>0$, and scaling the  quantities in the binary interaction 
\begin{equation}\label{eq:scaling2}
\begin{aligned}
&\eta=\varepsilon, \qquad \varsigma_F=\sqrt{\varepsilon} \sigma_F, \qquad \varsigma_L=\sqrt{\varepsilon} \sigma_L, \qquad  \varsigma_{FL}=\sqrt{\varepsilon} \sigma_{FL},\\
&\lambda_F=\dfrac{1}{c_F\varepsilon},\qquad \lambda_{FL}=\dfrac{1}{c_{FL}\varepsilon},\qquad \lambda_L=\dfrac{1}{c_L\varepsilon}, \qquad  \beta=\dfrac{2\varepsilon}{\kappa+2\varepsilon}.
\end{aligned}
\end{equation}
The scaled equation \eqref{boltzmann_equation} in the {\em quasi-invariant opinion limit} is well approximated by  
 a Fokker-Planck equation for the followers' opinion distribution
\begin{equation}\label{fokker-planck-follower}
\begin{aligned}
&\dfrac{\partial f_F}{\partial{t}}+\dfrac{\partial}{\partial w}\left(\left(\mathcal{P}[f_F](w)+\mathcal{S}_{}[f_L](w)\right)f_F(w)\right)=\dfrac{\partial ^2}{\partial w^2}\left(\mathcal{D}_F[f_F,f_L](w)f_F(w)\right),
\end{aligned}
\end{equation}
where 
\begin{equation*}
\begin{aligned}
&\mathcal{P}_{}[f_F](w)=\dfrac{1}{c_F}\int_I P(w,w_*)(w_*-w)f_F(w_*)dw_*,\\
&\mathcal{S}_{}[f_L](w)=\dfrac{1}{c_{FL}}\int_I S(w,v_*)(v_*-w)f_L(v_*)dv_*.\\
&\mathcal{D}_F[f_F,f_L](w) =\dfrac{\sigma_F^2}{2 c_F}D_F(w)^2+\dfrac{\sigma_{FL}^2\rho}{2c_{FL}}D_{FL}(w)^2,
\end{aligned}
\end{equation*}
and an equivalent Fokker-Planck equation for the leaders' opinion distribution 
\begin{equation*}\label{fokker-planck-leader}
\begin{aligned}
\dfrac{\partial f_L}{\partial t} + \dfrac{\partial}{\partial \tw}\left(\left(\mathcal{R}[f_L](\tw)+\mathcal{K}[f_L,f_F](\tw)\right)f_L(\tw)\right)=\dfrac{\partial ^2}{\partial \tw^2}\left(\mathcal{D}_L[f_L](\tw)f_L(\tw)\right),
\end{aligned}
\end{equation*}
where
\begin{equation*} 
\begin{aligned}
&\mathcal{R}[f_L](\tw)=\dfrac{\rho}{c_L}\int_I R(\tw,\tv)(\tv-\tw)f_L(\tv)d\tv,\quad \mathcal{D}_L[f_L](\tw) = \dfrac{{\sigma}_L^2\rho}{2c_L}{D}_L^2(\tw),\\
&\mathcal{K}[f_L,f_F](\tilde{w})=\frac{\psi}{\kappa c_L}\left(\tw+m_L(t)-2w_d\right)+\frac{\mu}{\kappa c_L}\left(\tw+m_L(t)-2m_F(t)\right).
\end{aligned}\end{equation*}
In some cases it is possible to recover explicitly the stationary states of the Fokker-Planck system  \eqref{fokker-planck-follower}, \eqref{fokker-planck-leader}. In the simplified case 
where every interaction function is constant and unitary, i.e. $P\equiv S\equiv R\equiv 1$, and 
$D_F(w)={D}_L(w)={D}_{FL}(w)=1-w^2$, we have
\begin{equation}
f_{F,\infty}=\dfrac{a_F}{(1-w^2)^2}\exp\left\{{-\dfrac{2}{b_F}\int_0^{w}\dfrac{z-w_d}{(1-z^2)^2}}dz\right\},\quad b_F=\dfrac{\sigma_{F}^2 c_{FL}+\sigma_{FL}^2 c_F\rho}{c_{FL}+c_F\rho}
\end{equation}
\begin{equation}
f_{L,\infty}=\dfrac{a_L}{(1-\tilde{w}^2)^2}\exp\left\{-\dfrac{2}{b_L}\int_0^{\tilde{w}}\left(\dfrac{z-w_d}{(1-z^2)^2}\right)dz\right\},\quad b_L=\dfrac{\sigma^2_L\rho \kappa}{2 c_L (\psi+\mu)},
\end{equation}
where $a_F$, $a_L$ are suitable normalization constants.
We refer to \cite{APZa} for further details.
%

\subsubsection{Numerical experiments}\label{sec:Numeric}
In this section we present some numerical results concerning the simulation of the Boltzmann type control model \eqref{boltzmann_equation}. All the results have been computed using the Monte Carlo method for the Boltzmann model developed in \cite{APb} in the Fokker-Planck regime $\varepsilon=0.01$ under the scaling (\ref{eq:scaling2}). 

In the numerical tests we assume that the $\rho_L = 0.05$, (five per cent of the population is composed by opinion leaders \cite{DMPW}). Note that, for clarity, in all figures the leaders' profiles have been scaled by a factor $10$. The random diffusion effects have been computed in the case of a uniform random variable with $\sigma_F^2=\sigma_{FL}^2=\varsigma_L^2=0.01$. It is easy to check that the above choice preserves the bounds in the numerical simulations. 
First we present some test cases with a single population of leaders. Then we consider the case of multiple populations of leaders with different time-dependent strategies. This leads to more realistic applications of our arguments, introducing the concept of competition between the leaders. For the sake of simplicity, the interaction functions $P(\cdot,\cdot)$ and $R(\cdot,\cdot)$ are assumed to be constant. The remaining computational parameters have been summarized in Table \ref{tab:par}. 
\begin{table}[h]
\centering
\caption{Computational parameters for the different test cases.}
\begin{tabular}{c|c|c|ccccc|ccccc}
\hline
\hline
Test & $S(\cdot,\cdot)$ & $c_F$& $\hat{c}_{FL}$&$\hat{c}_{L}$ & $\rho$ &$\psi$&$w_d$ \\
\hline
\hline
\#1 & eq. (\ref{eq:bc}) & 1& 0.1&0.1 & 0.05& 0.5& 0.5 & & & & &\\
\hline
\hline
~& $S(\cdot,\cdot)$ & $c_F$& $\hat{c}_{FL_1}$&$\hat{c}_{L_1}$ & $\rho_1$ &$\psi_1$&$w_{d_1}$& $\hat{c}_{FL_2}$&$\hat{c}_{L_2}$ & $\rho_2$ &$\psi_2$&$w_{d_2}$\\
\hline
\hline
\#2& 1 & 1& 0.1&0.1 & 0.05& eq. (\ref{psi_timedep}) & 0.5 & 1&0.1 & 0.05& eq. (\ref{psi_timedep})&-0.5\\
\hline
\end{tabular}
\label{tab:par}
\end{table}

\subsubsection*{Test 1. Leaders driving followers}
In the first test case we consider the system of Boltzmann equations \eqref{boltzmann_equation} with a single population of leaders driving followers.

We assume the initial distributions $f_F\sim U([-1,-0.5])$ and $f_L\sim N(w_d,0.05)$, where $U(\cdot)$ and $N(\cdot,\cdot)$ denote the uniform and the normal distributions respectively.  We consider constant interaction functions $P(\cdot,\cdot)$ and $R(\cdot,\cdot)$ and a bounded confidence-type function for the leader-follower interactions
\be
S(w,\tw)=\chi(|w-\tw|\leq \Delta),
\label{eq:bc}
\ee
with $\Delta = 0.5$. Other parameters are defined in Table \ref{tab:par}, where we used the compact notations $\hat{c}_{FL}={c_{FL}}/{\rho}$ and $\hat{c}_L={c_L}/{\rho}$.

\begin{figure}[ht]
\centering
\includegraphics[scale=.3]{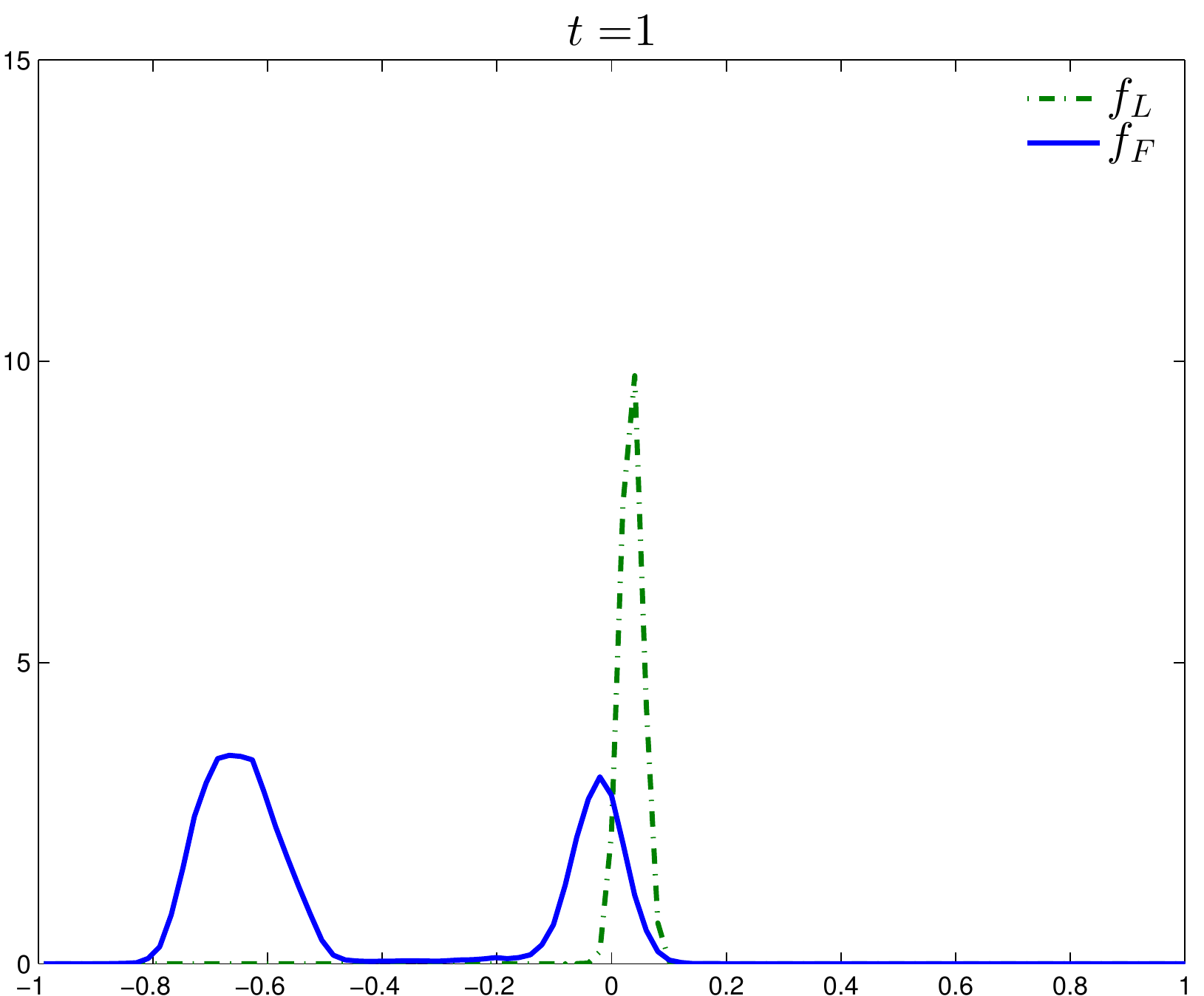}
\includegraphics[scale=.3]{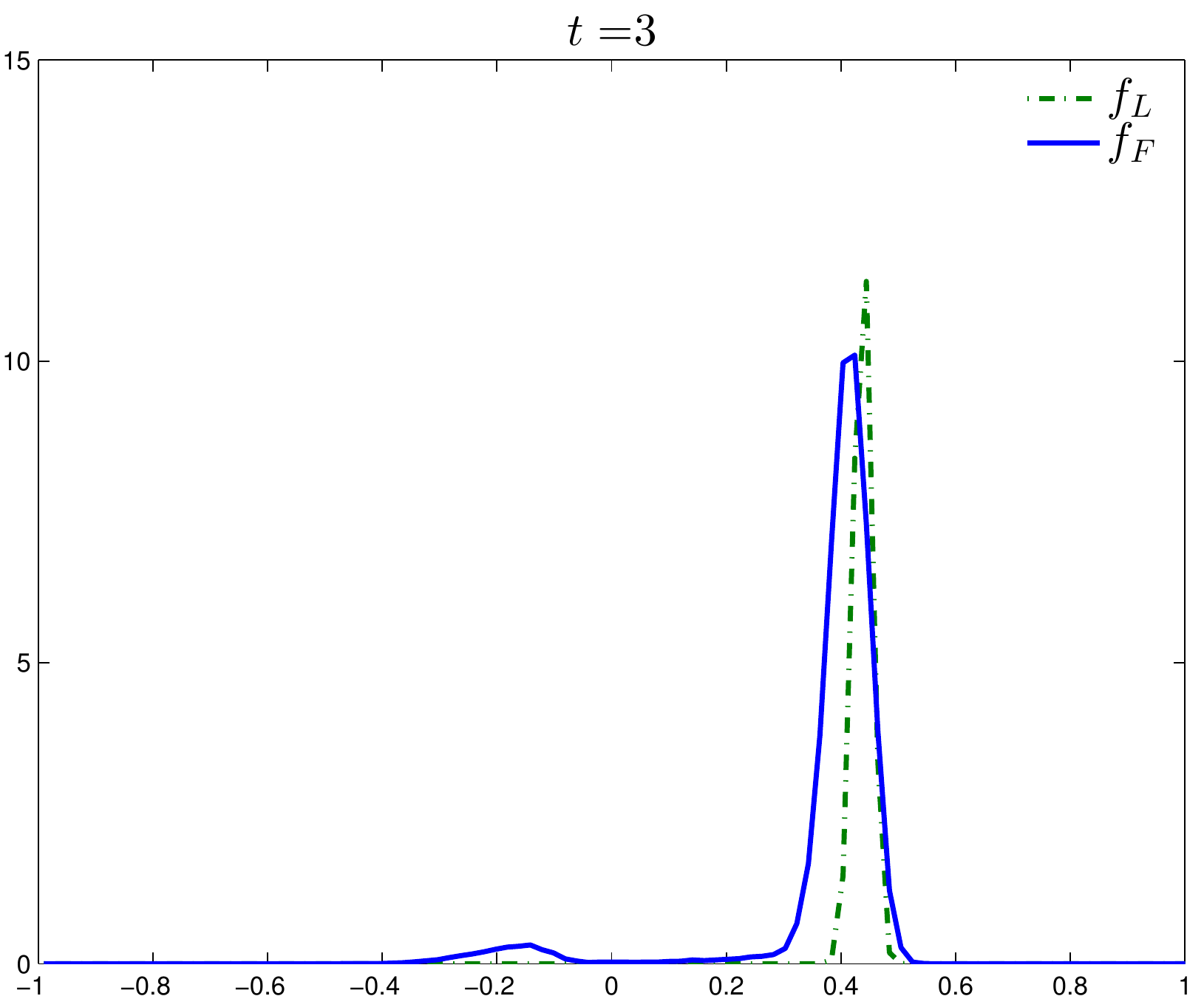}
\caption{Test \#1: Kinetic densities at different times for a single  population of leaders with bounded confidence interaction.
}
\label{fg:dc1}
\end{figure}

In Figure \ref{fg:dc1} we report the evolution, over the time interval $[0, 3]$, of the kinetic densities $f_F(w,t)$ and $f_L(\tw,t)$ .
The numerical experiment shows that the optimal control problem is able to generate a non monotone behavior of $m_L(t)$,  resulting from the combined leaders'strategy of a populists and radical behavior. In an electoral context, this is a characteristic which can be found in populist radical parties, which typically include non-populist ideas and their leadership generates through a dense network of radical movements \cite{MUD}. 


\subsubsection*{Test 2. The case of competing multi-leaders populations}
When more than one population of leaders is present, each one with a different strategy, we describe the evolution of the kinetic density of the system through a Boltzmann approach. Let $M> 0$ be the number of families of leaders, each of them described by the density $f_{L_j}, j=1,...,M$ such that
\begin{equation}
\int_I f_{L_j}(\tw)d\tw=\rho_{j}.
\end{equation}
If a unique population of followers is present, with density $f_F$, a follower interacts both with the others agents from the same population and with every leader of each $j$-th family. Given a suitable test function $\varphi$ the evolution of the densities is given by the system of Boltzmann equations 
\begin{equation}\begin{cases} \vspace{2 mm}
\dfrac{d}{dt}\displaystyle\int_I \varphi(w)f_F(w,t)dw=\left(Q_F(f_F,f_F),\varphi\right)+\displaystyle\sum_{k=1}^{M}\left(Q_{FL}(f_{L_k},f_F),\varphi\right), &\\ 
\dfrac{d}{dt}\displaystyle\int_I\varphi(\tilde{w})f_{L_j}(\tilde{w},t)d\tilde{w}=(Q_L(f_{L_j},f_{L_j}),\varphi), \qquad j=1,\ldots,M&.
\end{cases}\end{equation} 
By assuming that the leaders aim at minimizing cost functionals of the type (\ref{JL_cost}), the differences consist in two factors: in the target opinions $w_{d_j}$ and in the leaders' attitude towards a radical ($\psi_j\approx 1$) or populist strategy ($\mu_j\approx 1$).

\begin{figure}[t]
\centering
\includegraphics[scale=0.3]{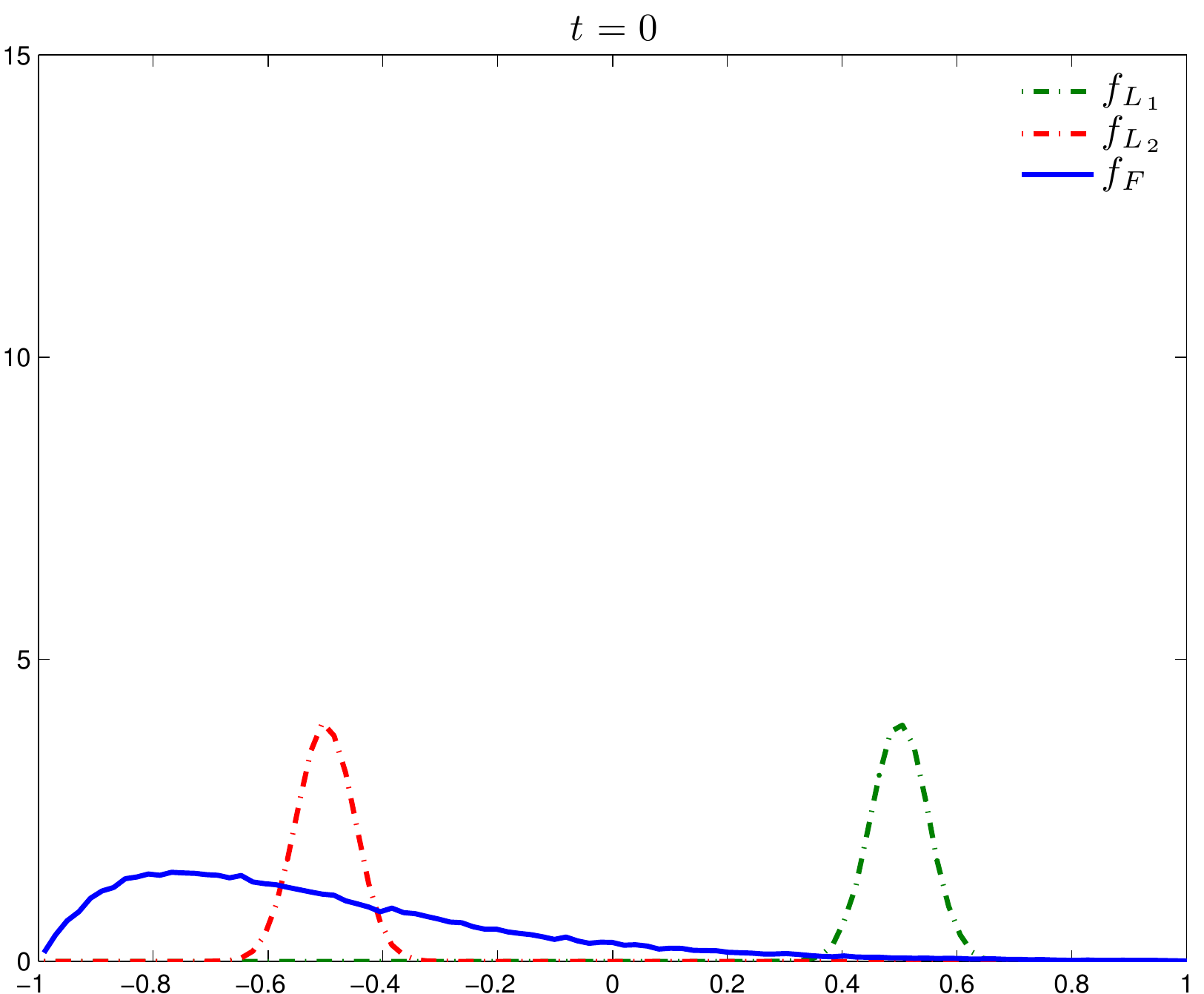}
\includegraphics[scale=0.3]{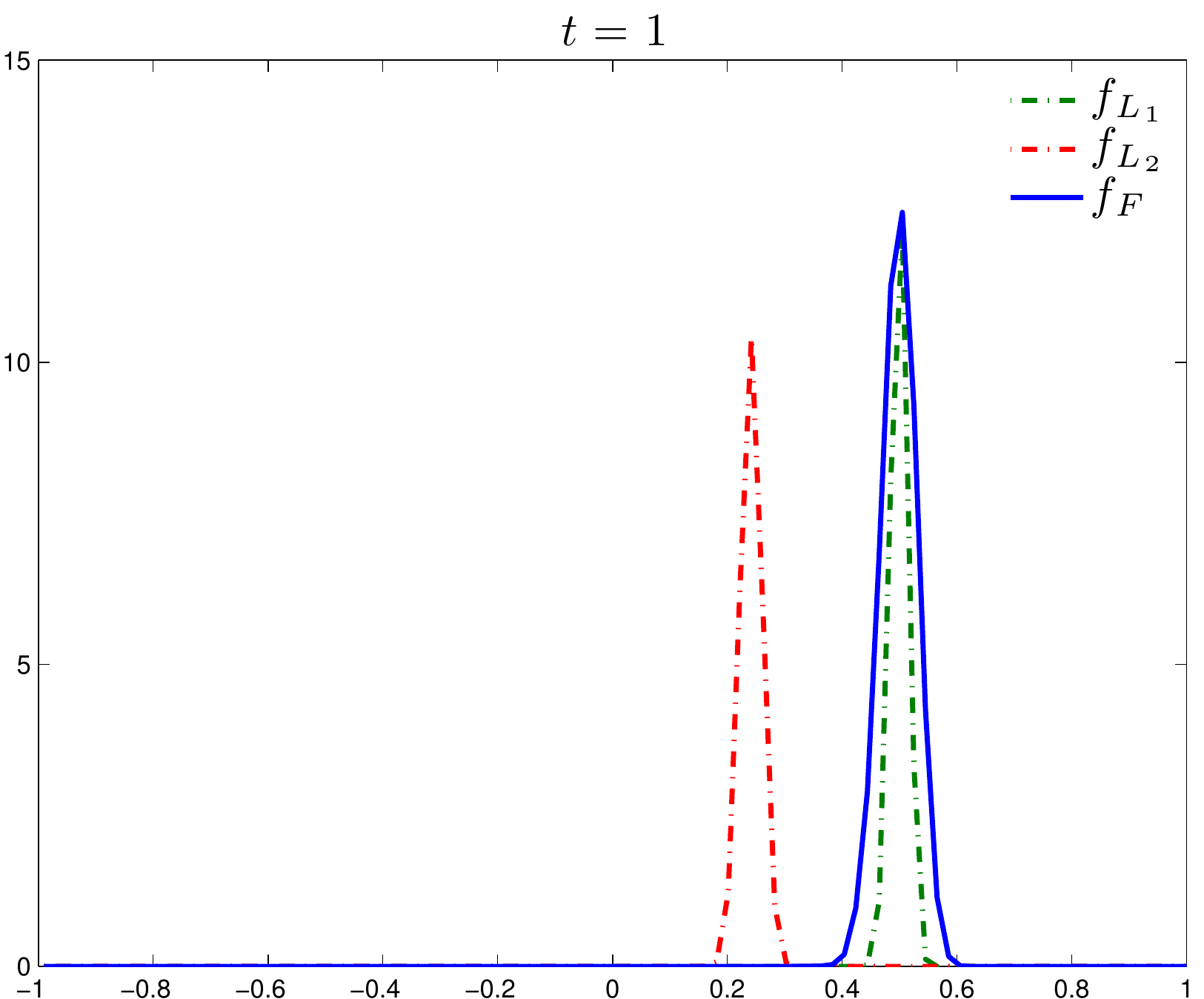}
\\
\includegraphics[scale=0.3]{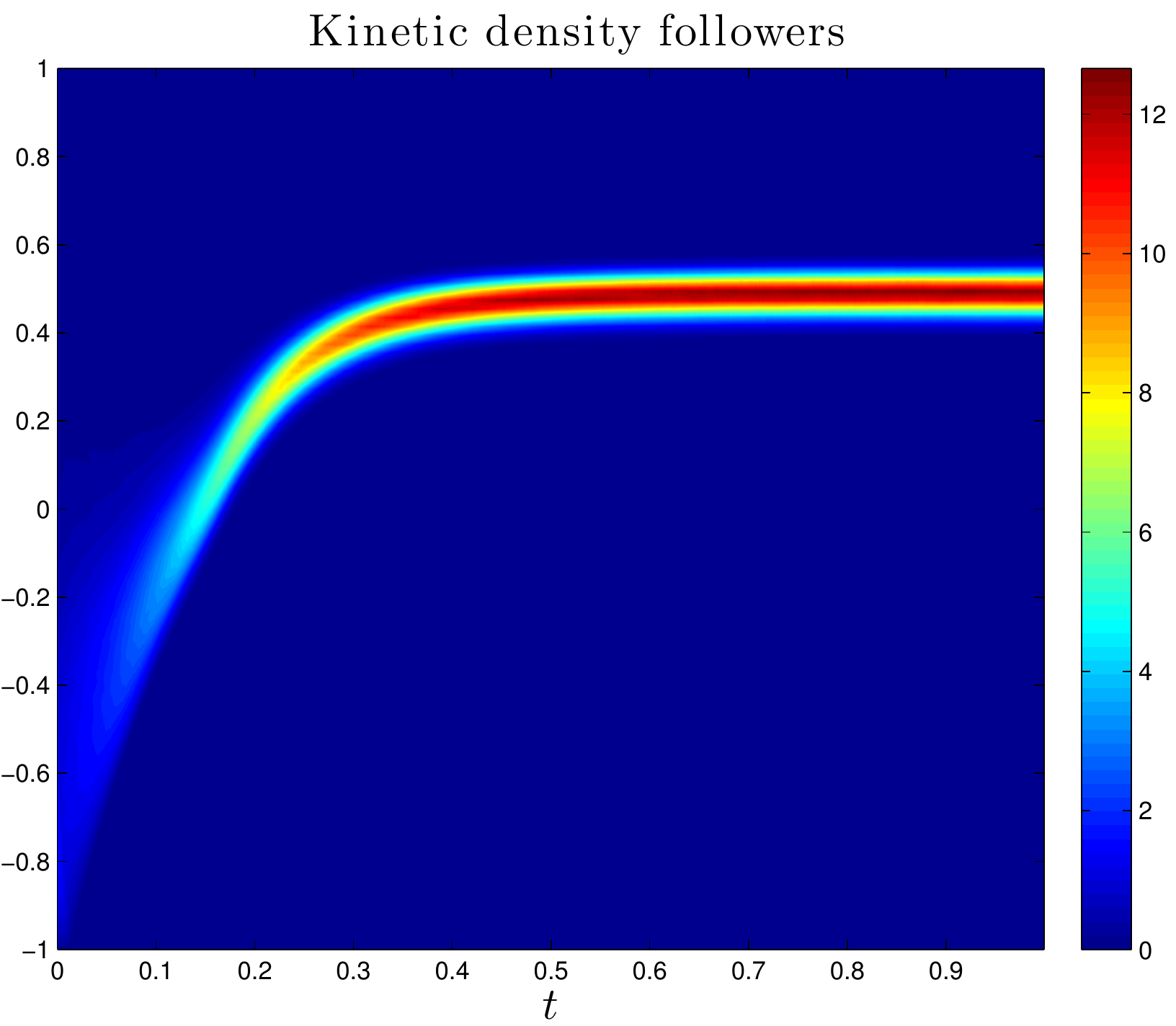}
\includegraphics[scale=0.3]{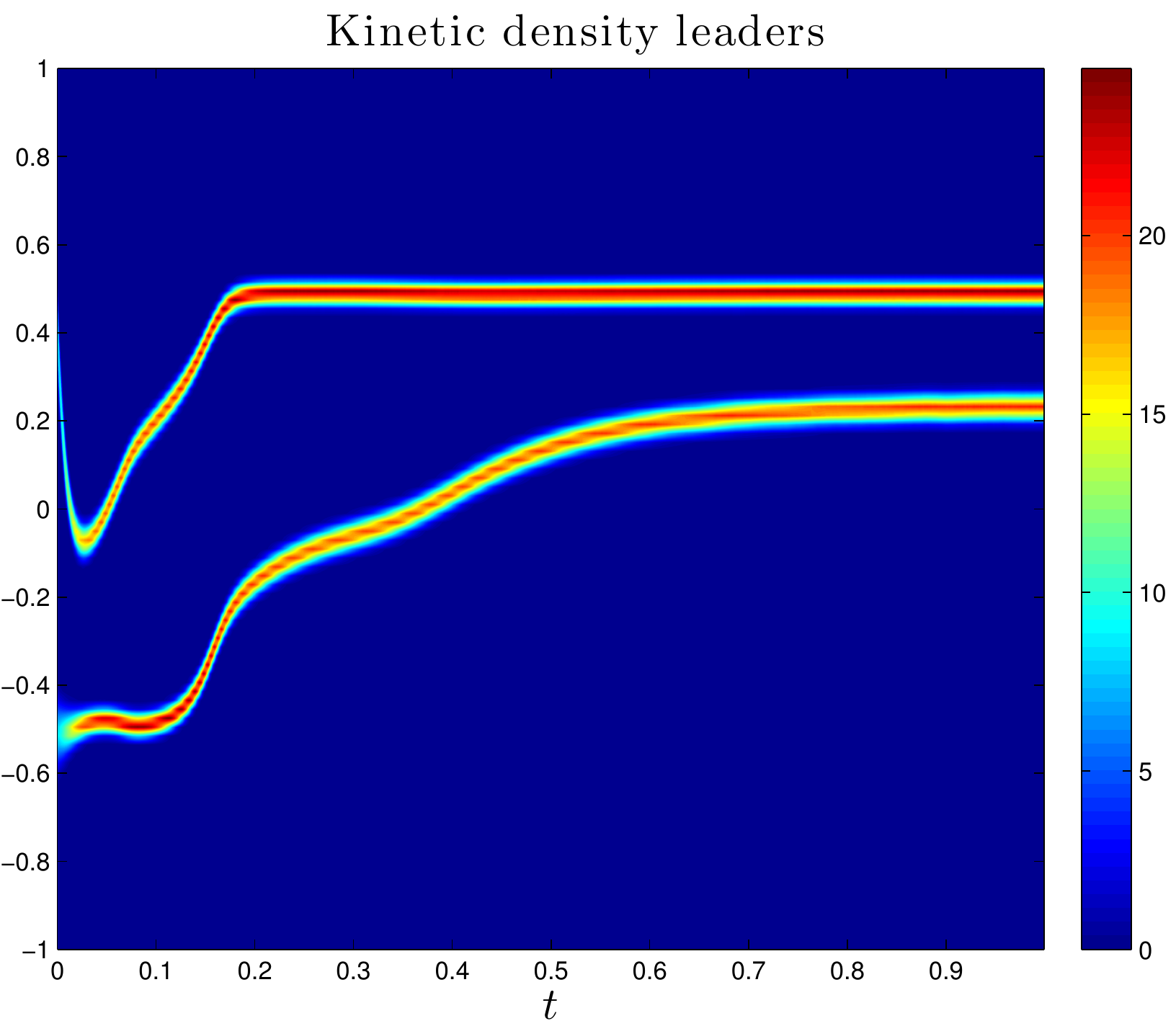}
\caption{Test \#2: Kinetic densities at different times for  for a two populations of leaders model with time dependent strategies (top row). Kinetic density evolution over the time interval $[0,1]$ (bottom row).}
\label{fig:2pop_time_dep}
\end{figure}

To include {\em competition} between different populations, we introduce time-dependent coefficients in the leaders'strategies. This approach leads to the concept of {\em adaptive strategy } for every family of leaders $j=1,...,M$.
Thus we assume that coefficients $\psi$ and $\mu$ which appear into the functional now evolve in time and are defined for each $t \in [0,T]$ as 
\begin{equation}\begin{split}\label{psi_timedep}
\psi_j(t)=&\dfrac{1}{2}{\int_{w_{d_j}-\delta}^{w_{d_j}+\delta}f_F(w)dw}+\dfrac{1}{2}\int_{{m_{L_j}}-\bar{\delta}}^{{m_{L_j}}+\bar{\delta}}f_F(w)dw, \qquad
\mu_j(t)=1-\psi_j(t)
\end{split}\end{equation}
where both $\delta,\bar{\delta} \in [0,1]$ are fixed and $m_{L_j}$ is the average opinion of the $j$th population of leaders. The introduced choice of coefficients is equivalent to consider a competition between the populations of leaders, where each leader try to adapt its populist or radical attitude accordingly to the success of the strategy. Note also that the success of the strategy is based on the local perception of the followers.  

In the numerical experiments reported in Figure \ref{fig:2pop_time_dep} 
we take into account two populations of leaders, initially normally distributed with mean values $w_{d_1}$ and $w_{d_2}$ and parameters $\delta=\bar{\delta}=0.5$, respectively, and a single population of followers, represented by a skewed distribution $f_F \sim \Gamma (2,\frac{1}{4})$ over the interval $[-1,1]$, where $\Gamma(\cdot,\cdot)$ is the Gamma distribution. Here the frequencies of interactions are assumed to be unbalanced since $\hat{c}_{FL_1}=0.1$ and $\hat{c}_{FL_2}=1$. In the test case we assume that the follo\-wers group has an initial natural inclination for the position represented by one leader but, thanks to communication strategies pursued by the minority leader, it is driven to different positions (see Figures \ref{fig:2pop_time_dep}). In a bipolar electoral context, an example of the described behavior would consist of a better use of the media in a coalition with respect to the opponents. 

\section{Multivariate models}\label{sec3}
In several recent works additional variables have been introduced quantifying relevant indicators for the spreading of opinions \cite{APZc,APZd,BrTo,DM,DW,PVZ}. In this class of  models  the opinion dynamics depends on an additional parameter, continuous or discrete, which influences the binary exchanges. 
We present in this section two kinetic multivariate models. The first takes into account a continuous variable called \emph{conviction} representing the strength of individuals in pursuing their opinions. Afterwards we develop a model for the dynamic of opinions in large evolving networks where the \emph{number of connections} of each individuals, a discrete variable, influences the dynamics. 
  
\subsection{The role of conviction}

Resembling the model for wealth exchange in a multi-agent society introduced in \cite{ChaCha}, this new model has an additional  parameter to quantify the personal \emph{conviction}, representing a measure of the influencing ability of individuals \cite{BrTo}. Individuals with high conviction
are resistant to change opinion, and have a prominent role in attracting other
individuals towards their opinions. In this sense, individuals with high conviction
play the role of leaders \cite{DMPW}.

The goal is to study the evolution of a multi-agent system
characterized by two variables, representing conviction and opinion,
where the way in which conviction is formed is
independent of the personal opinion. Then, the (personal) conviction parameter will
enter into the microscopic binary interactions for opinion formation considered in
Section \ref{standard}, to modify them in the compromise and self-thinking terms.  A typical and natural assumption
is that high conviction could act on the interaction process both to reduce the
personal propensity to compromise, and to reduce the self-thinking.
 Numerical investigation shows that the role of the
additional conviction variable is to bring the system towards a steady distribution in
which there is formation of clusters even in absence of bounded confidence hypotheses
\cite{BN,BNKR,BNKVR, HK}.

\subsubsection{The formation of conviction}\label{know6}

Let us briefly summarize the key points at the basis of  the model for conviction \cite{BrTo}.
 Each variation  is interpreted as an interaction where a fraction of the
conviction of the individual is lost by virtue of afterthoughts and insecurities,
while at the same time the individual can absorb  a certain amount of conviction
through the information achieved from the external background (the surrounding
environment). In this approach, the conviction of the individual is quantified in terms
of a scalar parameter $x$, ranging from zero to infinity.  Denoting with $z \ge 0$ the
degree of conviction achieved from the background, it is assumed that the new amount
of conviction  in a single interaction can be computed as
 \be\label{k1}
 x^* = (1-\lambda(x))x + \lambda_B(x) z + \vartheta H(x).
 \ee
In \eqref{k1} the functions $\lambda= \lambda(x)$ and $\lambda_B= \lambda_B(x)$
quantify, respectively, the personal amounts of insecurity and willingness to be
convinced by others, while $\vartheta$ is a random parameter which takes into account the
possible unpredictable modifications of the conviction process. We will in general fix
the mean value of $\vartheta$ equal to zero. Last, $H(\cdot)$ will denote an increasing
function of conviction. The typical choice is to take $H(x)= x^\nu$, with $0<\nu\le
1$. Since some insecurity is always present, and at the same time it can not exceed a
certain amount of the total conviction, it is assumed that $\lambda_- \le \lambda(x)
\le \lambda_+$, where $\lambda_- >0$, and $\lambda_+ < 1$. Likewise, we will assume an
upper bound for the willingness to be convinced by the environment. Then, $0 \le
\lambda_B(x)\le \bar\lambda$, where $\bar\lambda <1$.  Lastly, the random part is
chosen to satisfy the lower bound $\vartheta \ge - (1-\lambda_+)$. By these assumptions,
it is assured that the post-interaction value
 $x^*$ of the conviction is nonnegative.

Let $C(z)$, $z \ge 0$ denote the probability distribution of
degree of conviction of the (fixed) background. We will suppose that
$C(z)$ has a bounded mean, so that
 \be\label{ba1}
 \int_{\R_+} C(z) \, dz = 1; \quad \int_{\R_+} z\,C(z) \, dz = M
  \ee
We note that the distribution of the background will induce a certain policy of
acquisition of conviction. This aspect has been discussed in \cite{PTb}, from which
we extract the example that follow. Let us assume that the background is a random
variable uniformly distributed on the interval $(0, a)$, where $a > 0$ is a fixed
constant. If we choose for simplicity $\lambda(x) = \lambda_B(x)= \bar\lambda$, and
the individual has a degree of conviction $x>a$, in absence of randomness the
interaction  will always produce a value $x^* \le x$, namely a partial decrease of
conviction. In this case, in fact, the process of insecurity in an individual with
high conviction can not be restored by interaction with the environment.

The study of the
time-evolution of the distribution of conviction produced by binary
interactions  of type \eqref{k1}  can be obtained by
resorting to kinetic collision-like models \cite{PTa}. Let $F=
F(x,t)$ the density of agents which at time $t >0$ are represented by their
conviction $x \in \R_+$. Then, the time
evolution of $F(x, t)$  obeys to a
Boltzmann-like equation. This equation is usually written
in weak form. It corresponds to say that the solution $F(x,t)$
satisfies, for all smooth functions $\varphi(x)$ (the observable quantities)
 \begin{equation}
 \begin{aligned}
  \label{kine-w}
 &\frac{d}{dt}\int_{\R_+}F(x,t)\varphi(x)\,dx  =  \Big \langle \int_{\R_+^2} \bigl( \varphi(x^*)-\varphi(x) \bigr) F(x,t)C(z)
\,dx\,dz \Big \rangle,
\end{aligned}
 \end{equation}
where $x^*$ is the post-interaction conviction and $\langle \cdot \rangle$ denotes the expectation with respect to the random parameter $\vartheta$ introduced in \eqref{k1}.
Through the techniques analyzed in the previous sections of this work we can derive the asymptotic solution of the Fokker-Planck equation which follows from \eqref{kine-w} in the limit $\varepsilon\rightarrow 0$.

If $\lambda(x) = \lambda$ and $\lambda_B(x) = \lambda_B$ we get
the explicit form of the steady distribution of conviction \cite{BrTo,PTa}. We will
present two realizations of the asymptotic profile, that enlighten the consequences of
the choice of a particular function $H(\cdot)$. First, let us consider the case in
which $H(x)=x$. In this case, the Fokker--Planck equation coincides with the
one obtained in \cite{CPT}, related to the steady distribution of wealth in a
multi--agent market economy. One obtains
 \be\label{stead1}
 G_\infty(x) = \frac{G_0}{x^{2 + 2\lambda/\mu}} \exp \left\{ - \frac{2\lambda_B M}{\mu x}\right\},
 \ee
where the constant $G_0$ is chosen to fix the total mass of $G_\infty(x)$
equal to one. Note that the steady profile is heavy tailed, and the size of the
polynomial tails is related to both $\lambda$ and $\sigma$. Hence, the percentage of
individuals with high conviction is decreasing as soon as the parameter $\lambda$ of
insecurity is increasing, and/or the parameter of self-thinking is decreasing. It is
moreover interesting to note that the size of the parameter $\lambda_B$ is important
only in the first part of the $x$-axis, and contributes to determine the size of the
number of undecided. 

The second case refers to the choice $H(x) = \sqrt x$. Now, people with high
conviction is more resistant to change (randomly) with respect to the previous case.
On the other hand, if the conviction is small, $x < 1$, the individual is less
resistant to change. Direct computations now show that the steady profile is given by
 \be\label{stead2}
 H_\infty(x) = {H_0}\,{x^{-1 + (2\lambda_B M)/\mu}} \exp \left\{ - \frac{2\lambda}{\mu} x \right\},
 \ee
where the constant $H_0$ is chosen to fix the total mass of $H_\infty(x)$ equal to
one. At difference with the previous case, the distribution decays exponentially to
infinity, thus describing a population in which there are very few agents with a large
conviction. Moreover, this distribution describes a population with a huge number of
undecided agents. Note that, since the exponent of $x$ in $H_\infty(\cdot)$ is
strictly bigger than $-1$, $H_\infty(\cdot)$ is integrable for any choice of the
relevant parameters.

\subsubsection{The Boltzmann equation for opinion and conviction}\label{model}

 In its original formulation \eqref{ch6:weak boltz}
both the compromise  and the self-thinking intensities  were assigned in terms of the
universal constant $\eta$ and of the universal random parameters
$\xi,\xi_*$. Suppose now that these quantities in \eqref{ch6:trade_rule} could
depend of the personal conviction of the agent. For example, one reasonable assumption
would be that an individual with high personal conviction is more resistant to move
towards opinion of any other agent by compromise. Also, an high conviction could imply
a reduction of the personal self-thinking. If one agrees with these assumptions, the
binary trade \eqref{ch6:trade_rule} has to be modified to include the effect of
conviction. Given two agents $A$ and $B$ characterized by the pair $(x,w)$
(respectively $(y,\v)$) of conviction and opinion, the new binary trade between $A$ and
$B$ now reads
\begin{equation}\begin{aligned}
  \label{eq.cpt2}
& \pw =  w - \eta\,\Psi(x)P(w)( w- \v) + \Phi(x)\xi D(w), \\
&\null \\[-.25cm]
 & \pv  =  \v - \eta\,\Psi(y) P(\v)(\v- w) + \Phi(y)\xi_* D(\v) .
\end{aligned}\end{equation}
In \eqref{eq.cpt2} the personal compromise propensity and self-thinking of the agents
are modified by means of the functions $\Psi =\Psi(x)$ and $\Phi= \Phi(x)$, which
depend on the convictions parameters. In this way, the outcome of the interaction
results from a combined effect of (personal) compromise propensity, conviction and
opinion. Among other possibilities, one reasonable choice is to fix the functions
$\Psi(\cdot)$ and $\Phi(\cdot)$ as non-increasing functions. This reflects the idea
that the conviction acts to increase the tendency to remain of the same opinion. Among others,
a possible choice is
 \[
\Psi(x) = (1+(x-A)_+)^{-\alpha}, \quad \Phi(x) = (1+(x-B)_+)^{-\beta}.
 \]
Here  $A, B, \alpha, \beta$ are nonnegative constants, and $h(x)_+$ denotes the positive part of $h(x)$. By choosing $A>0$ (respectively $B>0$), conviction will start to influence the change of opinion only when $x >A$ (respectively $x>B$).
It is interesting to remark that the presence of the conviction parameter (through the
functions $\Psi$ and $\Phi$), is such that the post-interaction opinion of an agent
with high conviction remains close to the pre-interaction opinion. This induces a
mechanism in which the opinions of agents with low conviction are attracted towards
opinions of agents with high conviction.

Assuming the binary trade \eqref{eq.cpt2} as the microscopic binary exchange of
conviction and opinion in the system of agents, the joint evolution of these
quantities is described in terms of the density $f= f(x,w,t)$ of agents which at time
$t >0$ are represented by their conviction $x \in \R_+$ and wealth $w\in I$. The
evolution in time of the density $f$ is described by the following kinetic equation
(in weak form) \cite{PTa}
\begin{equation}
 \begin{aligned}  \label{kine-xv}
&\frac{d}{dt} \int_{\R_{+}\times I} \varphi(x,w) f(x,w,t)\,dx\, dw  = \\
&\qquad\qquad\frac 12 \Big \langle \int_{\R_+^2\times I^2} \bigl( \varphi(x',\pw) +
\varphi(y',\pv) -\varphi(x,w) - \varphi(y,v) \bigr) \\
&\qquad\qquad f(x,w,t)f(y,\v,t)C(z) \,dx\,dy\,dz\,dw\, d\v \Big \rangle.
\end{aligned}
 \end{equation}
In \eqref{kine-xv} the pairs $(x', w)$ and  $(y', \v)$ are obtained from the pairs
$(x,w)$ and $(y,\v)$  by \eqref{k1} and \eqref{eq.cpt2}.  Note that, by choosing $\varphi$
independent of $w$, that is $\varphi=\varphi(x)$, equation \eqref{kine-xv} reduces to
the equation \eqref{kine-w} for the marginal density of conviction $F(x,t)$.

To obtain analytic solutions to the Boltzmann-like equation \eqref{kine-xv} is
prohibitive. The main reason is that the unknown density in the kinetic equation
depends on two variables with different laws of interaction. In addition, while the
interaction for conviction does not depend on the opinion variable, the law of
interaction for the opinion does depend on the conviction. Also, at difference with
the one-dimensional models, passage to Fokker-Planck equations (cf. \cite{PTb} and
the references therein) does not help in a substantial way. For this reason, we will
resort to numerical investigation of \eqref{kine-xv}, to understand the effects of the
introduction of the conviction variable in the distribution of opinions.

\subsubsection{Numerical experiments}\label{nume}

This section contains a numerical description of the solutions  to the Boltzmann-type
equation \eqref{kine-xv}. For the numerical approximation of the Boltzmann equation we
apply a Monte Carlo method, as described in Chapter 4 of \cite{PTa}. If not otherwise
stated the kinetic simulation has been performed with $N=10^4$ particles.

The numerical experiments will help to clarify the role of conviction in the final
distribution of the opinion density among the agents. The numerical simulations
enhance the fact that the density $f(x,w,t)$ will rapidly converge towards a
stationary distribution \cite{PTa}. As usual in kinetic theory, this stationary
solution will be reached in an exponentially fast time.

The numerical experiments will report the joint density of conviction and opinion in
the agent system. The opinion variable will be reported on the horizontal axis, while the conviction variable will be reported on the vertical one. The color intensity will refer to the concentration of opinions. The following numerical tests have been considered.

\subsubsection*{Test 1}
In the first test we consider the case of a conviction interaction where the diffusion
coefficient in \eqref{k1} is linear, $H(x)=x$.  As described in Section \ref{know6} the
distribution of conviction in this case is heavy tailed, with an important presence of
agents with high conviction, and a large part of the population with a mean degree of
conviction. In \eqref{eq.cpt2} we shall consider
\[
\Phi(x)= \Psi(x) = \frac 1{1+(x-1)_+}.
\]
We further  take $\lambda=\lambda_B=0.5$ in \eqref{k1}, and $P(w)=1$, $D(w)
=\sqrt{1-w^2}$ in \eqref{eq.cpt2}.  We consider a population of agents with an initially uniformly distributed opinion and a conviction uniformly distributed on the interval $[0,5]$. We choose a time step of $\Delta t=1$ and a final
computation time of $t=50$, where the steady state is practically reached. 

\begin{figure}[t]
\centering
\includegraphics[scale=.30]{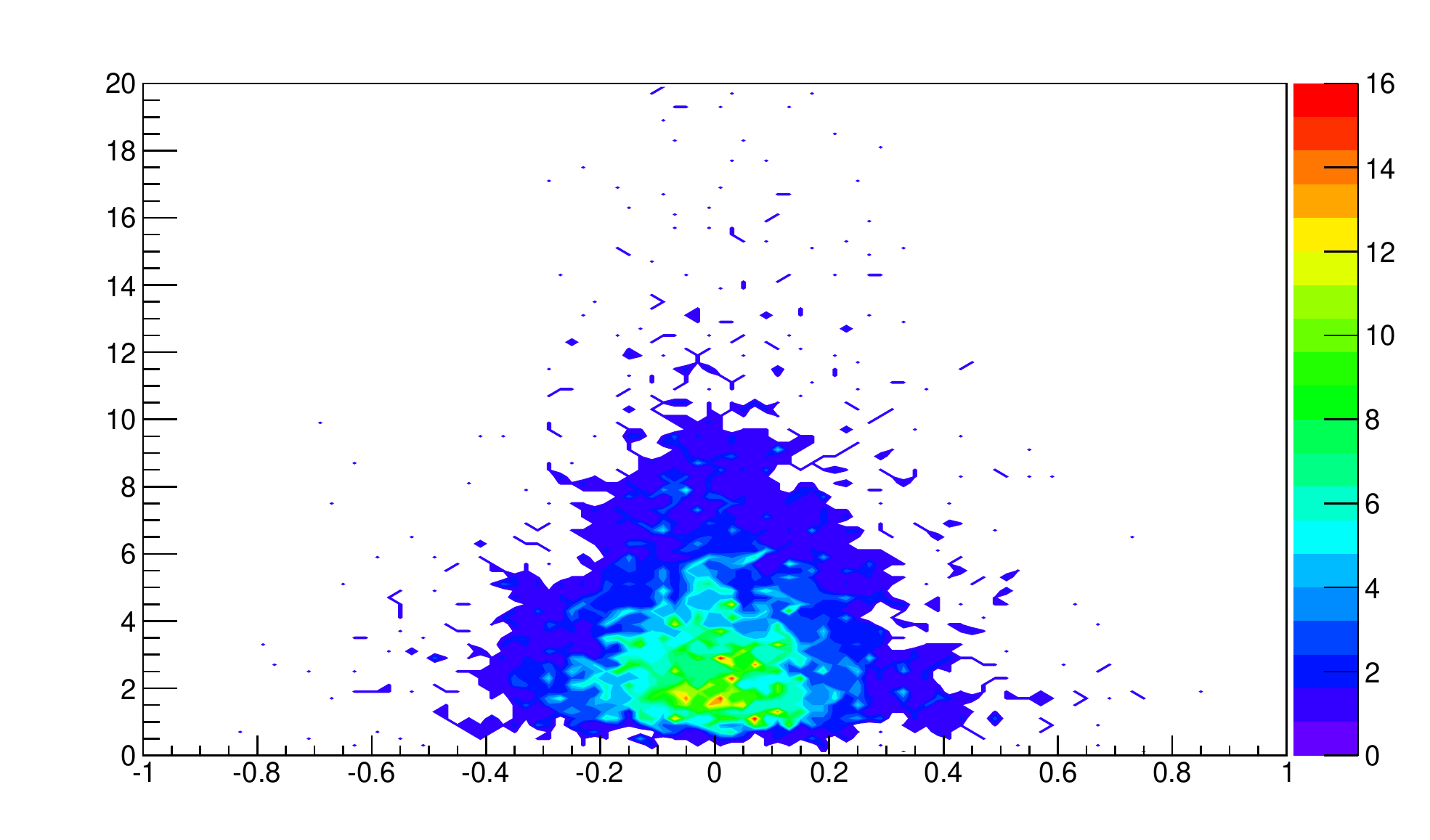}
\includegraphics[scale=.30]{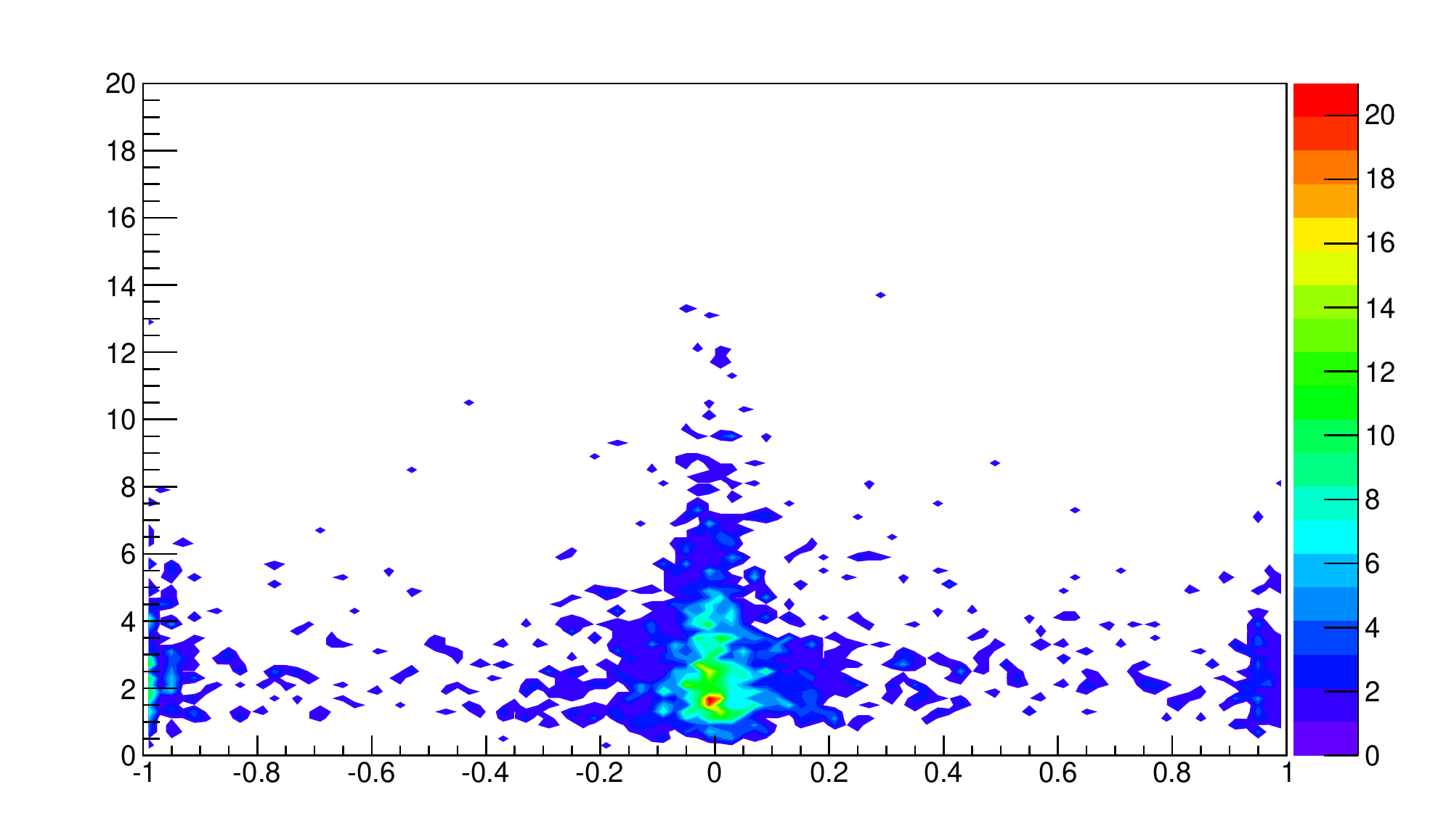}
\caption{Test 1: The particles solution with  $N=10000$ particles and linear $H$. High diffusion in conviction and reduced self-thinking (up) compared to low diffusion in conviction and high self-thinking (down) }
\label{fg:fig1}
\end{figure}

Since the
evolution of the conviction in the model is independent from the opinion, the latter
is scaled in order to fix the mean equal to $0$. We report the results for the
particle density corresponding to different values of $\mu$, $\eta$ and the
variance $\varsigma^2$ of the random variables $\xi$ and $\xi_*$ in Figure
\ref{fg:fig1}. This allows to verify the essential role of the diffusion processes in conviction and opinion formation.  

\subsubsection*{Test 2}
\begin{figure}[t]
\centering
\includegraphics[scale=.30]{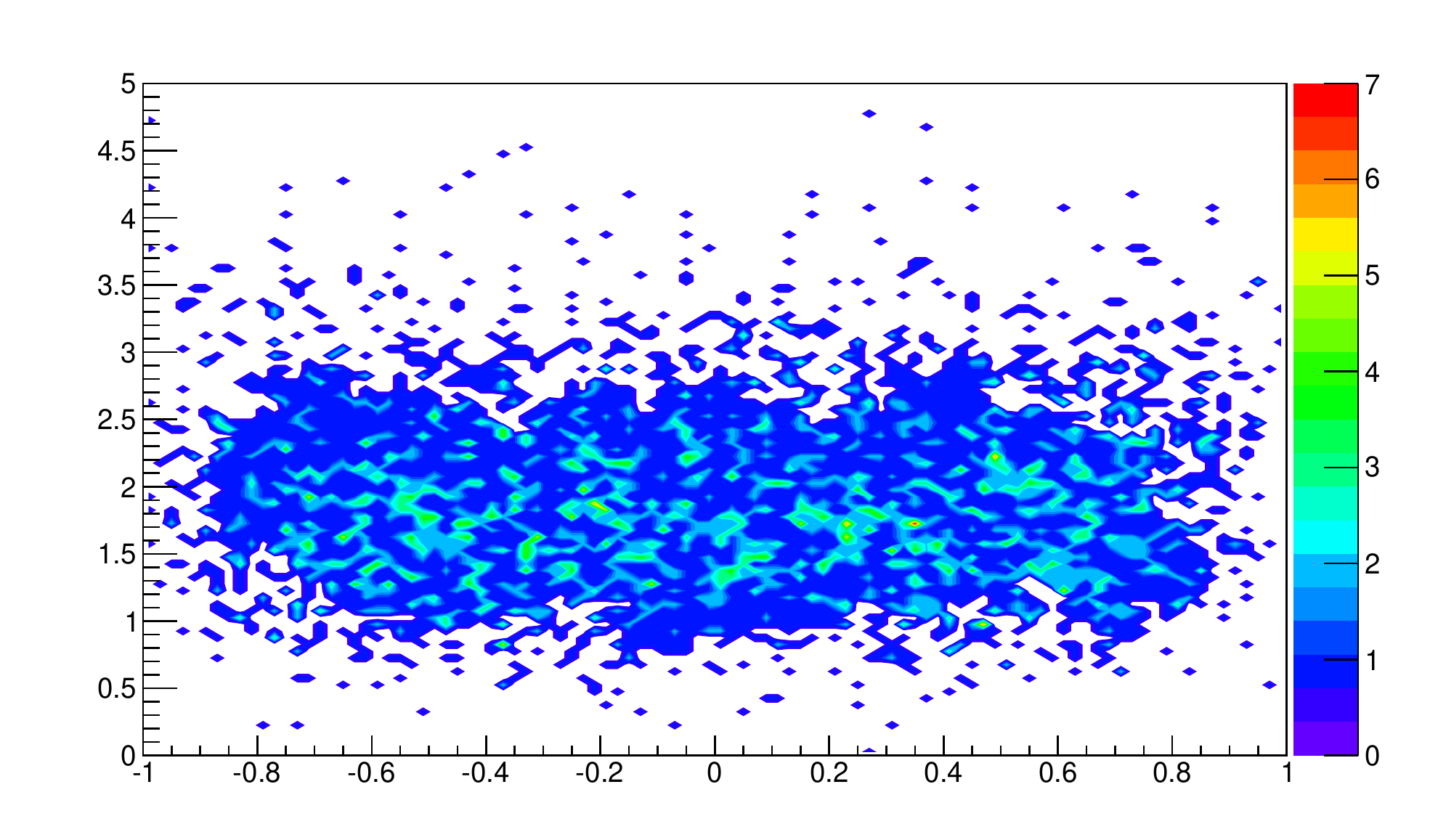}
\includegraphics[scale=.30]{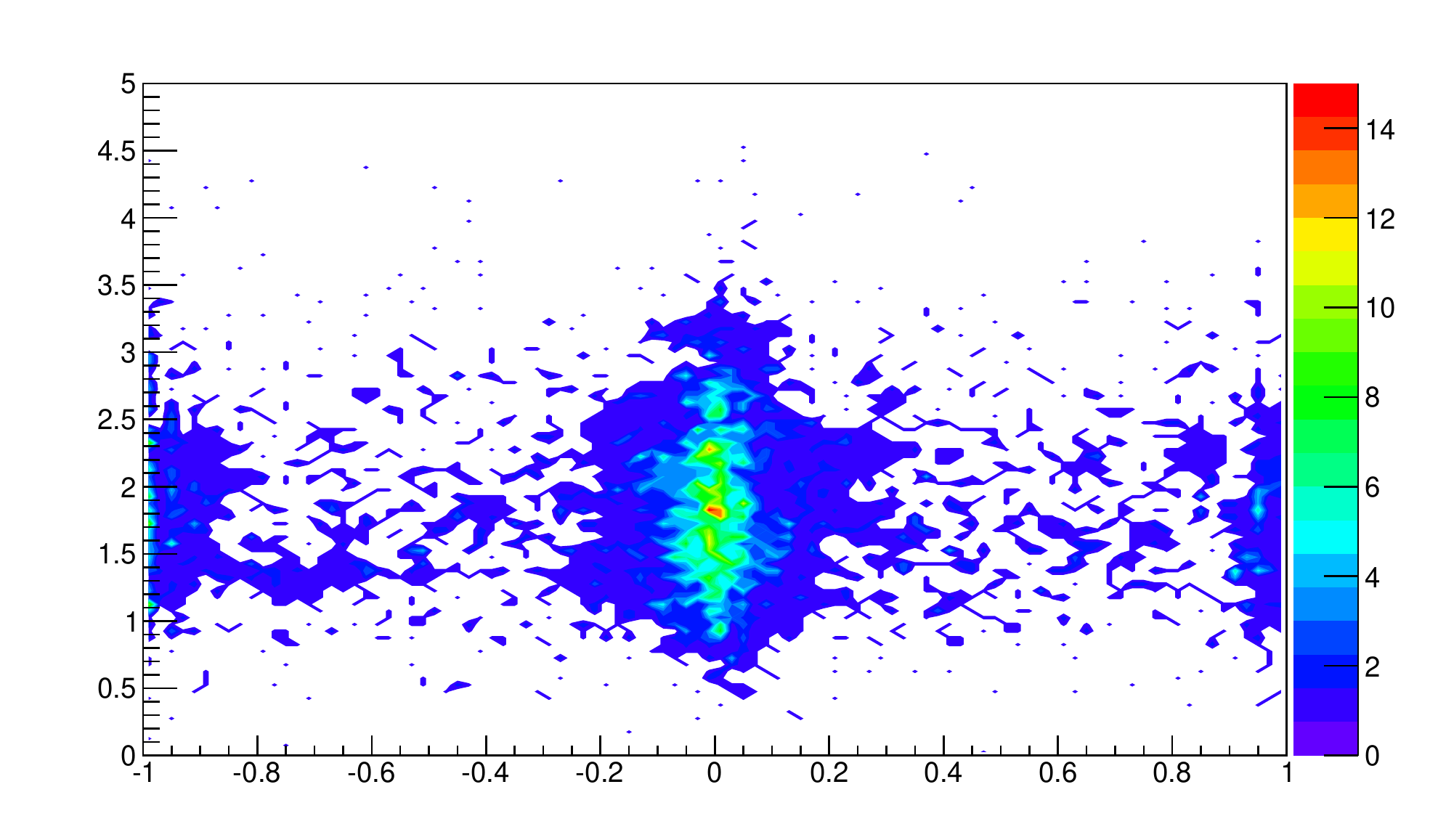}
\caption{Test 2: The particles solution with  $N=10000$ particles and $H(x) = \sqrt{x}$. High diffusion in conviction and reduced self-thinking (up) compared to low diffusion in conviction and high self-thinking (down).}
\label{fg:fig2}
\end{figure}

In this new test, we maintain the same values for the parameters, and we modify the
diffusion coefficient in \eqref{k1}, which is now assumed as $H(x)=\sqrt x$. Within this
choice, with respect to the previous test we expect the formation of a larger class on
undecided agents.  The results are reported in Figure \ref{fg:fig2} for the full
density. At difference with the results of Test 1, opinion is spread out almost uniformly among people with low conviction. It is remarkable that in this second test, as expected, conviction is essentially distributed in the interval $[0,5]$, at difference with Test 1, where agents reach a conviction parameter of $20$.

\begin{figure}[t]
\centering
\includegraphics[scale=0.30]{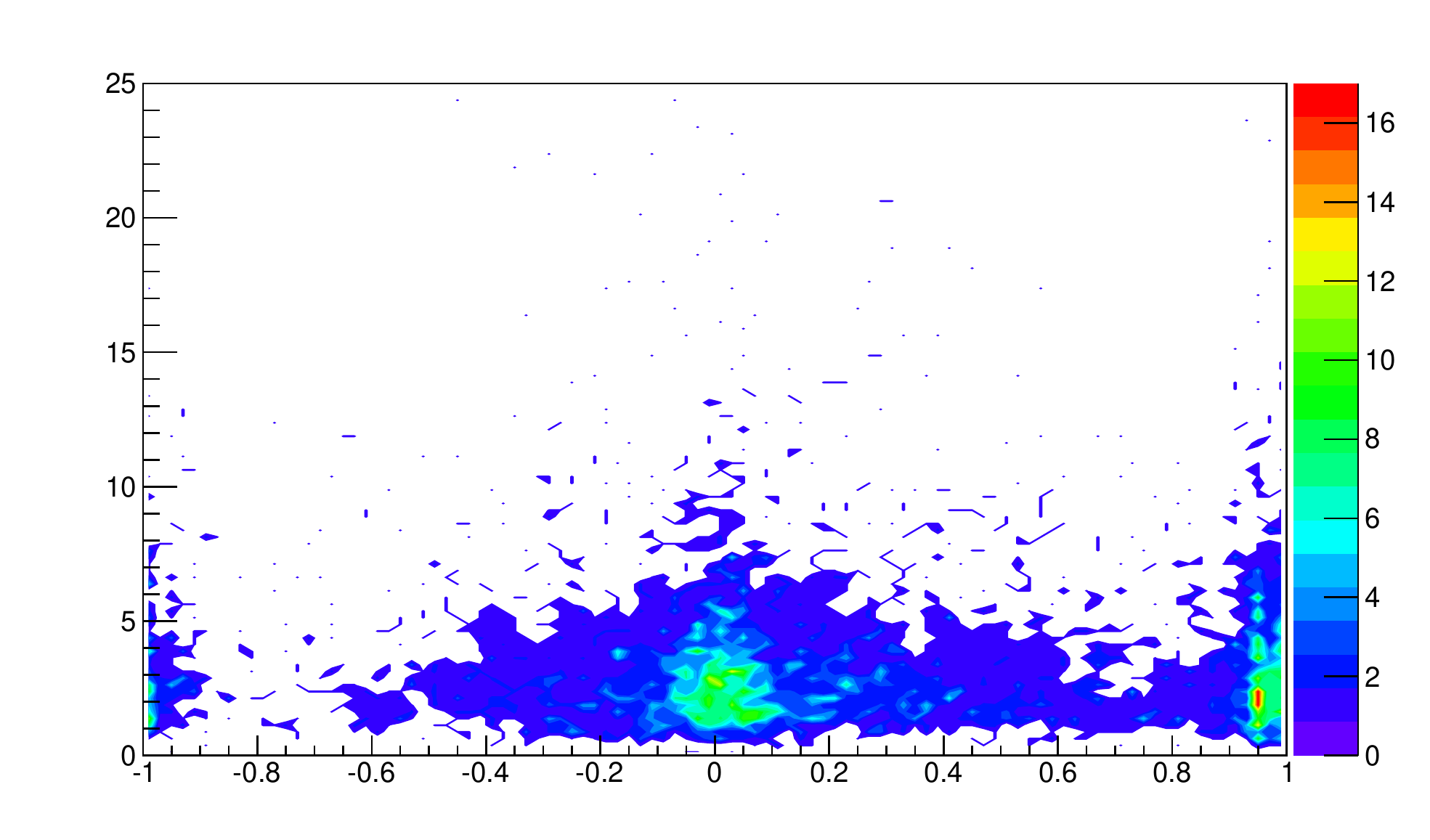}
\includegraphics[scale=0.30]{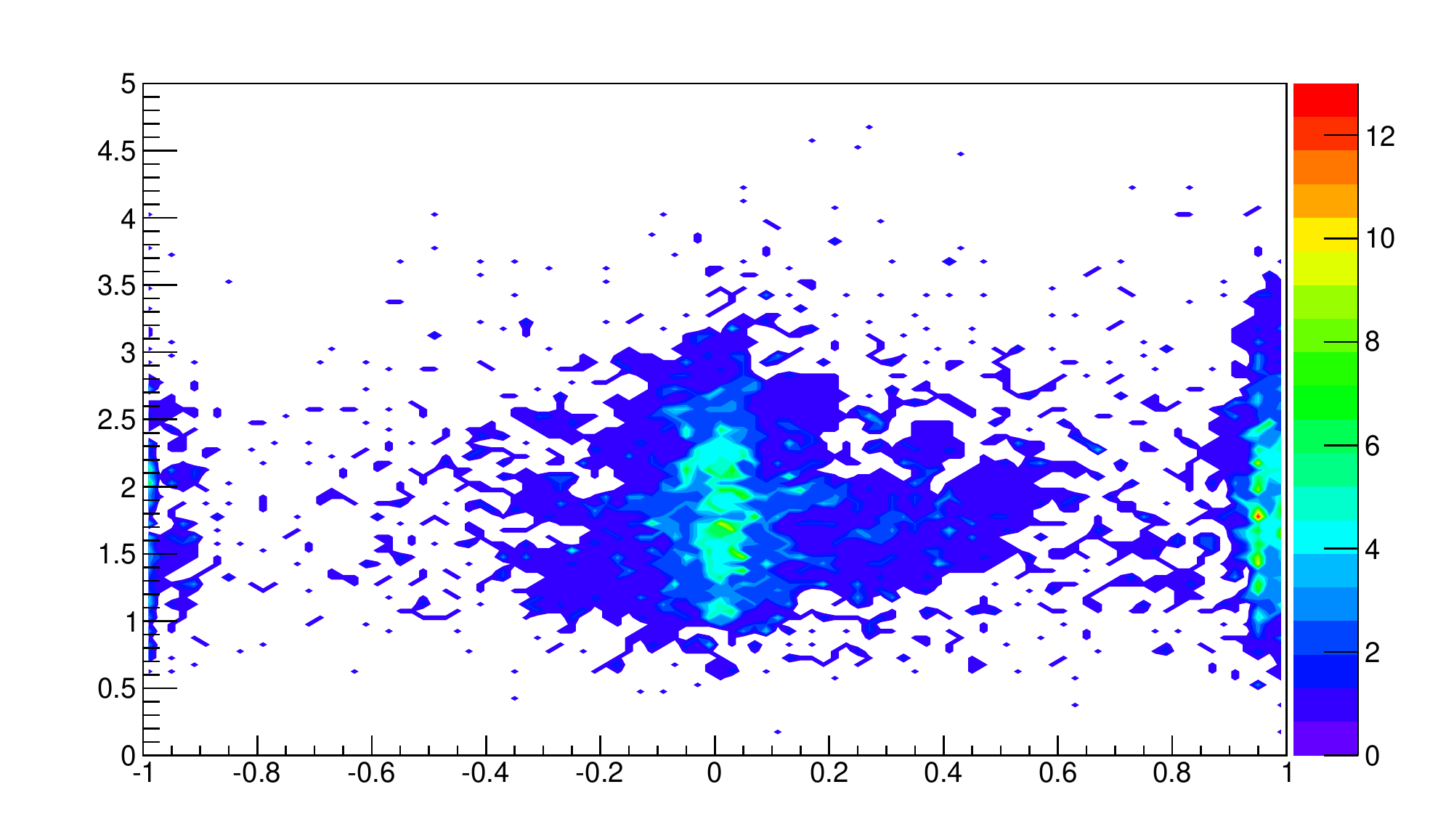}
\caption{Initial asymmetry in opinion leads to different opinion-conviction distributions. $H(x) = x$ (up), and $H(x) = \sqrt{x}$ (down).}
\label{fg:fig3}
\end{figure}

The same effect is evident in Figure \ref{fg:fig3}, which refers to both Tests 1 and 2 in which, to understand the evolution in case of asymmetry, the initial distribution of opinions was chosen  uniformly distributed on the positive part of the interval.

%
%

\subsection{Modeling complex networks}
The present setting takes into account large complex networks of interacting agents by introducing a kinetic model which couples an alignment dynamics with the underlying evolution of the network. The coupled evolution of opinions and network is described by a Boltzmann-type equation where the probability distribution of opinions depends on a second relevant variable called \emph{connectivity}.  In principle, the ideas proposed here are not limited to a particular kind of opinion dynamics and one can easily adapt to the same situation other models developed in the literature \cite{BMS, DMPW, SWS}.

\subsubsection{A Boltzmann-type model for opinion and number of connections}
Let us consider a large system of agents interacting through a given network. 
We associate to each agent an opinion $w$, which varies continuously in $I = [-1,1]$, and his number of connections $c$, a discrete variable varying between $0$ and the maximum number of connections allowed by the network. Note that this maximum number typically is a fixed value which is several orders of magnitude smaller then the size the network.

We are interested in the evolution of the density function 
\begin{equation}\label{eq:def_f}
f=f(w,c,t), \qquad f: I\times {\mathcal C}\times \mathbb{R}^+ \rightarrow \mathbb{R}^+
\end{equation}
where $w\in I$ is the opinion variable, $c\in {\mathcal C}=\{0,1,2,\ldots,\cm\}$ is the discrete variable describing the number of connections and $t\in\RR^+$ denotes as usual the time variable. For each time $t\ge 0$ the marginal density
\begin{equation}\label{eq:integration_wc}
\p(c,t)=\int_I f(w,c,t)dw, 
\end{equation}
 defines the evolution of the number of connections of the agents or equivalently the degree distribution of the network. 
In the sequel we assume that the total number of agents is conserved, i.e. 
$
\sum_{c=0}^{\cm}\p(c,t)=1.
$
The overall opinion distribution is defined likewise as the  marginal density function
\begin{equation}
g(w,t)=\sum_{c=0}^{\cm} f(w,c,t).
\end{equation}
We express the evolution of the opinions by a binary interaction rule. From a microscopic point of view we suppose that the agents modify their opinion through binary interactions which depend on opinions and number of connections. If two agents with opinion and number of connections $(w,c)$ and $(w_*,c_*)$ meet, their post-interaction opinions are given by
\begin{equation}\begin{cases}\label{eq:binary}
w' &=w-\eta P(w,w_*;c,c_*)(w-w_*)+\xi D(w,c), \\
w_*' &=w_*-\eta P(w_*,w;c_*,c)(w_*-w)+\xi_* D(w_*,c_*),
\end{cases}\end{equation}
Note that, in the present setting the compromise function $P$ depends both on the opinions and on the number of connections of each agent. In \eqref{eq:binary} all the other quantities are defined as in \eqref{ch6:trade_rule}.  We will consider by now a general interaction potential such that $0\le P(w,w_*,c,c_*)\le 1$. In absence of diffusion $\xi,\xi_*\equiv 0$, and from \eqref{eq:binary} we have 
\begin{equation}
|w'-w_*'| = |1-\eta(P(w,w_*;c,c_*)+P(w_*,w;c_*,c))||w-w_*|.
\end{equation}
Hence the post-exchange distances between agents are diminishing if we consider $\eta\in (0,1)$ and $0\le P(w,w_*,c,c_*)\le 1$. Similarly to  Section \ref{sec:AHP},  Proposition \ref{proposition:bounds}, we can require the conditions on the noise term to ensure that the post-interaction opinions do not leave the reference interval interval. 

The evolution in time of the density function $f(w,c,t)$ is described by the following integro-differential equation of Boltzmann-type
\begin{equation}\label{eq:boltz_lin}
\dfrac{d}{d t}f(w,c,t)+\N[f(w,c,t)]=Q(f,f)(w,c,t),
\end{equation}
where $\N[\cdot]$ is an operator which is related to the evolution of the connections in the network and $Q(\cdot,\cdot)$ is the binary interaction operator. It is convenient to define $Q$ in weak form as follows
\begin{equation}\begin{split}\label{eq:collisional_op}
 \int_I & Q(f,f)\varphi(w)dw = \lambda \sum_{c_*=0}^{\cm}\Big<\int_{I^2 } (\varphi(w')-\varphi(w)) f(w_*,c_*)f(w,c) dw dw_* \Big>.
\end{split}\end{equation}
Consequently the equation \eqref{eq:boltz_lin}  in weak form reads
\begin{equation}\begin{split}\label{eq:boltz_weak2}
& \dfrac{d}{dt}\int_I f(w,c)\varphi(w)dw+ \int_I \N[ f(w,c)]\varphi(w)dw = \\
&\frac{\lambda}{2} \sum_{c_*=0}^{\cm}\left<\int_{I^2} \left(\varphi(w')+\varphi(w'_*)-\varphi(w)-\varphi(w_*)\right)f(w_*,c_*)f(w,c)dwdw_*
\right>.
\end{split}
\end{equation}

\subsubsection{Evolution of the network}\label{sec:main}


The operator $\N[\cdot]$ is defined through a combination of preferential attachment and uniform processes describing the evolution of the connections of the agents by removing and adding links in the network. These processes are strictly related to the generation of stationary scale-free distributions \cite{BA,XZW}. More precisely, for each $c=1,\ldots,\cm-1$ we define
\begin{equation}\begin{split}\label{eq:master2}
\N[f(w,c,t)] =& -\dfrac{2\Ur(f;w)}{\gamma+\beta}\left[(c+1+\beta)f(w,c+1,t)-(c+\beta)f(w,c,t)\right]\\
&-\dfrac{2\Ua(f;w)}{\gamma+\alpha}\left[(c-1+\alpha)f(w,c-1,t)-(c+\alpha)f(w,c,t)\right],
\end{split}\end{equation} 
where
$\gamma=\gamma(t)$ is the mean density of connectivity defined as 
\begin{equation}
\gamma(t) = \sum_{c=0}^{\cm}c\p(c,t),
\label{eq:gammad}
\end{equation}
$\alpha, \beta>0$ are attraction coefficients, and $\Ur(f;w)\geq 0$, $\Ua(f;w)\geq 0$ are characteristic rates of the removal and adding steps, respectively. The first term in \eqref{eq:master2} describes the net gain of $f(w,c,t)$ due to the connection removal between agents whereas the second term represents the net gain due to the connection adding process. The factor $2$ has been kept in evidence since connections are removed and created pairwise.  At the boundary we have the following equations
\begin{equation}\label{eq:BD_master2}
\begin{split}
\N[f(w,0,t)] =&  -\dfrac{2\Ur(f;w)}{\gamma+\beta}(\beta+1)f(w,1,t)+\dfrac{2\Ua(f;w)}{\gamma+\alpha}\alpha f(w,0,t), \\
\N[f(w,\cm,t)] =& \dfrac{2\Ur(f;w)}{\gamma+\beta}(\cm+\beta) f(w,\cm,t)\\
& -\dfrac{2\Ua(f;w)}{\gamma+\alpha}(\cm-1+\alpha)f(w,\cm-1,t),
\end{split}
\end{equation}
which are derived from (\ref{eq:master2}) taking into account the fact that, in the dynamics of the network, connections cannot be removed from agents with $0$ connections and cannot be added to agents with $\cm$ connections.


The evolution of the connections of the network can be recovered taking $\varphi(w)=1$ in  \eqref{eq:boltz_weak2} 
\begin{equation}\label{eq:Ldef2}
\dfrac{d}{dt}\p(c,t) + \int_I \N[f(w,c,t)]\,dw=0.
\end{equation}
From the  definition of the network operator $\N[\cdot]$ it follows that 
\begin{equation}
\dfrac{d}{dt}\sum_{c=0}^{\cm}\p(c,t)=0. 
\label{eq:tnc}
\end{equation}
Then, with  the collisional operator defined in \eqref{eq:collisional_op} and  of $\N[\cdot]$ in \eqref{eq:master2} the total number of agents is conserved. \\

Let us take into account the evolution of the mean density of connectivity $\gamma$ defined in (\ref{eq:gammad}). For each $t\ge0$ 
\begin{equation}
\begin{split}
&\dfrac{d}{dt}\gamma(t)=-2\int_I\Ur(f;w)\frac{\gamma_f+\beta g(w,t)}{\gamma+\beta}dw+2\int_I\Ua(f;w)\frac{\gamma_f+\alpha g(w,t)}{\gamma+\alpha}dw\\
&\,\,\,\,\,+\dfrac{2\beta}{\gamma+\beta}\int_I \Ur(f;w) f(w,0,t)\,dw
-\dfrac{2(\cm+\alpha)}{\gamma+\alpha}\int_I \Ua(f;w) f(w,\cm,t)\,dw.
\end{split}
\label{eq:mdc}
\end{equation}
Therefore $\gamma(t)$ is in general not conserved. 
The explicit computations for the conservation of the total number of connections and for the evolution of the mean density of connectivity are reported under specific assumptions in \cite{APZd}. 

When $\Ua$ and $\Ur$ are constants,  the operator $\N[\cdot]$ is linear and will be denoted by $\L[\cdot]$. In this case, the evolution of the network of connections is independent from the opinion and one gets the closed form  
\begin{equation}\label{eq:Ldef}
\dfrac{d}{dt}\p(c,t) + \L[\p(c,t)]=0,
\end{equation}
where 
\begin{equation}\begin{split}\label{eq:master}
\L[\p(c,t)] =& -\dfrac{2\Ur}{\gamma+\beta}\left[(c+1+\beta)\p(c+1,t)-(c+\beta)\p(c,t)\right]\\
&-\dfrac{2\Ua}{\gamma+\alpha}\left[(c-1+\alpha)\p(c-1,t)-(c+\alpha)\p(c,t)\right],
\end{split}\end{equation}
with the boundary conditions
\begin{equation}\label{eq:BD_master}
\begin{split}
\L[\p(0,t)]& =  -\dfrac{2\Ur}{\gamma+\beta}(\beta+1)\p(1,t)+\dfrac{2\Ua}{\gamma+\alpha}\alpha \p(0,t), \\
\L[\p(\cm,t)]& = \dfrac{2\Ur}{\gamma+\beta}(\cm+\beta) \p(\cm,t)-\dfrac{2\Ua}{\gamma+\alpha}(\cm-1+\alpha)\p(\cm-1,t).
\end{split}
\end{equation}
Note that  in (\ref{eq:master}) the dynamic corresponds to a combination of preferential attachment processes ($\alpha,\beta\approx 0$) and uniform processes ($\alpha,\beta \gg 1$) with respect to the probability density of connections $\rho(c,t)$. 
Concerning the large time behavior of the network of connections, in the linear case with $\Ur=\Ua$, $\beta=0$ and  denoting by $\gamma$ the asymptotic value of the density of connectivity it holds
\begin{proposition}
\label{pr:1}
For each $c\in {\mathcal C}$ the stationary solution to (\ref{eq:Ldef}) or equivalently 
\be\label{prop:st}
(c+1)\ps(c+1) = \dfrac{1}{\gamma+\alpha}\left[(c(2\gamma+\alpha)+\gamma\alpha)\ps(c)-\gamma(c-1+\alpha)\ps(c-1)\right],
\ee
is given by 
\be\label{prop:sol}
\ps(c) = \left(\dfrac{\gamma}{\gamma+\alpha}\right)^c \dfrac{1}{c!}\alpha(\alpha+1)\cdots (\alpha+c-1)\ps(0),
\ee
where
\be
\ps(0) = \left(\dfrac{\alpha}{\alpha+\gamma}\right)^{\alpha}.
\ee
\end{proposition}
A detailed  proof is given in \cite{APZd}. Further approximations are possible if $\alpha\gg 1$ or $\alpha\approx 0$. For large $\alpha$ the preferential attachment process described by the master equation \eqref{eq:master} is destroyed and the network approaches  a random network, whose degree distribution coincides with the Poisson distribution. In fact, in the limit $\alpha\rightarrow+\infty$ we have $(\alpha+\gamma)^c\approx \alpha(\alpha+1)\cdots(\alpha+c-1),$ and
\be
\ps(c) =\lim_{\alpha\rightarrow+\infty} \left(1+\dfrac{\gamma}{\alpha}\right)^{-\alpha}\gamma^c = \dfrac{e^{-c}}{c!}\gamma^c.
\ee
In the second case, for $\gamma\ge1$ and small values of $\alpha$, the distribution can be correctly approximated with a truncated power-law with unitary exponent
\be
\ps(c) = \left(\dfrac{\alpha}{\gamma}\right)^{\alpha}\dfrac{\alpha}{c}.
\ee

\subsubsection{Fokker-Planck modelling}
\label{sec:FP}
Similarly to Section \ref{sec:AHP} we can derive a Fokker-Planck equation  through the quasi-invariant opinion limit. 
Let us introduce the scaling parameter $\varepsilon>0$ and consider the scaling 
\begin{equation}\label{eq:scaling}
\eta=\varepsilon, \qquad \lambda=\dfrac{1}{\varepsilon}, \qquad \varsigma^2=\varepsilon\sigma^2.
\end{equation}
In the limit $\epsilon\rightarrow 0$ we obtain
the Fokker-Planck equation for the evolution of the opinions' distribution through the evolving network 
\begin{equation}\begin{split}\label{eq:FP}
\dfrac{\partial}{\partial t}f(w,c,t)+\N\left[f(w,c,t)\right]=\dfrac{\partial }{\partial w}\mathcal{P}[f]f(w,c,t)+\dfrac{\sigma^2}{2}\dfrac{\partial^2}{\partial w^2}(D(w,c)^2 f(w,c,t)),
\end{split}\end{equation}
where
\begin{equation}\label{eq:K}
\mathcal{P}[f](w,c,t) = \sum_{c_*=0}^{\cm}\int_I P(w,w_*;c,c_*)(w_*-w)f(w_*,c_*,t)dw_*.
\end{equation}

\begin{figure}[ht]
\centering
\includegraphics[scale= 0.3]{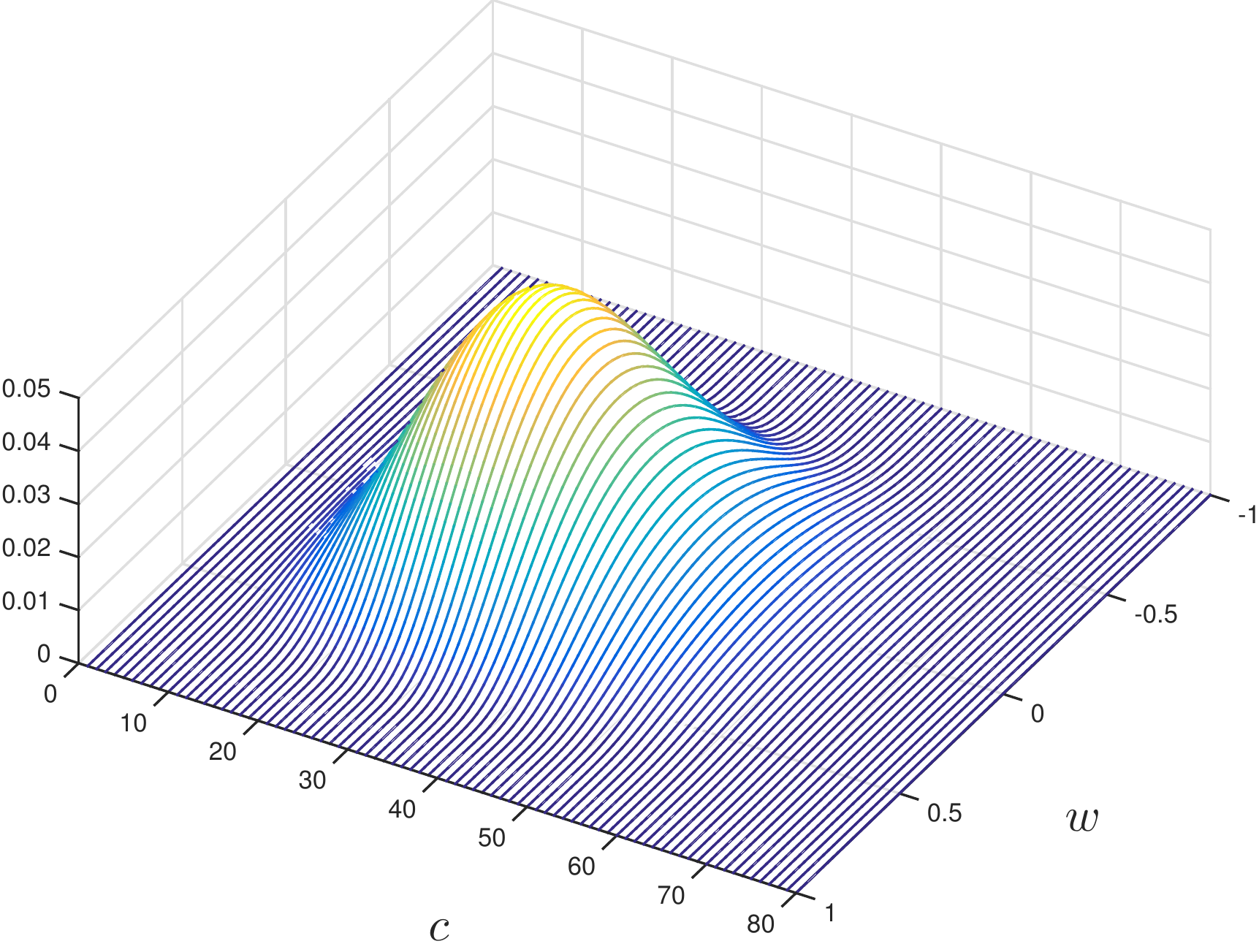}
\hskip +1cm
\includegraphics[scale= 0.3]{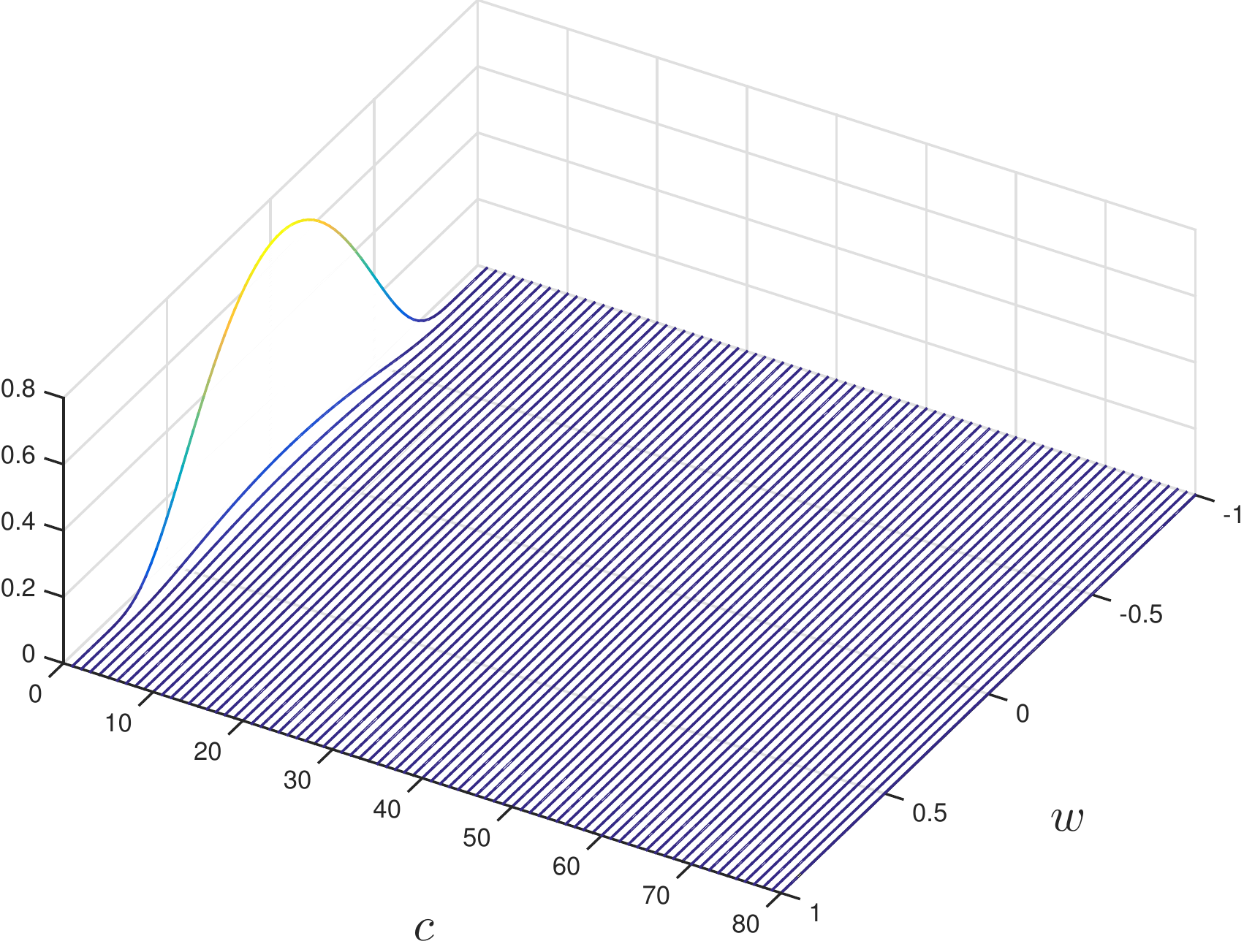}
\caption{Stationary solutions of type $f_\infty(w,c) = g_\infty(w)p_\infty(c)$, where $ g_\infty(w)$ is given by \eqref{eq:stat_1} with $\kappa =1$, $m_w = 0$, $\sigma^2 = 0.05$ and $p_\infty(c)$ defined by \eqref{prop:sol}, with $\Ur=\Ua=1$, and $\alpha = 10$ on the left and $\alpha = 0.1$ on the right.
 }\label{Fig:stat1}
\end{figure}

In some case it is possible to compute explicitly the steady state solutions of the Fokker-Planck system \eqref{eq:FP}. We restrict to linear operators $\L[\cdot]$ and asymptotic solutions of the following form 
\begin{equation}\label{eq:steady}
f_{\infty}(w,c)=g_{\infty}(w)\ps(c),
\end{equation} 
where $\ps(c)$ is the steady state distribution of the connections (see Proposition \ref{pr:1}) and 
\begin{equation}
\int_I f_{\infty}(w,c)dw=\ps(c), \qquad \qquad \sum_{c=0}^{\cm}f_{\infty}(w,c)=g_{\infty}(w).
\end{equation}
From the definition of the linear operator $\L[\cdot]$ we have $\L[\ps(c)]=0$. Hence the stationary solutions of type \eqref{eq:steady} satisfy the  equation
\begin{equation}
\label{eq:steadye}
\dfrac{\partial }{\partial w}\mathcal{P}[f_{\infty}]f_{\infty}(w,c)+\dfrac{\sigma^2}{2}\dfrac{\partial^2}{\partial w^2}(D(w,c)^2 f_{\infty}(w,c))=0.
\end{equation}
Equation \eqref{eq:steadye} can be solved explicitly in some case \cite{APZa,T}. 
If $P$ is in the form 
\begin{equation}\label{eq:factorHK}
P(w,w_*;c,c_*)=H(w,w_*)K(c,c_*),
\end{equation}
the operator $\mathcal{P}[f_\infty]$ can be written as follows 
\begin{equation}\label{eq:K2}
\begin{split}
\mathcal{P}[f_\infty](w,c)& =  \left(\sum_{c_*=0}^{\cm}{K}(c,c_*)\ps(c_*)\right)\left(\int_I H(w,w_*)(w_*-w)g_\infty(w_*)dw_*\right).
\end{split}
\end{equation}
We further assume that $K(c,c_*)={K}(c_*)$ is independent of $c$ and denote
\[
\kappa=\sum_{c_*=0}^{\cm}{K}(c_*)\ps(c_*),\quad \bar m_w=\sum_{c=0}^{\cm} m_w(c,t),\quad m_w (c,t) =  \int_I w f(w,c,t)dw.
\] 
\begin{itemize}
 \item[a)] In the case $H\equiv 1$ and $D(w)=1-w^2$ the steady state solution $g_{\infty}$ is given by 
\begin{equation}\label{eq:stat_1}
g_{\infty}(w)= C_0(1+w)^{-2+\bar m_w\kappa/\sigma^2}(1-w)^{-2-m_w\kappa/\sigma^2}\exp\Big\{-\dfrac{\kappa(1-\bar m_w w)}{\sigma^2(1-w^2)}\Big\},
\end{equation}
\item[b)] For $H(w,w_*)=1-w^2$ and $D(w)=1-w^2$ the steady state solution $g_{\infty}$ is given by 
\begin{equation}\label{eq:stat_2}
g_{\infty}(w)=C_0(1-w)^{-2+(1-\bar m_w){\kappa}/{\sigma^2}}(1+w)^{-2+(1+\bar m_w){\kappa}/{\sigma^2}},
\end{equation}
\end{itemize}
In Figure \ref{Fig:stat1}  we report the stationary solution $f_\infty(w,c) = g_\infty(w)\rho_\infty(c)$, where $ g_\infty(w)$ is given by \eqref{eq:stat_1} with $\kappa =1$, $m_w = 0$, $\sigma^2 = 0.05$ and $p_\infty(c)$ defined by \eqref{prop:sol}, with $\Ur=\Ua=1$, $\gamma = 30$ and $\alpha = 10$ on the left and $\alpha = 0.01$ on the right.

\subsubsection{Numerical Experiments}
In this section we perform some numerical experiments to study the behavior of the new kinetic models.
We focus on the case $\alpha< 1$, since it represents the most relevant case in complex networks \cite{AB, XZW}. Within this range of the parameter we have emergence of power law distributions for  network's  connectivity. In the tests that follow the opinion dynamics evolves according to \eqref{eq:FP}. The compromise function $P$ and the local diffusion function $D$ in the various tests will be specified in each test. The different tests are summarized in Table \ref{tab:par}, where other parameters are introduced and additional details are reported.
\begin{table}[htb]
\caption{Parameters in the various test cases}
\label{tab:all_parameters}
\begin{center}
\begin{tabular}{cccccccccc}
\hline
Test & $\sigma^2$ & $\sigma^2_F$ & $\sigma^2_L$ & $\cm$  & $\Ur$ &$\Ua$ &  $\gamma_0$ & $\alpha$ & $\beta$ \\
\hline
\hline
\#1 & $5\times10^{-2}$  & $6\times10^{-2}$ & $-$ & $250$   & $1$& 1 &  $30$ & $1\times10^{-1} $ & $0$ \\
\hline
\#2 & $5\times10^{-3}$  & $4\times10^{-2}$ & $2.5\times10^{-2}$ & 250  & $1$ & $1$& 30 & $1\times10^{-4}$ & $0$\\
\hline
\#3 & $1\times10^{-3}$  & $-$ & $-$ & 250  & $1$ & $1$& 30 & $1\times10^{-1}$ & $0$ \\
\hline
\end{tabular}\label{tab:par}
\end{center}
\end{table}
In Test 1 a Monte Carlo method is used to solve the Boltzmann model \eqref{eq:boltz_lin} we refer to \cite{APb, PTa} and to the Appendix for a description on these class of methods. In Test 2, 3, 4 the Fokker-Planck system \eqref{eq:FP} is solved via the steady-state preserving Chang-Cooper scheme, see the Appendix and \cite{BCDS,BD,CC,LLPS,MB} for further details. 

\subsubsection*{Test 1}
We first consider the one dimensional setting to show the convergence of the Boltzmann model \eqref{eq:boltz_lin} to the exact solution of the Fokker-Planck system \eqref{eq:FP}, via Monte Carlo methods.
We simulate the dynamics with the linear interaction kernel, $P(w,w_*;c,c_*) = 1$, and $D(w,c) = 1-w^2$, thus we can use the results \eqref{eq:stat_1}, to compare the solutions obtained through the numerical scheme with the analytical one, the other parameters of the model are reported in Table \ref{tab:all_parameters} and we define the following initial data
\begin{align}\label{eq:g0}
g_0(w)      = \frac{1}{2\sqrt{2\pi\sigma_F^2 }}(\exp\{-(w+1/2)^2/{(2\sigma_F^2)}\}+\exp\{-(w-1/2)^2/{(2\sigma_F^2)}\}).
\end{align}

\begin{figure}[ht]
\centering
\includegraphics[scale= 0.3]{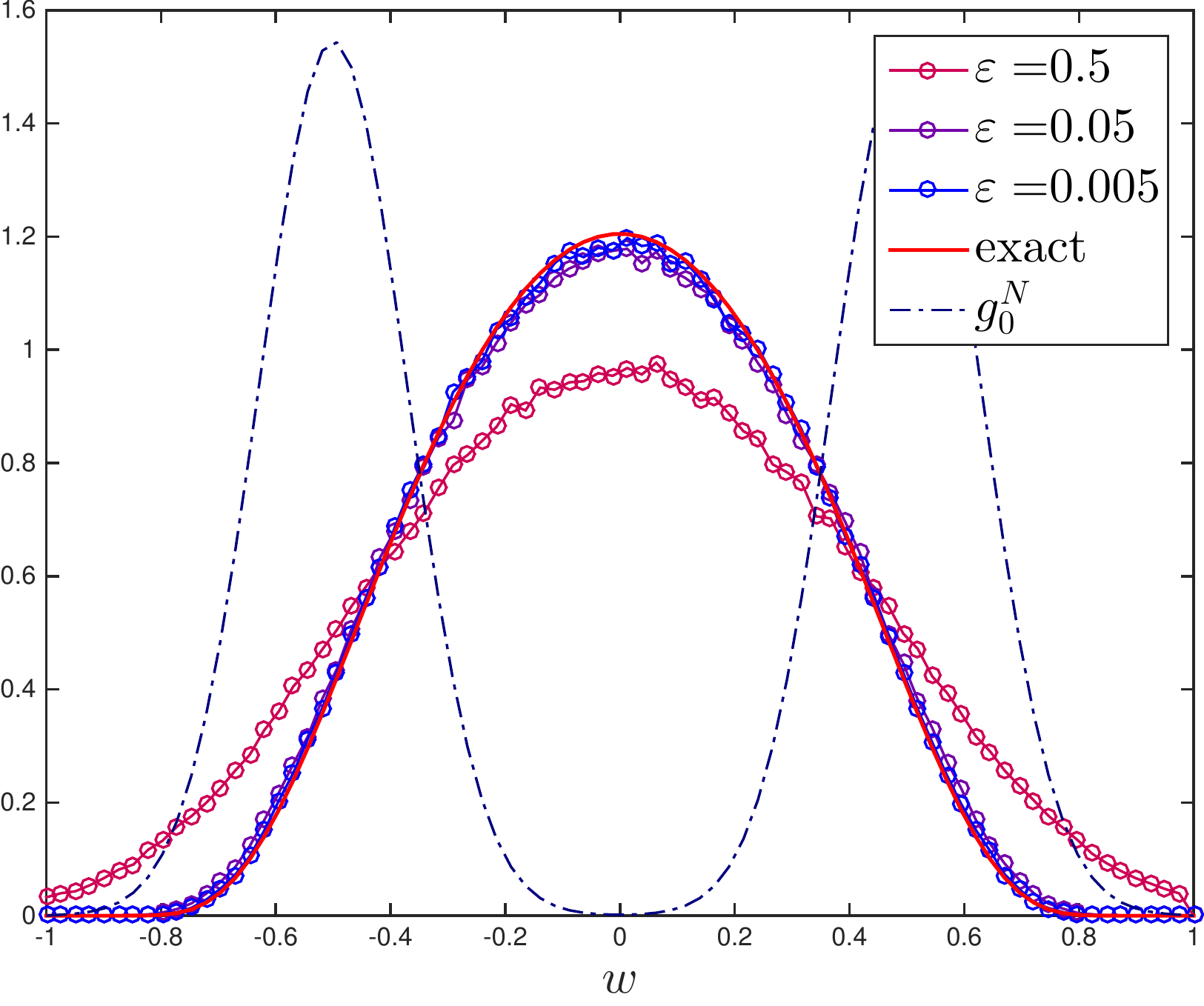}
\hskip +0.2cm
\includegraphics[scale= 0.3]{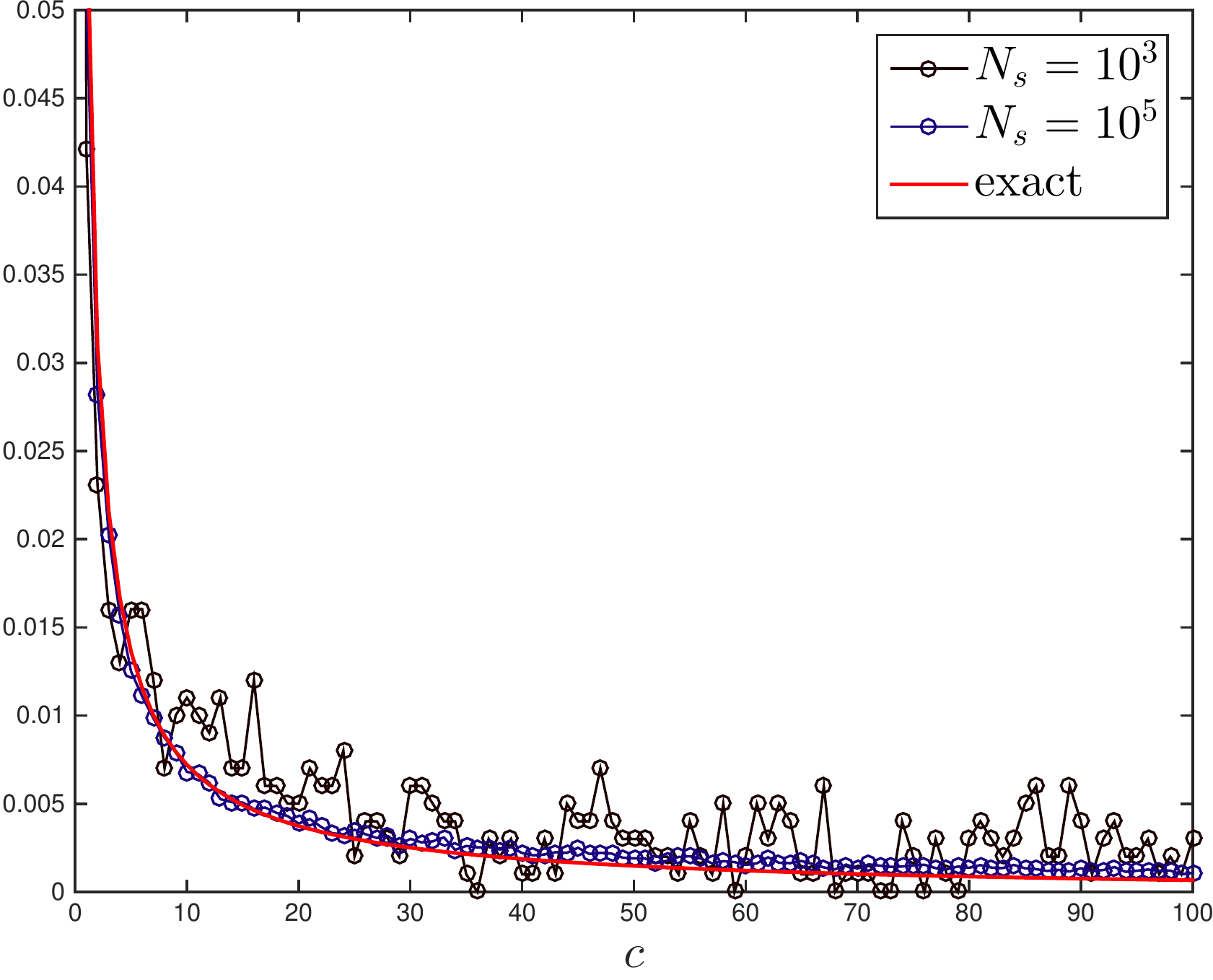}
\caption{{\bf Test 1.}
One-dimensional setting: on the left, convergence of \eqref{eq:boltz_lin} to the stationary solution \eqref{eq:stat_1}, of the Fokker-Planck equation, for decreasing values of the parameter $\varepsilon$, $g^N_0$ represent the initial distribution. On the right, convergence of the Monte-Carlo Algorithm \ref{MCmaster}, see the Appendix , to the reference solution \eqref{prop:st}  for increasing value of the the number of  samples $N_s$.
 }\label{Fig:MC}
\end{figure}

In Figure \ref{Fig:MC}, on the left hand-side, we report the qualitative convergence of the Binary Interaction algorithm, \cite{APb}, where we consider $N_s = 10^5$ samples to reconstruct the opinion's density, $g(w,t)$, on a grid of $N = 80$ points. The figure shows that for decreasing values of the scaling parameter $\varepsilon = \{0.5, 0.05, 0.005\}$, we have convergence to the reference solutions, \eqref{eq:stat_1} of the Fokker-Planck equation.
On the right we report the convergence to the stationary solution of the connectivity distribution, \eqref{prop:st}, for $\alpha = 0.1$ and $V = 1$ and with $\cm = 250$. In this case we show two different qualitative behaviors for an increasing number of samples $N_s=\{10^3, 10^5\}$ and for sufficient large times, obtained through the stochastic Algorithm \ref{MCmaster}.

\subsubsection*{Test 2}
In the second test we analyze the influence of the connections over the opinion dynamics, for a compromise function of the type \eqref{eq:factorHK}
where $H(w,w_*) = 1-w^2$ and  $K$  defined by
 \begin{align}\label{eq:Kc}
\quad K(c,c_*) = \left(\frac{c}{\cm}\right)^{-a} \left(\frac{c_*}{\cm}\right)^{b},
\end{align}
for $a,b >0$. This type of kernel assigns higher relevance into the opinion dynamic to higher connectivity, and low influence to low connectivity.
The diffusivity is weighted by $D(w,c) = 1-w^2$. We perform a first computation with the initial condition
\begin{align}\label{eq:T2f0}
f_0(w,c) =C_0
\begin{cases}
 \rho_\infty(c)\exp\{-(w+\frac{1}{2})^2)/(2\sigma_F^2)\},&\quad \textrm{ if }  0\leq c\leq 20,\\
 \rho_\infty(c)\exp\{-(w-\frac{3}{4})^2/(2\sigma_L^2)\},&\quad \textrm{ if }   60\leq c\leq 80,\\
 0, &\quad \textrm{ otherwise}.
\end{cases}
\end{align}
The values of the parameters are reported in the third line of Table \ref{tab:par}. In the interaction function $K(\cdot,\cdot)$ in \eqref{eq:Kc} we choose  $a=b=3$.
The evolution is performed through the Chang-Cooper type scheme with $\Delta w = 2/N$ and $N = 80$. The evolution of the system is studied in the time interval $[0,T]$, with $T=2.5$.

\begin{figure}[ht]
\centering
\subfigure[t=0]{\includegraphics[scale= 0.3]{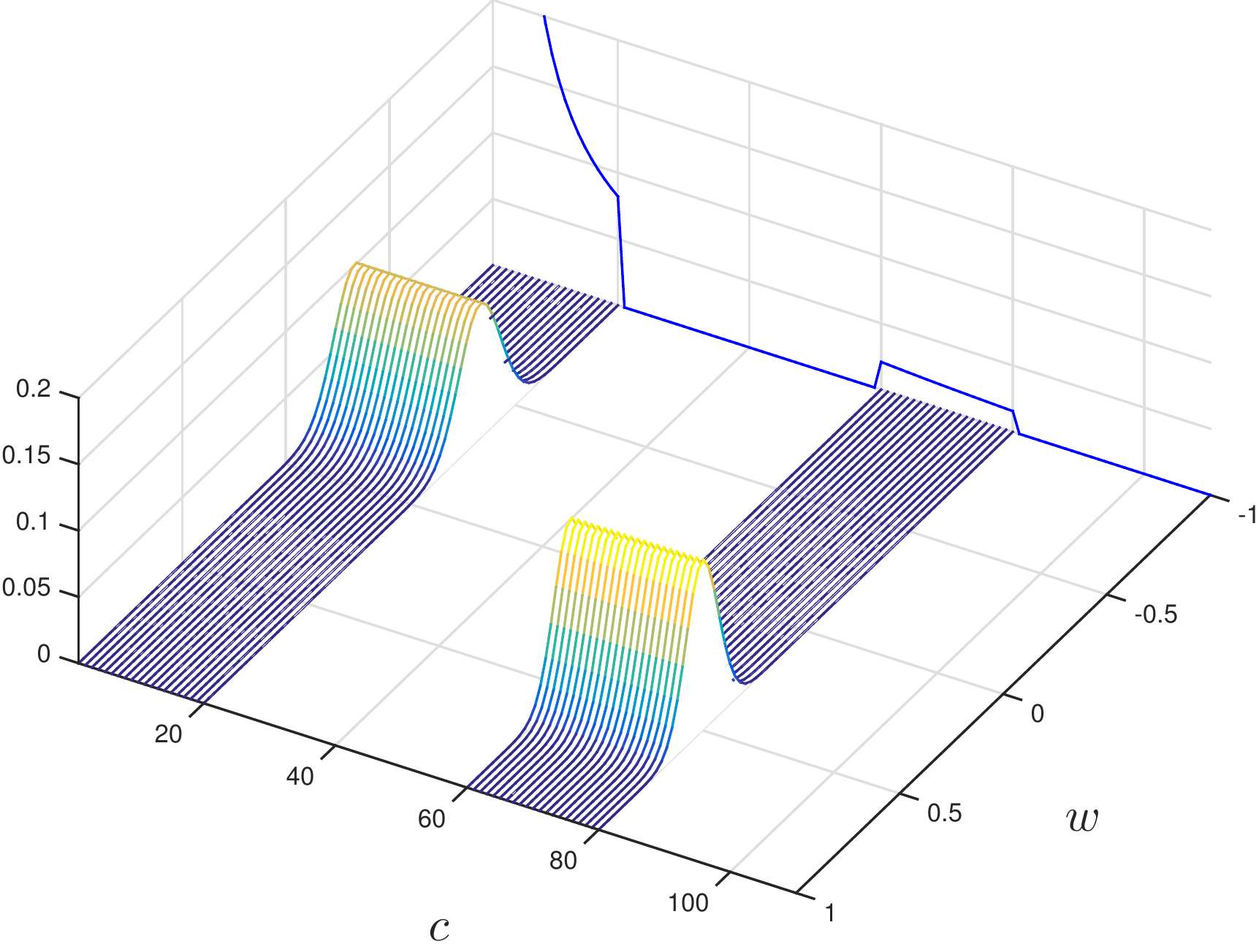}}
\hskip +0.2cm
\subfigure[t=0.2]{\includegraphics[scale= 0.3]{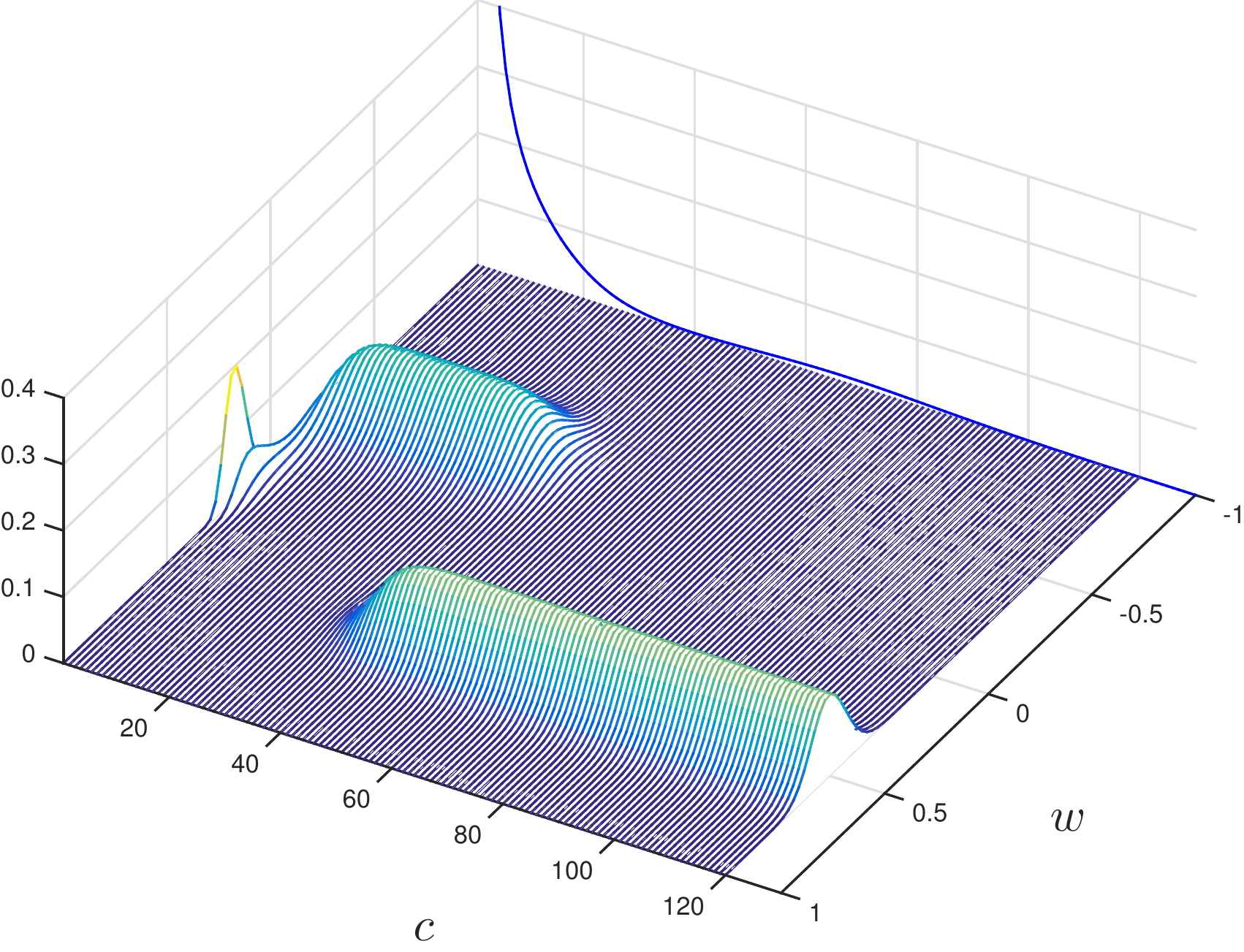}}
\\
\subfigure[t=1.5]{\includegraphics[scale= 0.3]{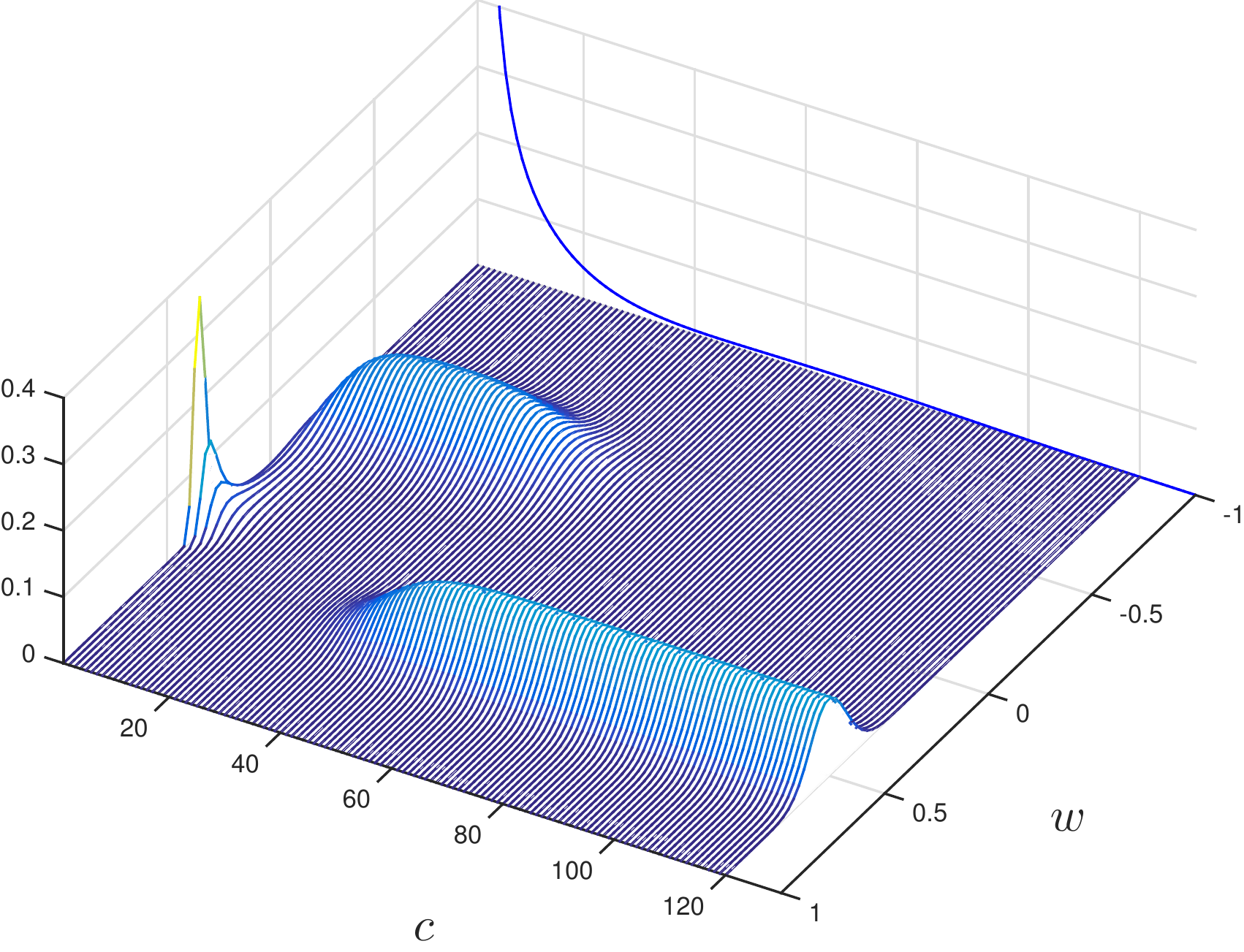}}
\hskip +0.2cm
\subfigure[t=2.5]{\includegraphics[scale= 0.3]{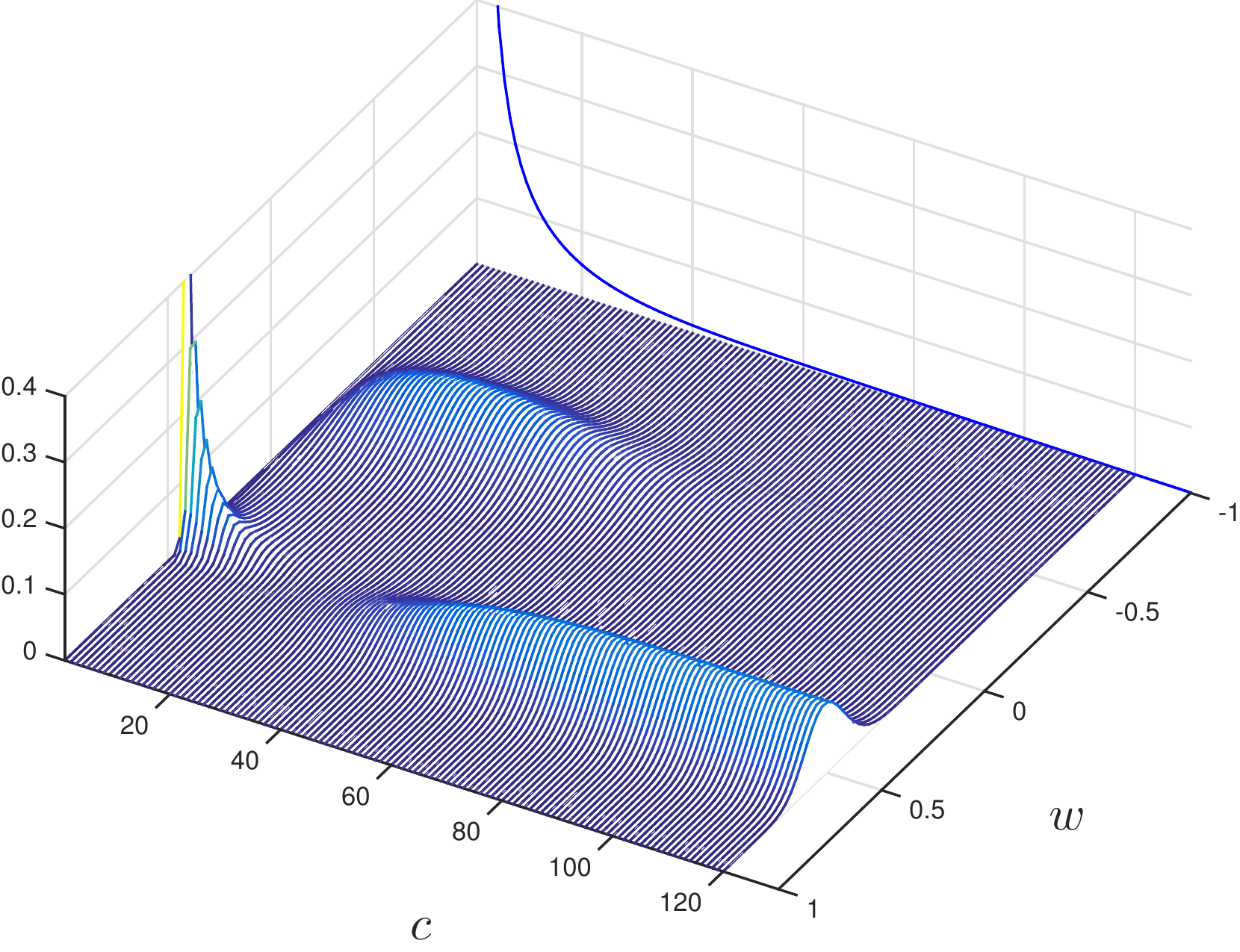}}
\caption{Test \#2. Evolution of the initial data \eqref{eq:T2f0} in the time interval $[0,T]$, with $T = 2.5$. The evolution shows how a small portion of density with high connectivity can bias the majority of the population towards their position. 
(Note: The density is scaled according to the marginal distribution $\rho(c,t)$ in order to better show its evolution. The actual marginal density $\rho(c,t)$ is depicted in the background, scaled by a factor 10).
}\label{Fig:F3}
\end{figure}

In Figure \ref{Fig:F3} we report the result of the simulation. On the first plot the initial configuration is split in two parts, the majority concentrated around the opinion $\bar{w}_F = - 1/2$ and only a small portion concentrated around $\bar{w}_L= 3/4$. 
We observe that, because of the anisotropy induced by $K(c,c_*)$, the density with a low level of connectivity is influenced by the small concentration of density around $w_L$ with a large level of connectivity. 

\subsubsection*{Test 3}
\begin{figure}[ht]
\centering
\subfigure[t=0]{\includegraphics[scale= 0.3]{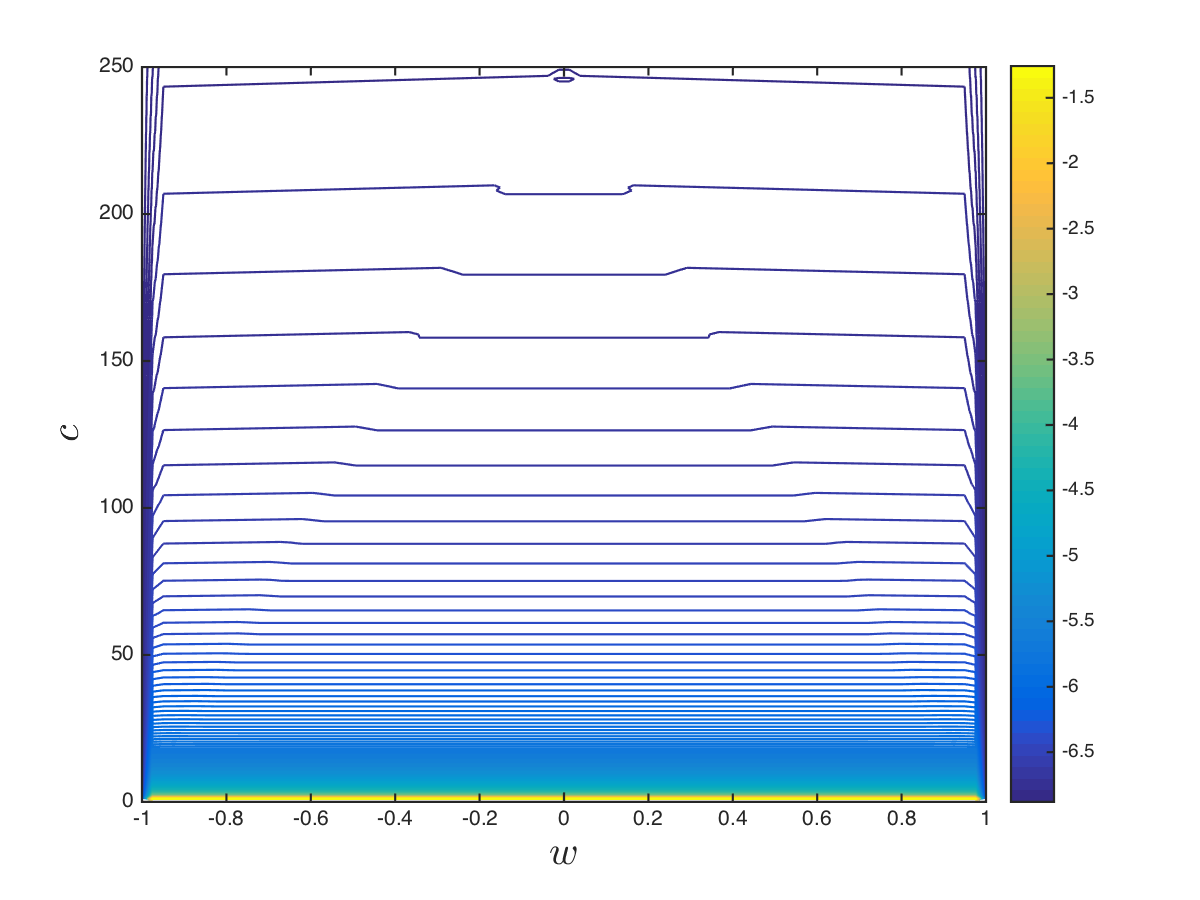}}
\hskip +0.2cm
\subfigure[t=10]{\includegraphics[scale= 0.3]{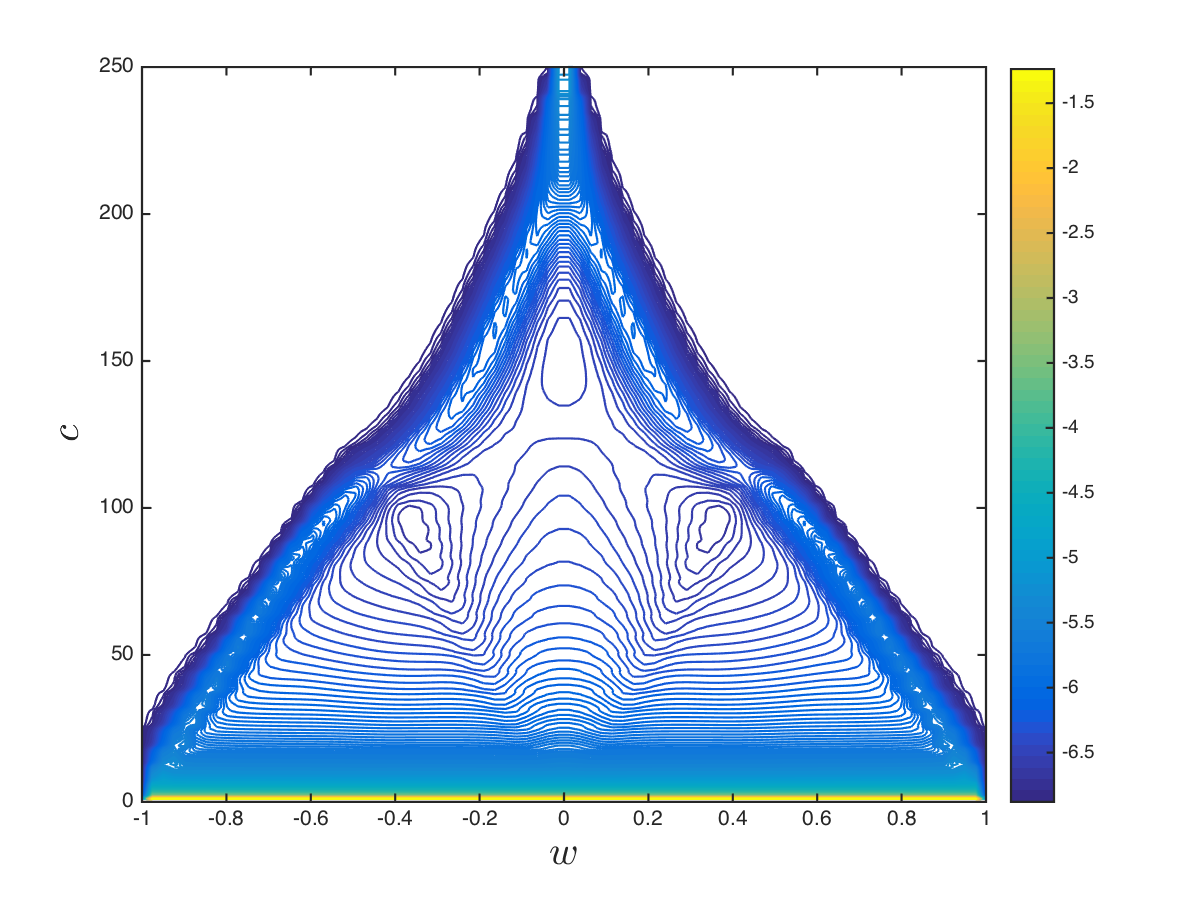}}
\\
\subfigure[t=50]{\includegraphics[scale= 0.3]{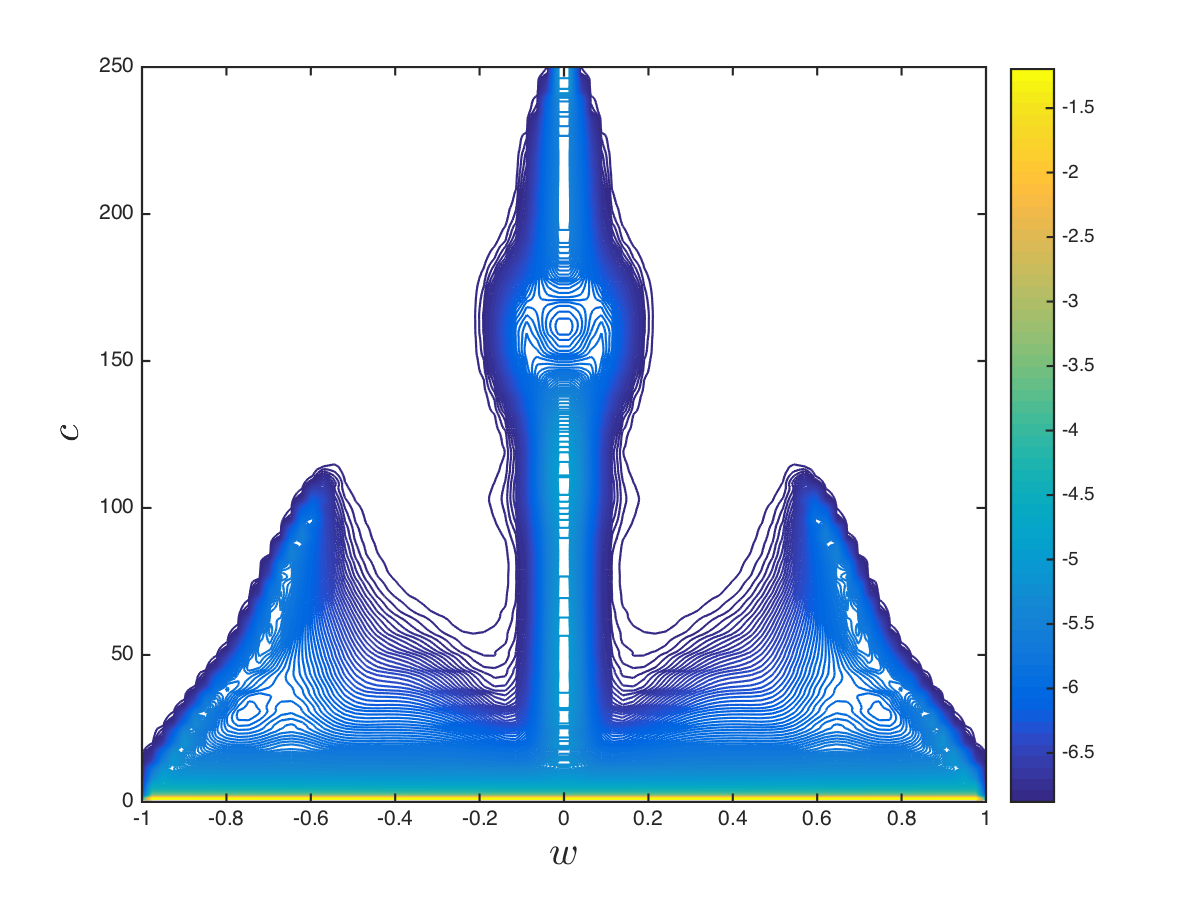}}
\hskip +0.2cm
\subfigure[t=100]{\includegraphics[scale= 0.3]{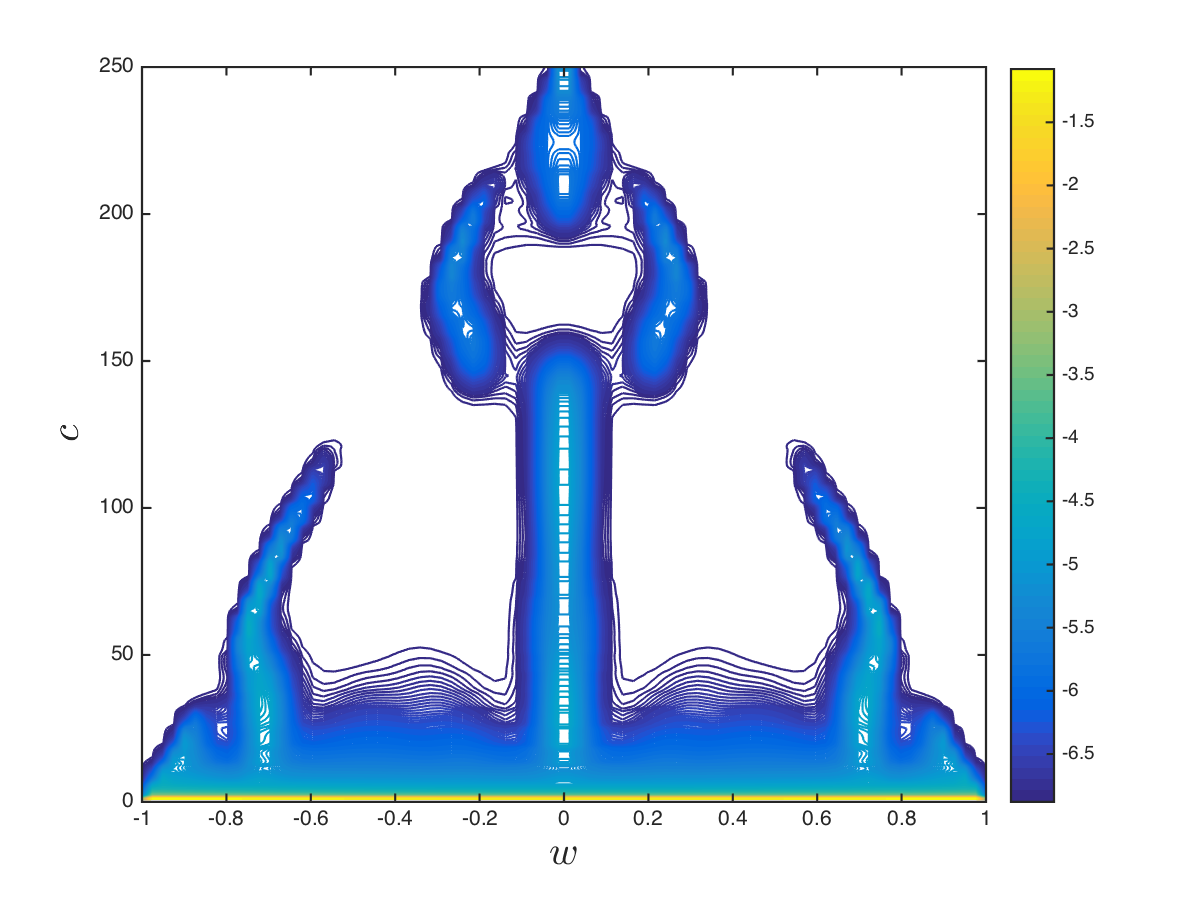}}
\caption{Test \#3. Evolution of the solution of the Fokker-Planck model \eqref{eq:FP}, for the bounded confidence model, with $\Delta(c)=d_0c/\cm$, and $d_0=1.01$, in the time frame $[0,T]$, with $T = 100$. The choice of $\Delta(c)$ reflects in the heterogeneous emergence of clusters with respect to the connectivity level: for higher level of connectivity consensus is reached, instead for lower levels of connectivity multiple opinion clusters are present. 
(Note: In order to better show its evolution, we represent the solution as $\log(f(w,c,t)+\epsilon)$, with $\epsilon = 0.001$.)
}\label{Fig:F6}
\end{figure}

Finally, we consider the Hegselmann-Krause model, \cite{HK}, known also as bounded confidence model, where agents interact only with agents whose opinion lays within a certain range of confidence. Thus we define the following compromise function
\begin{align*}
P(w,w_*;c,c_*) = \chi_{\{|w-w_*|\leq\Delta(c)\}}(w_*)\quad \textrm{  with }\quad \Delta(c) = d_0\frac{c}{\cm},
\end{align*}
where the confidence level, $\Delta(c)$, is assumed to depend on the number of connections, so that agents with higher number of connections are prone to larger level of confidence.
We define the initial data 
\begin{align}\label{eq:initD}
f_0(w,c) = \frac{1}{2}\rho_\infty(c),
\end{align}
therefore the opinion is uniformly distributed on the interval $I=[-1,1]$ and it decreases along $c\in[0,\cm]$ following $\rho_\infty(c)$, as in \eqref{prop:st}, with parameters defined in Table \ref{tab:all_parameters} and $D(w,c) = 1-w^2$. 
%
Figure \ref{Fig:F6} shows the evolution of \eqref{eq:initD}, where $\Delta(c)$ creates an heterogeneous emergence of clusters with respect to the connectivity level: for higher level of connectivity consensus is reached, since the bounded confidence level is larger, instead for lower levels of connectivity multiple clusters appears, up to the limiting case $c=0$, where the opinions are not influenced by the consensus dynamics.

\section{Final considerations}
The mathematical modeling of opinion formation in multi-agent systems is nowadays a  well studied field of application of kinetic theory. Starting from some basic models \cite{PTa, T} we tried to enlighten some recent improvements, in which the models have been enriched by adding further aspects with the goal to better reflect various facts of our daily life.  
Particular emphasis has been done  to control strategies on opinion formation, a theme of paramount importance with several potential applications. Testimonials in advertising a product or opinion leaders during elections may lead the group of interacting agents towards a desired state and practically can modify the way a society behaves and is ruled by a government.
Furthermore, conviction  has been shown to be important in order to achieve a final personal opinion,  in view of the fact that a society with a high number of stubborn people clearly behaves very differently from a society composed by very susceptible persons. Last, the recent development and increasing importance of social networks made the study of opinions in this area a crucial and actual research theme. 

Finally, we mention here that the idea to study the role of opinion leaders in this kinetic setting has been first studied in \cite{DMPW} and subsequently improved by considering another independent variable which quantifies leadership qualities in \cite{DW}. 
Clearly, our point of view and the material we presented here gives only a selected partial view of the whole research in the field.  The interested reader can however find an almost exhaustive list of references which could certainly help to improve his knowledge on this interesting subject. 


\section*{Acknowledgement}
This work has been written within the activities of the National Groups of Scientific Computing (GNCS) and Mathematical Physics (GNFM)  of the National Institute of High Mathematics of Italy (INDAM). GA acknowledges the ERC-Starting Grant project High-Dimensional Sparse Optimal Control (HDSPCONTR). GT acknowledges the partial support of the MIUR project \emph{Optimal mass transportation, geometrical and functional inequalities with applications}.

\appendix
\section*{Appendix: Numerical simulation methods}
\addcontentsline{toc}{section}{Appendix}
In this short appendix we sketch briefly some particular numerical technique used to produce the various simulation results presented in the manuscript. We omit the description of the Monte Carlo simulation approach for the Boltzmann equation describing the opinion exchange dynamics addressing the interested reader  to \cite{PTa}. For the development of Monte Carlo methods that works in the Fokker-Planck regime we refer to \cite{APb}.

We first summarize the Monte Carlo approach used to deal with the evolution of the social network and then the steady state preserving finite-difference approach used for the mean-field models. More details can be found in \cite{APZd}.

\subsection*{Monte Carlo algorithm for the evolution of the network}\label{App:MC}
The evolution of the network is given by
\begin{equation*}\begin{cases}\vspace{0.5em}
\dfrac{d}{dt} f(w,c,t) + \N[f(w,c,t)]=0, \\
 f(w,c,0)=f_0(w,c).
\end{cases}\end{equation*}
Let $f^n = f(w,c,t^n)$ the empirical density function for the density of agents at time $t^n$ with opinion $w$ and connections $c$. For any given opinion $w$ we approximate the solution of the above problem at time $t^{n+1}$ by
\begin{equation*}\label{eq:MCmaster}
\begin{split}
f^{n+1}(w,c) =&\left(1-\Delta t\dfrac{{\Ur}(c+\beta)}{\gamma^n+\beta}-\Delta t\dfrac{ {\Ua}(c+\alpha)}{\gamma^n+\alpha}\right)f^n(w,c)\\
&+\Delta t  \dfrac{{\Ur}(c+\beta)}{\gamma^n+\beta}f^n(w,c-1)+\Delta t\dfrac{ {\Ua}(c+\alpha)}{\gamma^n+\alpha}f^n(w,c+1),
\end{split}\end{equation*}
with boundary conditions 
\begin{equation*}\begin{split}
f^n(w,0) & = \left(1-\Delta t\dfrac{ {\Ua}(c+\alpha)}{\gamma^n+\alpha}\right)f^n(w,0)+\Delta t\dfrac{ {\Ua}(c+\alpha)}{\gamma^n+\alpha}f^n(w,1), \\
f^n(w,\cm) & = \left(1-\Delta t\dfrac{{\Ur}(c+\beta)}{\gamma^n+\beta}\right)f^n(w,\cm)+\Delta t  \dfrac{{\Ur}(\cm+\beta)}{\gamma^n+\beta}f^n(w,\cm-1),
\end{split}\end{equation*}
and temporal discretization such that 
\begin{equation}\tag{A3}\label{eq:timeL}
\Delta t\le \min \left\{ \dfrac{\gamma^n+\beta}{{\Ur}(\cm+\beta)},\dfrac{\gamma^n+\alpha}{{\Ua}(\cm+\alpha)} \right\}.
\end{equation}
The algorithm to simulate the above equation reads as follows

\begin{alg}\label{MCmaster}~
  \begin{enumerate}
  \item Sample $(w^0_i,c^0_i)$, with $i=1,\ldots,N_s$, from the distribution $f^{0}(w,c)$.
  \item \texttt{for} $n=0$ \texttt{to} $n_{tot}-1$  
  \begin{enumerate}
  \item Compute   $\gamma^n =\frac{1}{N_s} \sum_{j=1}^{N_s}c^n_j$;
  \item Fix $\Delta t$ such that condition \eqref{eq:timeL} is satisfied.
   \item \texttt{for} $k=1$ \texttt{to} $N_s$  
     \begin{enumerate}
     \item Compute the  following probabilities rates 
     \[ p_k^{(a)} =\frac{\Delta t V_a(c_k^n+\alpha)}{\gamma^n+\alpha},\qquad p_k^{(r)} = \frac{\Delta t V_r(c_k^n+\beta)}{\gamma^n+\beta},\]
     \item Set $c^*_k =c_k^n$.
     \item \texttt{if} \,\, $0 \leq c_k^*\leq \cm-1$,\\ \quad with probability $p_k^{(a)}$ add a connection: $c_k^* = c_k^* +1$; 
      \item \texttt{if} \,\, $1 \leq c_k^*\leq \cm$,\\ \quad  with probability $p_k^{(r)}$ remove a connection: $c_k^* = c_k^* -1$;
     \end{enumerate}
     \item[]\texttt{end for}
  \item set $c^{n+1}_i  = c^*_i$, for all $i= 1,\ldots, N_s$.
\end{enumerate}
\item[]\texttt{end for}
  \end{enumerate}
  \label{ANMCS}
\end{alg}
\subsection*{Chang-Cooper type numerical schemes}\label{App:CC}
In the domain $(w,c)\in I\times{\mathcal C}$ we consider the Fokker-Planck system 
\begin{equation}\tag{A4}\label{eq:FP2}
\dfrac{\partial}{\partial t}f(w,c,t)+\N\left[f(w,c,t)\right]=\dfrac{\partial }{\partial w}\mathcal{F}[f],
\end{equation}
with zero flux boundary condition on $w$, initial data $f(w,c,0)=f_0(w,c)$ and
\begin{equation*}\label{eq:F}
\mathcal{F}[f] = \left(\mathcal{P}[f] + \sigma^2 D'(w,c)D(w,c)\right)f(w,c,t)+\dfrac{\sigma^2}{2}D(w,c)^2 \dfrac{\partial}{\partial w}f(w,c,t),
\end{equation*}
where $\mathcal P[f]$ is given by (\ref{eq:K}). Let us introduce a uniform grid $w_{i}=-1+i\Delta w$, $i=0,\ldots,N$ with $\Delta w = 2/N$, we denote by $w_{i \pm 1/2}=w_i \pm \Delta w/2$ and define
\[
f_{i}(c,t)=\frac{1}{\Delta w}\int_{w_{i+1/2}}^{w_{i-1/2}} f(w,c,t)\,dw.
\]
Integrating equation (\ref{eq:FP2}) yields
\begin{equation*}\begin{split}\label{eq:FPd}
\dfrac{\partial}{\partial t}f_{i}(c,t)+\N\left[f_{i}(c,t)\right]=\frac{\mathcal{F}_{i+1/2}[f]-\mathcal{F}_{i-1/2} [f]}{\Delta w},
\end{split}\end{equation*} 
where $\mathcal{F}_{i}[f]$ is the flux function characterizing the numerical discretization. We assume the Chang-Cooper flux function
\be\nonumber
\begin{split}
\label{eq:flux}
\mathcal{F}_{i+1/2}[f]=&\left((1-\delta_{i+1/2})(\mathcal{P}[f_{i+1/2}] + \sigma^2 D'_{i+1/2}D_{i+1/2})+\frac{\sigma^2}{2\Delta w}D^2_{i+1/2} \right)f_{i+1}\\
&+\left(\delta_{i+1/2} (\mathcal{P}[f_{i+1/2}] + \sigma^2 D'_{i+1/2}D_{i+1/2})-\frac{\sigma^2}{2\Delta w} D^2_{i+1/2} \right)f_{i},
\end{split}
\ee  
where $D_{i+1/2}=D(w_{i+1/2},c)$ and $D'_{i+1/2}=D'(w_{i+1/2},c)$. The weights $\delta_{i+1/2}$ have to be chosen in such a way that a steady state solution is preserved. Moreover this choice permits also to preserve nonnegativity of the numerical density.
The choice
\be\tag{A5}
\label{eq:weights}
\delta_{i+1/2}=\frac1{\lambda_{i+1/2}}+\frac{1}{1-\exp(\lambda_{i+1/2})}, 
\ee 
where
\be\nonumber
\lambda_{i+1/2}=\frac{2\Delta w}{\sigma^2}\frac{1}{D^2_{i+1/2}}\left(\mathcal{P}[f_{i+1/2}] + \sigma^2 D'_{i+1/2}D_{i+1/2}\right),
\ee
leads to a second order Chang-Cooper nonlinear approximation of the original problem. Note that here, at variance with the standard Chang-Cooper scheme \cite{CC}, the weights depend on the solution itself as in \cite{LLPS}. Thus we have a nonlinear scheme which preserves the steady state with second order accuracy. In particular, by construction, the weight in (\ref{eq:weights}) are nonnegative functions with values in $[0,1]$. 

Higher order accuracy of the steady state can be recovered using a more general numerical flux \cite{APZd}.

\end{document}